\documentclass[aps,prb,twocolumn,10pt]{revtex4-1}

\usepackage{graphicx, bm, amsmath, amsfonts,amssymb,amsthm, dsfont}

\usepackage{float}
\usepackage[usenames, dvipsnames]{color}
\usepackage{ulem}
\usepackage{amsmath, amssymb}
\usepackage{relsize}
\usepackage{footmisc}

\usepackage{hyperref}    
\hypersetup{linktocpage}

\newcommand{\ep}{\epsilon}

\newcommand{\Z}{\mathbb{Z}}

\newcommand{\lan}{\langle}
\newcommand{\ran}{\rangle}

\newcommand{\hatG}{{\hat{G}}}

\newcommand{\hatg}{{\hat{g}}}
\newcommand{\hath}{{\hat{h}}}

\newcommand{\hatU}{{\hat{U}}}
\newcommand{\hatW}{{\hat{W}}}
\newcommand{\hatP}{{\hat{P}}}

\newcommand{\hatX}{{\hat{X}}}
\newcommand{\hatZ}{{\hat{Z}}}
\newcommand{\hatS}{{\hat{S}}}
\newcommand{\hattau}{{\hat{\tau}}}
\newcommand{\hats}{{\hat{s}}}
\newcommand{\hatV}{{\hat{V}}}

\newcommand{\mz}{\mathbb{Z}}

\newcommand{\la}{\langle}
\newcommand{\ra}{\rangle}

\newcommand{\tG}{\tilde{G}}
\newcommand{\ts}{\tilde{s}}
\newcommand{\gammabar}{{\overline{\gamma}}}

\newcommand{\id}{{1}}

\newcommand{\Hstd}{{\hat{H}'_{\text{b}}}}
\newcommand{\Hspt}{{\hat{H}_{SPT}}}

\newcommand{\edge}[1]{{\lan #1 \ran}}

\graphicspath{{figures/}}

\begin{document}


\title{Disentangling interacting symmetry protected phases of fermions in two dimensions}

\author{Tyler D. Ellison}
\author{Lukasz Fidkowski}
\affiliation{Department of Physics, University of Washington, Seattle WA 98195, USA}

\begin{abstract}

We construct fixed point lattice models for group supercohomology symmetry protected topological (SPT) phases of fermions in $2+1$D.  A key feature of our approach is to construct finite depth circuits of local unitaries that explicitly build the ground states from a tensor product state.  We then recover the classification of fermionic SPT phases, including the group structure under stacking, from the algebraic composition rules of these circuits.  Furthermore, we show that the circuits are symmetric, implying that the group supercohomology phases can be many body localized.  Our strategy involves first building an auxiliary bosonic model, and then fermionizing it using the duality of Chen, Kapustin, and Radicevic.  One benefit of this approach is that it clearly disentangles the role of the algebraic group supercohomology data, which is used to build the auxiliary bosonic model, from that of the spin structure, which is combinatorially encoded in the lattice and enters only in the fermionization step.  In particular this allows us to study our models on $2$d spatial manifolds of any topology, and to define a lattice-level procedure for ungauging fermion parity.

\end{abstract}

\maketitle

\tableofcontents

\section{Introduction}

A major goal in understanding symmetry protected topological (SPT) phases is their classification, i.e. the identification and enumeration of the possible phases.  Essential to a classification scheme is the construction of microscopic models for each phase, as well as the identification of quantized many-body invariants which discriminate between the different phases.  For bosonic SPT phases in 2+1D with unitary onsite symmetries, the classification is well understood in terms of the framework of group cohomology theory.  The algebraic data of group cohomology is used both in the construction of exactly solvable lattice models [\onlinecite{Chen2013}] and in the identification of quantized invariants, where group cohomology classes appear in the universal statistics of the symmetry flux excitations [\onlinecite{LevinGu}].

In contrast, despite much recent progress [\onlinecite{Bhardwaj,GK,Gu_Wen,Chenjie_Gu,Cheng15, Bultinck17, Gu_Wang}], the classification of fermionic SPT phases is not as well understood.  A mathematical structure analogous to group cohomology - termed group supercohomology - was introduced in the pioneering work of Ref. [\onlinecite{Gu_Wen}] to describe a subset of fermionic SPT phases.  However, group supercohomology has yet to be as directly connected to explicit lattice Hamiltonians or to universal quantized observables.  While certain lattice fermionic Hamiltonians were, in fact, written down in terms of group supercohomology data in Ref. [\onlinecite{Gu_Wen}], these intricate constructions rely on seemingly arbitrary choices and cannot straightforwardly be put on spatial manifolds of general topology.  In a space-time path integral formalism, these arbitrary choices have since been interpreted as choices of spin structure [\onlinecite{GK}] -- now understood to be a crucial ingredient in constructing fixed point fermionic SPT models.  Progress has been made in incorporating spin structures directly in a Hamiltonian formalism [\onlinecite{Tarantino2015, Ware,GWW}], in particular in Ref. [\onlinecite{Gu_Wang}], where ground state wavefunctions incorporating spin structure were defined implicitly in terms of constraints that involve different lattice structures related by local deformations.  However, there is still no general prescription for turning group supercohomology data and a choice of spin structure into a fermionic Hamiltonian on a fixed lattice in a general spatial geometry.


Group supercohomology classes have also not yet been directly connected to quantized many-body invariants of gapped, lattice Hamiltonians.  It has been shown [\onlinecite{KapustinThorngren}], that the supercohomology data can be interpreted as quantized topological terms in the effective space-time action for a combination of the global symmetry and fermion parity gauge fields. That being the case, these space-time observables should in principle be encoded in the joint braiding statistics of symmetry and fermion parity fluxes, but such statistics have only been studied in the continuum [\onlinecite{CTW,Cheng15}].  For bosonic SPT phases, the underlying group cohomology data can be extracted using a well defined lattice minimal coupling gauging procedure that maps the SPT system to a system with topological order. An analogous lattice Hamiltonian procedure has so far been missing on the fermionic side - making it difficult to argue that group supercohomology classes are quantized invariants of lattice fermionic SPT Hamiltonians.

In this paper, we solve both of these problems in the case of 2+1 dimensions and finite unitary on-site symmetry $G \times \Z_2^{\text{f}}$, where $\mz_2^{\text{f}}$ is fermion parity. Specifically, we accomplish the following:

\vspace{2mm}
{\bf{(1)}} We construct a representative fermionic lattice SPT Hamiltonian for every choice of group supercohomology data, 2d oriented spatial manifold $M$, and spin structure on $M$.  Moreover, we write down an explicit finite depth quantum circuit of local unitaries that constructs its ground state from a trivial product state.

\vspace{2mm}
{\bf{(2)}} Using these finite depth circuits, we recover the group structure of our SPT phases under stacking.  We also find that two different sets of group supercohomology data can lead to circuits that differ only by a product of symmetric local unitaries, and hence define the same phase.  This leads to a natural equivalence relation on group supercohomology data, which matches that of previous works.  Conversely, we prove that for inequivalent group supercohomology data, the corresponding Hamiltonians are in distinct phases.

\vspace{2mm}
A choice of group supercohomology data is encoded in a pair $(n,\nu)$, where $n$ and $\nu$ are certain $\mz_2$ and $U(1)$-valued functions of $G$ variables, respectively (defined precisely in section \ref{gsd} below).  Given the data $(n,\nu)$, the construction of our fermionic lattice SPT Hamiltonian, inspired by the work of Ref. [\onlinecite{Bhardwaj}], proceeds in $3$ steps.

\vspace{1mm}
(i) We use $n$ and $\nu$ to construct an auxiliary {\textit{bosonic}} SPT Hamiltonian with enlarged symmetry group $\tG$, where $\tG$ is the extension of $G$ by $\Z_2$ determined by $n$.  $\tG$ contains $\Z_2$ as a subgroup and $G$ as a quotient: $G=\tG / \Z_2$, so the auxiliary bosonic SPT model has a global $\Z_2$ symmetry but is not in general $G$-symmetric.  Being a group cohomology bosonic SPT, it can be put on any spatial manifold $M$ with a triangulation and branching structure [\onlinecite{Chen2013}].

\vspace{1mm}
(ii) We gauge the $\Z_2 \subset \tG$ by minimally coupling the auxiliary bosonic SPT to a $\Z_2$ lattice gauge field and imposing a Gauss's law constraint.  By choosing an appropriate basis of gauge invariant operators, this gauge theory can be interpreted as an unconstrained bosonic model - which we refer to as the `shadow' model following Ref. [\onlinecite{Bhardwaj}] -  with global symmetry $G=\tG/\Z_2$ and toric code topological order.  Specifically, the shadow model has generalized $G$-spin vertex degrees of freedom, which transform under the $G$ symmetry in the standard way, and spin-$\frac{1}{2}$ link degrees of freedom, which encode a toric code topological order.

\vspace{1mm}

(iii) Finally, we obtain our fermionic SPT by applying the fermionization duality of Ref. [\onlinecite{Yu-an17}] (reviewed below) to trade the bosonic spin-$\frac{1}{2}$ link degrees of freedom in the shadow model for spinless complex fermions located on the triangular faces.  The underlying idea behind this fermionization is to represent the fermion as the bound state of a toric code charge and flux excitation [\onlinecite{LW2003, VC}].  The fermionization procedure is not unique, however, as it requires a choice of spin structure.  Spin structure enters our construction only here, encoded combinatorially in a certain subset of links $\cal{E}$. This step can be thought of as effectively `un-gauging' fermion parity symmetry [\onlinecite{Aasen17}], resulting in a model defined in a fermionic Fock space.
\vspace{1mm}

\begin{figure*}
\centering
\includegraphics[scale=.5,trim={.3cm 6cm 0cm 4cm},clip]{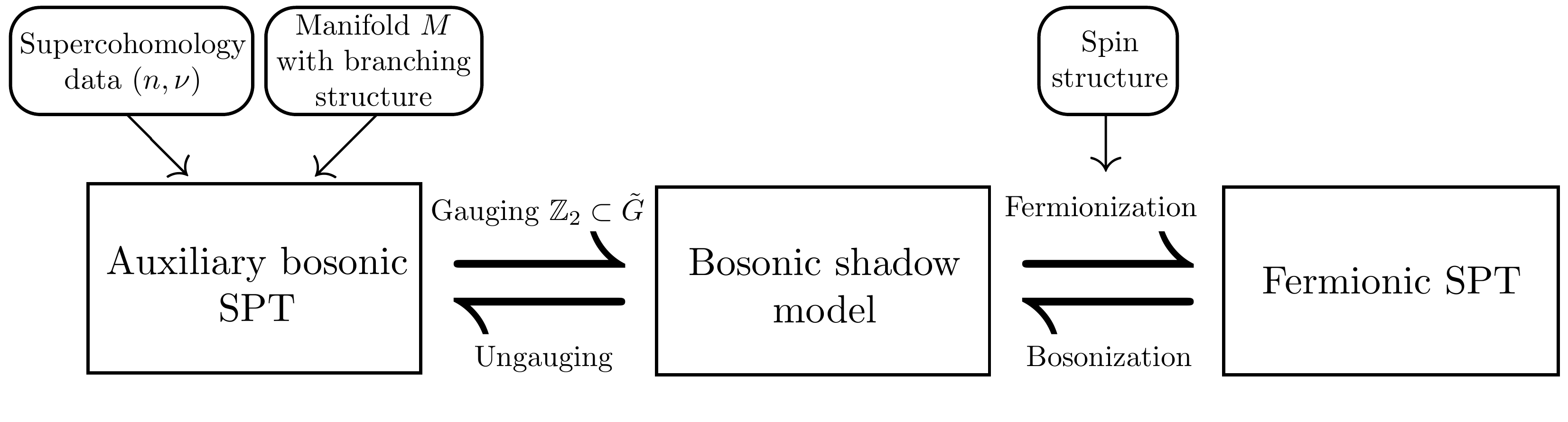}
\caption{Our construction of a fermionic SPT Hamiltonian begins with the input of supercohomology data and a choice of manifold ${M}$ with branching structure. This data is used to build an auxiliary bosonic SPT protected by a $\tG$ symmetry. Then we gauge the $\mz_2$ subgroup of $\tG$ to obtain the bosonic shadow model - a $G$ symmetry enriched toric code.  Finally, with a choice of spin structure, we fermionize the bosonic shadow model to arrive at the $G$ protected fermionic SPT.}
\label{fig:flowchart}
\end{figure*}

This three-step construction highlights one important advantage of our approach: it clearly disentangles the roles of group supercohomology data and spin structure in fermionic SPT models.  One needs just the group supercohomology data to construct the bosonic shadow model (steps $(i)$ and $(ii)$), whereas the spin structure enters only in the fermionization duality that maps this shadow model to the desired fermionic SPT (step $(iii)$).

A key part of our approach is the construction of finite depth quantum circuits of local unitaries\footnote{A finite depth quantum circuit of local unitaries is a unitary operator that can be expressed in the form \unexpanded{$\left ( \prod_j U_{n,j} \right ) \cdots \left( \prod_j U_{1,j} \right)$} where the unitaries \unexpanded{$U_{i,j}$} satisfy the following properties.  First, each \unexpanded{$U_{i,j}$} acts as the identity everywhere except on spins located in a disk of finite radius.  In this sense, \unexpanded{$U_{i,j}$} is a local unitary. Furthermore, for each $j \neq k$, $U_{i,j}$ has non-overlapping support with \unexpanded{$U_{i,k}$}.  The collection of unitaries sharing the first index define a `layer' of the quantum circuit.  That is, \unexpanded{$\prod_j U_{i,j}$} is the \unexpanded{$i^{\text{th}}$} layer of the quantum circuit. `Finite depth' means that the number of layers remains finite in the thermodynamic limit of large system size.} [\onlinecite{Chen_localunitaries}], which build the fermionic SPT ground states from a trivial product state.  Access to these finite depth circuits has several benefits.  First, they give us explicit representations of the corresponding ground states in terms of $G$ domain models decorated with fermions (as opposed to ground state wave functions that are only defined implicitly via constraints).  Second, we show that composing these circuits is equivalent to stacking the corresponding fermionic SPT phases, allowing us to extract the stacking group law for supercohomology data just by multiplying circuits.  Third, we show that equivalent group supercohomology data gives rise to circuits that differ by a product of symmetric local unitaries, and hence correspond to the same phase.  Conversely, by bosonizing our models and using well established classification results for bosonic symmetry enriched toric code phases [\onlinecite{Maissam2014,Tarantino2015,Teo_Hughes}], we show that inequivalent group supercohomology data always lead to inequivalent phases.

An intriguing feature of the finite depth circuit that builds our supercohomology fermionic SPT ground state is that, as a unitary operator, it is $G$-symmetric.  This is despite the fact that, when the SPT phase in question is nontrivial, the local unitaries that make it up cannot all be individually $G$-symmetric.  This is a property that the supercohomology models share with bosonic group cohomology models, but not with the so-called `beyond group cohomology' models (see {\textit{e.g.}} appendix C of Ref. [\onlinecite{PVF}]).  One consequence of this property is that the supercohomology phases can be many-body localized [\onlinecite{Basko, Pal, 1d_MBL_SPT_1, 1d_MBL_SPT_2, Chandran1, Bauer1, Potter_Vishwanath}].  This is done by disordering the couplings in a trivial commuting projector parent Hamiltonian for the trivial product state and then conjugating by the circuit.

The rest of this paper is structured as follows.  In section \ref{sec:bosonic}, we focus on the construction of the bosonic shadow model described in steps $(i)$ and $(ii)$ above.  In section \ref{sec:duality}, we review the bosonization duality of Ref. [\onlinecite{Yu-an17}] and complete step $(iii)$ of our construction.  In section \ref{sec:classification}, we study the group structure of fermionic SPT phases using the finite depth circuits that build their ground states. In particular, we derive a notion of equivalence of group supercohomology data (in agreement with Ref. [\onlinecite{Gu_Wang, Cheng15, Bhardwaj}]) such that equivalent data gives rise to models in the same phase and inequivalent data necessarily yields inequivalent phases.  We conclude in section \ref{sec:discussion} with some comments about many-body localizability for our models, possible future extensions of our work, and comparisons with other work.  Throughout the paper we illustrate our results for the simple case of $G=\Z_2$ (i.e. total symmetry $\Z_2 \times \Z_2^{\text{f}}$).  In Appendices \ref{shadowgsderivation}-\ref{ap:trivialcircuit}, we provide detailed derivations of the results in the main text.

\vspace{2mm}

As we were completing this work, we learned of a related preprint by N. Tantivasadakarn and A. Vishwanath [\onlinecite{Nat18}], which also constructs a many-body localizable model for the $\Z_2 \times \Z_2^{\text{f}}$ group supercohomology SPT.

\section{Bosonic shadow model from group supercohomology data} \label{sec:bosonic}

\begin{figure}
\centering
\includegraphics[scale=.27,trim={1cm 1.5cm 2cm 1.5cm},clip]{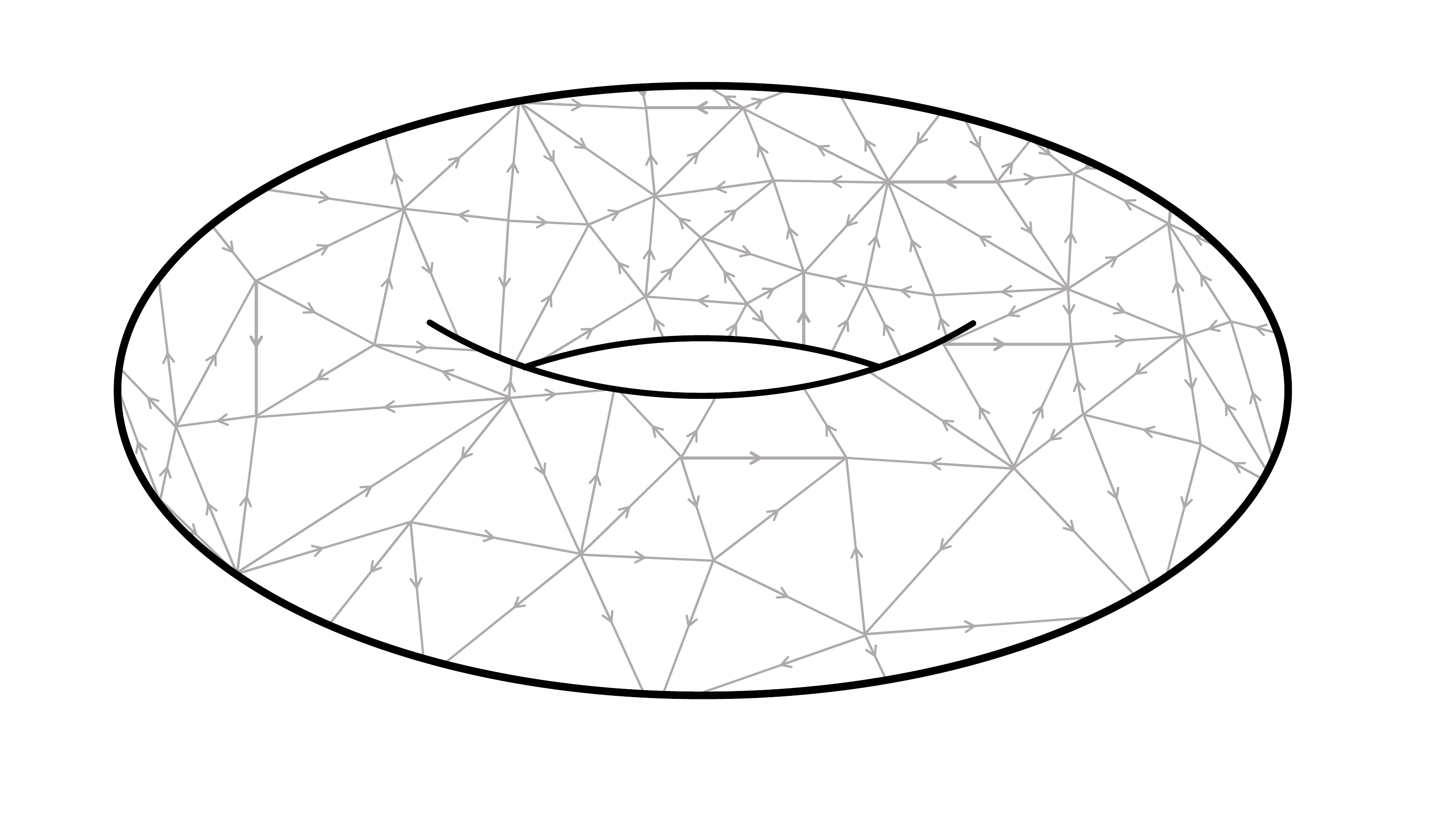}
\caption{All of the models constructed in this paper may be defined on an arbitrary triangulation of an orientable 2d manifold with a branching structure.  Note that a triangulation is a planar graph in which all faces are triangular.  Also, recall that a branching structure is an assignment of an orientation to each link such that there are no cycles around any of the triangles.}
\label{fig:torustriangulation}
\end{figure}

\begin{figure}
\centering
\includegraphics[scale=.28,trim={5cm 3cm 4cm 2cm},clip]{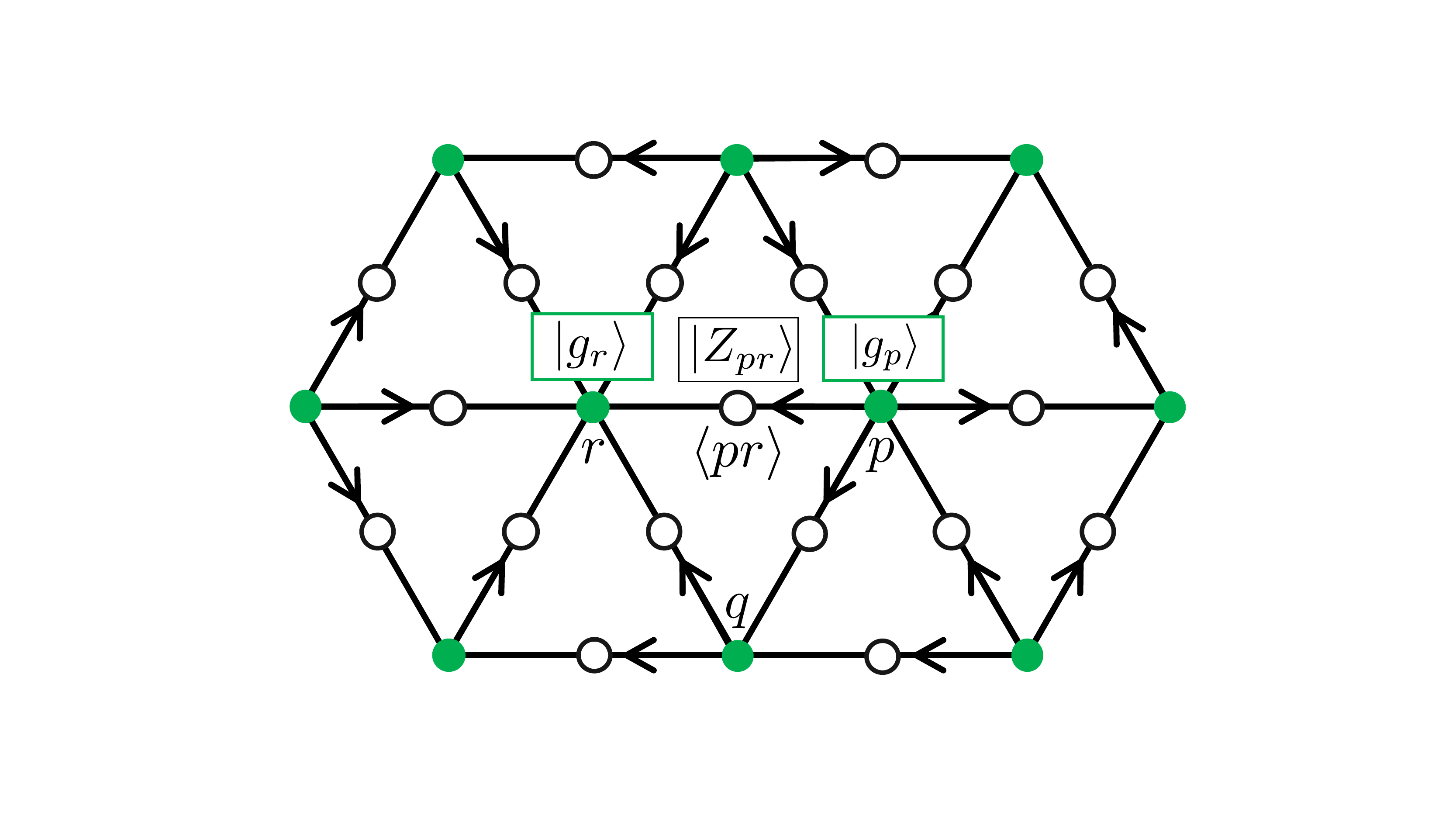}
\caption{The bosonic shadow model is defined on a Hilbert space with generalized $G$-spin degrees of freedom at vertices and spin-$\frac{1}{2}$ degrees of freedom on links. This gives the total Hilbert space: $\left( \bigotimes_p \mathbb{C}_p^{|G|} \right)\otimes \left( \bigotimes_{\la pq \ra} \mathbb{C}^2_{pq} \right)$.  A basis is given by configuration states $|\{ g_p \},\{ Z_{pq} \}\ra$, i.e. product states for which $g_p\in G$ is chosen for each vertex $p$ and $Z_{pq}=\pm 1$ is chosen for each link $\la pq \ra$.}
\label{fig:shadowdof}
\end{figure}

In this section we will show how to use group supercohomology data associated to a finite group $G$ to construct a purely bosonic Hamiltonian lattice model, which, in agreement with Ref. [\onlinecite{Bhardwaj}], we refer to as the shadow model.  The model is defined on a triangulation of a 2d manifold - i.e. a planar graph consisting of vertices $p$ and links $\la pq \ra$, all of whose faces are triangular - with branching structure (FIG. \ref{fig:torustriangulation}).  Recall that a branching structure is an assignment of an orientation to each link with the property that there are no cycles around any triangle.  The notation $\la pq \ra$ always denotes a link oriented from $p$ to $q$.  The Hilbert space will consist of generalized $G$-spin degrees of freedom $|g_p\ra$ at vertices $p$ and spin-$\frac{1}{2}$ degrees of freedom on links $\lan pq \ran$, with Pauli algebra generated by ${\hat X}_{pq}, {\hat Z}_{pq}$ (see FIG. \ref{fig:shadowdof}).

Before delving into the construction of the shadow model Hamiltonian, let us first provide some intuition for why a bosonic model built on such a Hilbert space can encode the physics of a fermionic SPT.  This intuition is based on interpreting the spin-$\frac{1}{2}$ link degrees of freedom as the Hilbert space of the usual commuting projector toric code Hamiltonian:
\begin{align}\label{toriccode0}
    \hat{H}^{\text{t.c.}}=-\sum_{p}\prod_{\la st \ra \ni p}\hatX_{st}-\sum_{\la pqr \ra}\hatZ_{pq} \hatZ_{qr} \hatZ_{pr},
\end{align}
where the product in the first sum above is over all oriented links $\la st \ra$ that contain the vertex $p$ (i.e. either $s=p$ or $t=p$).  A basis for this toric code Hilbert space can be obtained by specifying, for each basis state, the locations of all the vertex (`$e$') and triangular plaquette (`$m$') excitations, which are violations of the first and second terms in (\ref{toriccode0}), respectively.  The key idea is that the bound state of an $e$ and an $m$ excitation is a fermion, so a fermionic Hilbert space can effectively be constructed by restricting to the subspace where all of the $e$ excitations have been bound up with $m$ excitations into fermions.

\begin{figure}
\centering
\includegraphics[scale=.28,trim={3cm 4cm 2cm 3cm},clip]{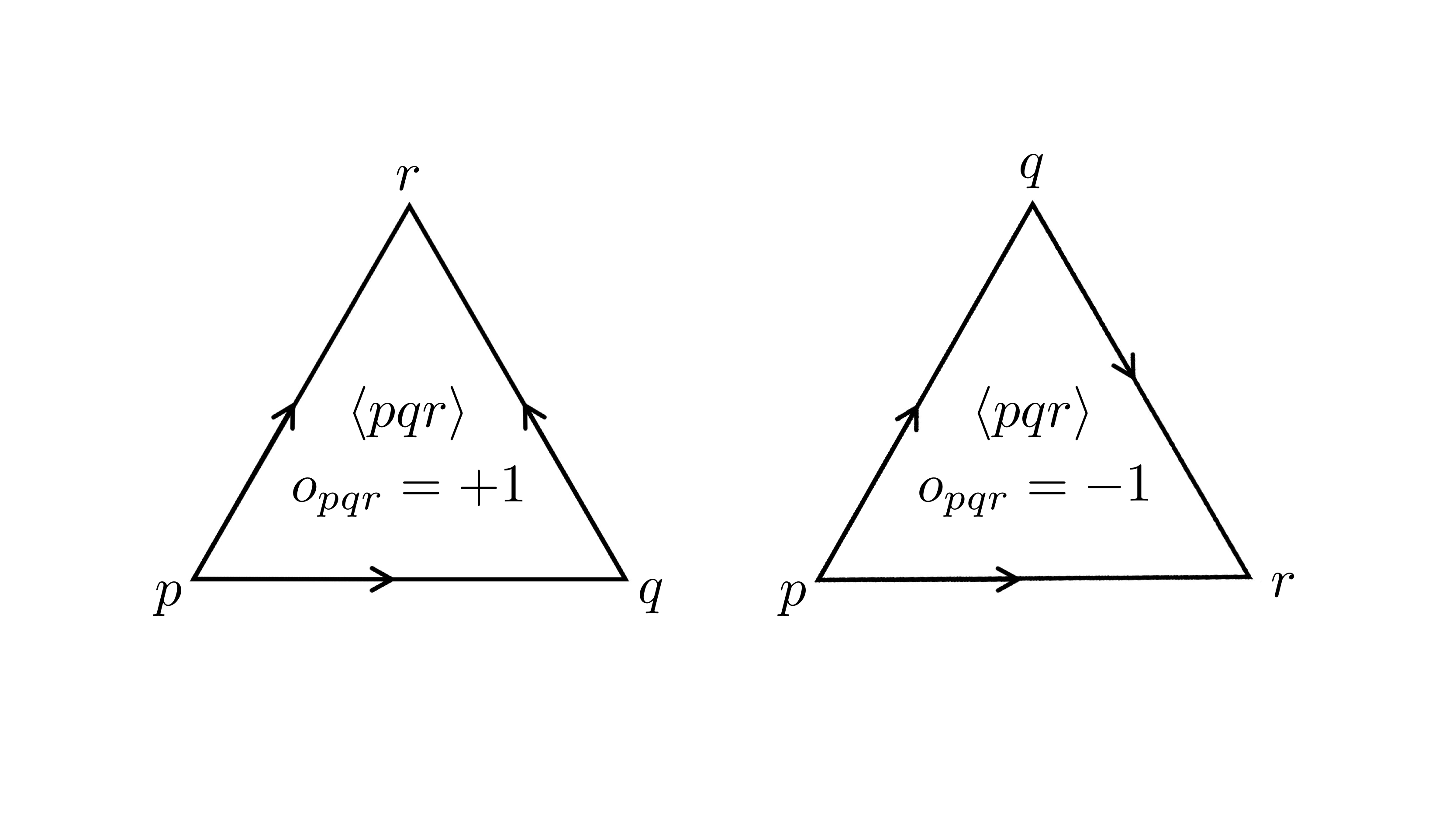}
\caption{The vertices around a triangle are ordered by the number of links pointing towards the vertex. $o_{pqr}$ is $-1$ or $+1$ depending on the orientation of the triangle relative to a choice of orientation for the manifold.}
\label{fig:orientations}
\end{figure}

Because the $e$ excitations live on vertices and the $m$ excitations live on plaquettes, there is some arbitrariness in defining their fermionic bound state.  This arbitrariness can be resolved by using the branching structure. Following Ref. [\onlinecite{Yu-an17}], we define a fermion on triangle $\la pqr \ra$ to be the bound state of an $m$ excitation on $\la pqr \ra$ with an $e$ excitation on its first vertex $p$.  Here the ordering $p,q,r$ of the vertices is specified uniquely by the branching structure (see FIG \ref{fig:orientations}).  The condition that all the $e$ excitations have been bound up with $m$ excitations into fermions in this way can then be stated as follows.  At each vertex $p$, the $\Z_2$ charge (i.e. number of $e$ excitations modulo $2$) measured at $p$ must be equal to the total $\Z_2$ flux (i.e. number of $m$ excitations modulo $2$) on all triangles $\la pqr \ra$ for which $p$ is the first vertex according to the branching structure.  Defining
\begin{align} \label{defW}
\hatW_{pqr}\equiv \hatZ_{pq} \hatZ_{qr} \hatZ_{pr}
\end{align}
\noindent to be the operator that measures the $\Z_2$ flux on $\la pqr \ra$, this is then just the condition that the state be in the $+1$ eigenspace of each operator 
\begin{align}\label{def:modifiedgausslaw}
{\hat G}_p \equiv \prod_{\substack{\la tqr \ra \\ t=p}} \hatW_{tqr} \prod_{\la st \ra \ni p} {\hat X}_{st}.
\end{align}
\noindent The first product above is over all triangles whose first vertex is $p$.  We thus expect that the shadow model Hamiltonian will commute with all of the $\hatG_p$ and that its ground states will lie in the $+1$ eigenspace of each $\hatG_p$.  Because of the second product in (\ref{def:modifiedgausslaw}), $\hatG_p$ resembles a Gauss's law constraint.  In accordance with Ref. [\onlinecite{Yu-an17}], we will refer to it as a `modified Gauss's law'.

Our construction of the shadow model Hamiltonian proceeds in two steps.  First, we use group supercohomology data to construct an auxiliary bosonic SPT, with an enlarged symmetry group $\tG$ equal to a certain $\Z_2$ extension of $G$.  Second, we gauge the global $\Z_2$ subgroup of $\tG$ in this auxiliary bosonic SPT to end up with our desired bosonic shadow model.  Again, we emphasize that because all of these constructions are bosonic, the spin structure does not enter into them at all.  To begin, we briefly review group supercohomology.

\subsection{Group supercohomology data} \label{gsd}
For a finite group $G$, group supercohomology data consists of a pair $(n,\nu)$, where $n : G \times G \times G \to \mz_2$ is a $\Z_2$ valued function of $3$ group variables, and $\nu: G \times G \times G \times G \to U(1)$ is a $U(1)$ valued function of $4$ group variables, satisfying the following two properties:  \\

\noindent 
1) $n$ is a homogeneous cocycle, where homogeneity means
\begin{align} \label{neq}
    n(gg_0,gg_1,gg_2)=n(g_0,g_1,g_2)
\end{align} for all $g$, and the cocycle property is \footnote{Let $h$ be a map from $G^k$ to $\mz_2$. The coboundary of $h$ is given by 
\begin{align}
    \unexpanded{\delta h(g_0,...,g_k)=\sum_{j=0}^k (-1)^j h(g_0,...,\widehat{g_j},...,g_k)} 
\end{align} where \unexpanded{$\widehat{g_j}$} means that $g_j$ is omitted.  Let $f$ be a map from $G^k$ to $U(1)$. Then the coboundary of $f$ is given by 
\begin{align}
    \unexpanded{\delta f(g_0,...,g_k)=\prod_{j=0}^k f(g_0,...,\widehat{g_j},...,g_k)^{(-1)^j}.}
\end{align}}
\begin{align} 
    \delta n=0.
\end{align}

\noindent 2) $\nu$ is homogeneous, i.e.
\begin{align} \label{nueq}
    \nu(gg_0,gg_1,gg_2,gg_3)=\nu(g_0,g_1,g_2,g_3),
\end{align}
for all $g$ and satisfies
\begin{align} \label{Steenrod}
    \delta \nu(g_0,g_1,g_2,g_3,g_4)=(-1)^{n(g_0,g_1,g_2)n(g_2,g_3,g_4)}.
\end{align}
Just as for ordinary group cocycles, there is an equivalence relation on group supercohomology data.  Rather than defining it now, we will postpone the discussion of this equivalence relation to section \ref{sec:superequivalencerelation}, where we identify it through physical arguments.  Group supercohomology classes will then be defined as equivalence classes of group supercohomology data modulo this relation. 


For convenience, in our constructions below, we will always take $n$ to be a normalized cocycle.  This is to say, we choose $n$ such that 
\begin{align} \label{normalized}
    n(g,g,h)=n(g,h,h)=0
\end{align}
\noindent for all $g,h$.  There is no loss of generality in restricting to normalized cocycles, because each equivalence class of group supercohomology data has a representative $(n,\nu)$ with $n$ normalized.



\subsection{Auxiliary bosonic SPT}

\begin{figure}
\centering
\includegraphics[scale=.28,trim={3cm 3cm 3cm 2cm},clip]{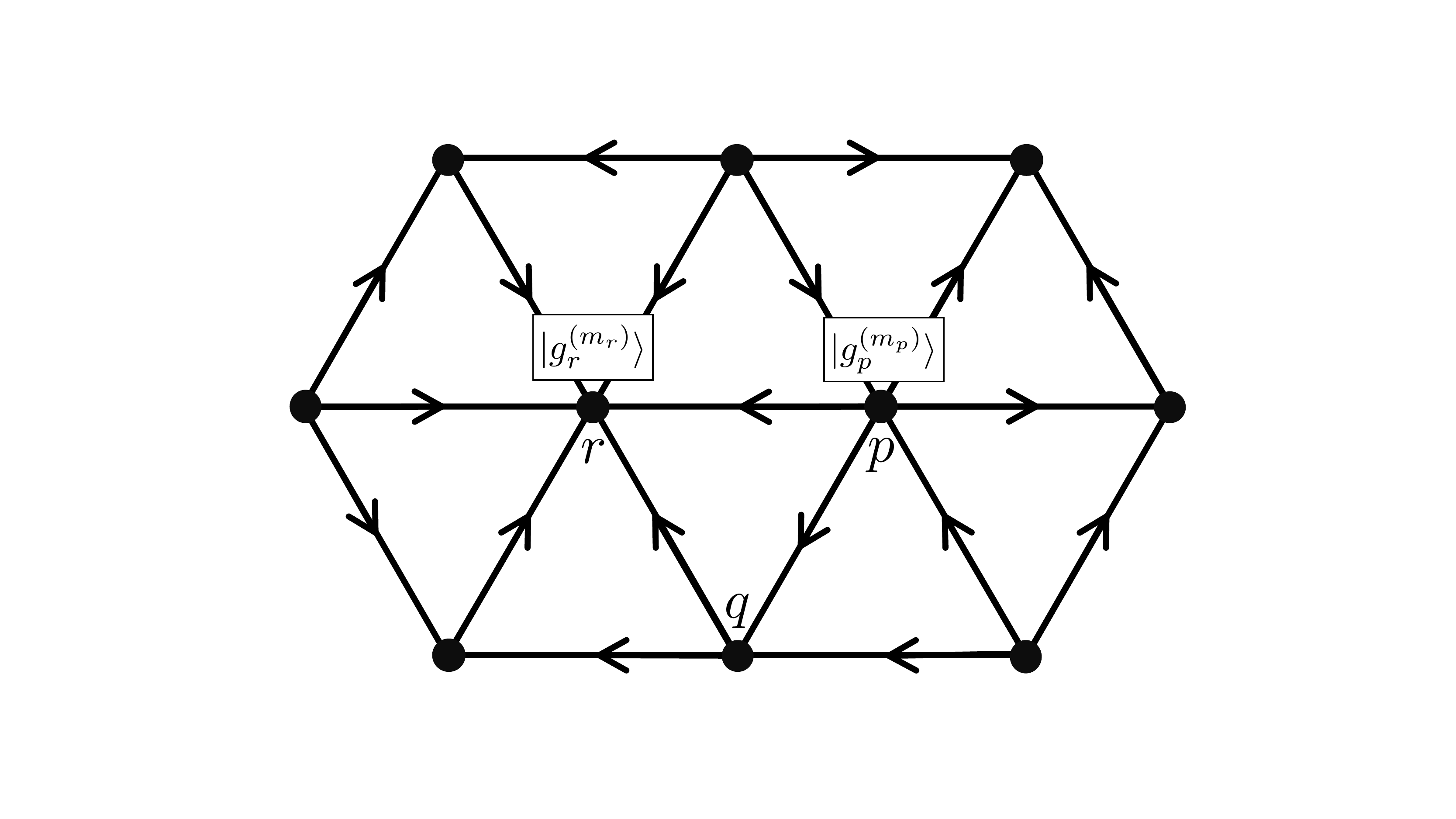}
\caption{The auxiliary bosonic SPT has $\tG$ degrees of freedom at each vertex.  Specifically, at each vertex, we attach a Hilbert space $\mathbb{C}^{|\tG|}$
with a basis labeled by elements of $\tG$. A natural basis for the total Hilbert space $\bigotimes_p \mathbb{C}_p^{|\tG|}$ is then a set of product states for which at each vertex an element of $\tG$ is chosen.  We refer to this basis of product states as the configuration basis. An arbitrary element of the configuration basis may be written as $\big|\big\{ g^{(m_p)}_p \big\} \big\ra$.}
\label{fig:auxiliarySPTdof}
\end{figure}

The auxiliary bosonic SPT is again defined on a triangulation of an orientable two dimensional spatial manifold $M$ together with a branching structure.  The symmetry group $\tG$ of the auxiliary bosonic SPT is the $\Z_2$ extension of $G$ determined by $n$.  Explicitly, $\tG$ consists of $2|G|$ elements ${g^{(m)}}$, where $g\in G$ and $m \in \mz_2 = \{0,1\}$, obeying the group law: 
\begin{align} \label{grouplaw}
{g^{(m)}} {h^{(\ell)}} = {(gh)^{(m+\ell+n(1,g,gh))}}.
\end{align}

\noindent The degrees of freedom in the auxiliary model are generalized $\tG$-spins $\big|g^{(m_p)}_p\big\ra$ living on the vertices of the triangulation, and the standard bosonic SPT construction of Ref. [\onlinecite{Chen2013}] allows us to write down the following SPT ground state wave function in the $\big\{g^{(m_p)}_p\big\}$ configuration basis:
\begin{align} \label{gsSPT}
    \Psi_{\text{SPT}}\big(\big\{g^{(m_p)}_p\big\}\big)=& \big \la \{ g^{(m_p)}_p \} \big| \Psi_{\text{SPT}} \big \ra =  \\ \nonumber = &\prod_{\la pqr \ra}\alpha\big(g^{(m_p)}_p,g^{(m_q)}_q,g^{(m_r)}_r,1\big)^{o_{pqr}}.
\end{align}
Here, as below, we do not keep track of the irrelevant overall normalization factor of the ground state wave function.  The product in (\ref{gsSPT}) is over ordered triangles $\la pqr \ra$, with the ordering determined by the branching structure. $o_{pqr}$ is $+1$ if the orientation of the triangle $\la pqr \ra$ is aligned with the orientation of the manifold and $-1$ otherwise (see FIG. \ref{fig:orientations}).  Finally, $\alpha$ is defined in terms of the group supercohomology data as [\onlinecite{Bhardwaj}]: 
\begin{align} \label{def_alpha}
\alpha\big({g^{(m_0)}_0}&, {g^{(m_1)}_1}, {g^{(m_2)}_2}, {g^{(m_3)}_3}\big) \equiv \\
& \nu(g_0,g_1,g_2,g_3) (-1)^{\ep\big(\big({g^{(m_0)}_0}\big)^{-1} {g^{(m_1)}_1}\big) n(g_1,g_2,g_3)}, \nonumber 
\end{align}
\noindent where we have defined the projector 
\begin{align}
\ep \big({g^{(m)}}\big) \equiv m.
\end{align}
\noindent One can explicitly verify that $\alpha$ is homogenous and a cocycle ($\delta \alpha=0$) by using equations (\ref{neq}) and (\ref{nueq}) along with the group law (\ref{grouplaw}) of $\tG$, as well as the normalization property (\ref{normalized}). Thus (\ref{gsSPT}) is a bosonic SPT ground state.  The seemingly complicated cocycle $\alpha$ is designed to produce a shadow model wave function that lies in the $\hatG_p=+1$ Hilbert space, as we will see in the next subsection.

\subsection{Bosonic shadow model wave function}\label{tGSPT}

We now construct the bosonic shadow model by gauging the $\Z_2$ subgroup of $\tG$ in the auxiliary bosonic SPT.  This is done in the standard way by introducing a lattice $\Z_2$ gauge field $\mu^z_{pq}=\pm 1$ and performing the usual minimal coupling procedure [\onlinecite{LevinGu}], so we relegate the details to Appendix \ref{shadowgsderivation}.  A complete set of commuting gauge invariant observables in the resulting gauge theory is given by $\{g_p,Z_{pq} \}$, where $g_p$ is the $G$ component of the $\tG$ degree of freedom $g^{(m_p)}_p$ at vertex $p$, and
\begin{align}
Z_{pq}&=\mu^z_{pq} (-1)^{\ep\big(\big(g^{(m_p)}_p\big)^{-1} g^{(m_q)}_q\big)}
\end{align}
can be thought of as the $\Z_2$ part of the lattice gauge covariant derivative of the $\tG$ `matter' fields.  We explicitly demonstrate in Appendix \ref{shadowgsderivation} that this gauge theory Hilbert space is isomorphic, via a duality transformation, to the unconstrained Hilbert space of generalized $G$-spin degrees of freedom $|g_p\ra$ at vertices $p$ and spin-$1/2$ degrees of freedom on links $\lan pq \ran$, with Pauli algebra generated by ${\hat X}_{pq}, {\hat Z}_{pq}$.

\begin{figure*}[t]
\centering
\includegraphics[scale=.5,trim={.6cm 13cm 0cm 2.5cm},clip]{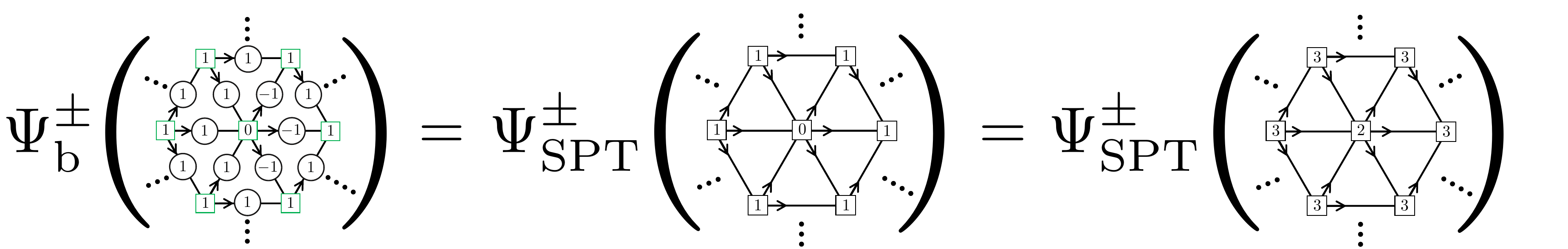}
\caption{Pictured above is an example of the amplitude $\Psi^\pm_\text{b}(\{ g_p \},\{ Z_{pq} \})$ for the $G=\mz_2$ case with nontrivial $n$, defined in (\ref{setgs2}).  Here, the argument of $\Psi^\pm_\text{b}$ is a particular configuration with a single $|0\ra$ vertex and all other vertices in the $|1\ra$ state.  The link degrees of freedom are $|1\ra$ everywhere besides the three $|-1\ra$ valued links illustrated in the figure.  The ellipses denote the fact that $\Psi^\pm_\text{b}$ is dependent on the global configuration despite the fact that we have only shown a local portion of the configuration. The amplitude $\Psi^\pm_\text{b}(\{ g_p \},\{ Z_{pq} \})$ is equivalent to $\Psi^\pm_\text{SPT}\big( \big \{ g^{(m_p)}_p \big \} \big)$ if there exists $\big \{ g^{(m_p)}_p \big \}$ such that $Z_{pq}=(-1)^{\ep\big( \big(g^{(m_p)}_p \big)^{-1} g^{(m_q)}_q \big)}$ for all $\la pq \ra$.  In this example, with $\tG = \Z_4$, $\ep(0)=\ep(3)=0$, and $\ep(1)=\ep(2)=1$, two such configurations exist. One has a single $|0\ra$ vertex with all other vertices $|1\ra$ while the other has a single $|2\ra$ vertex with all other vertices $|3\ra$. These two configurations differ by the square of the $\mz_4$ global symmetry generator, so due to the fact that $\Psi^\pm_\text{SPT}\big( \big \{ g^{(m_p)}_p \big \} \big)$ is $\Z_4$-symmetric, they give the same amplitude.}
\label{fig:z2amplitude}
\end{figure*}

A ground state wave function $\Psi_{\text{b}}$ of the gauged theory can be obtained by setting the amplitude $\Psi_{\text{b}}(\{g_p, Z_{pq}\})$ of any configuration $\{g_p, Z_{pq}\}$ equal to $\Psi_{\text{SPT}}\big(\big\{g_p^{(m_p)}\big\}\big)$ if there exists $\big \{g_p^{(m_p)} \big\}$ for which
\begin{align} \label{zpq}
Z_{pq}&=(-1)^{\ep\big(\big(g^{(m_p)}_p\big)^{-1} g^{(m_q)}_q\big)}=(-1)^{m_p+m_q+n(1,g_p,g_q)}
\end{align}
and zero otherwise (see FIG. \ref{fig:z2amplitude} and \ref{fig:z2amplitude2} for an example).  Such $\big\{g^{(m_p)}_p\big\}$, if it exists, is ambiguous only up to a global $\Z_2$ transformation, i.e. a shift $m_p\rightarrow m_p+1$,\footnote{Here we use the fact that $n$ is a normalized $2$-cocycle.} and since $\Psi_{\text{SPT}}$ is invariant under this shift, $\Psi_{\text{b}}$ is well defined.  Explicitly,
\begin{align}\label{gsHstd}
     &\Psi_{\text{b}}\left(\{g_p, Z_{pq}\}\right)= \\ \nonumber
     &\prod_{\la pqr \ra} \nu(g_p,g_q,g_r,1)^{o_{pqr}} Z_{pq}^{n(g_q,g_r,1)} \\ \nonumber
     \times  &  \left(\prod_{\la pqr \ra}  \delta_{Z_{pq}Z_{qr}Z_{pr},(-1)^{n(g_p,g_q,g_r)}}\right) h(\{Z_{pq}(-1)^{n(1,g_p,g_q)}\}), 
\end{align}
as can be verified by observing that we recover the auxiliary bosonic SPT ground state wave function amplitude by inserting (\ref{zpq}) in (\ref{gsHstd}).  Again, we do not keep track of the irrelevant overall normalization of the wave function.  The function $h(\{Z_{pq}(-1)^{n(1,g_p,g_q)}\})=0,1$ is a constraint that enforces trivial $\mu^z$-holonomy around each topologically nontrivial cycle in the geometry.  Specifically, it is equal to a product of delta functions over all nontrivial cycles, which enforce the constraint that the product of $Z_{pq}(-1)^{n(1,g_p,g_q)}$ along the links of the cycle is equal to $1$.  These holonomy constraints, together with the delta functions in (\ref{gsHstd}), ensure that the amplitude of a given configuration $\{g_p, Z_{pq}\}$ is nonzero if and only if there exists $\big\{g_p^{(m_p)}\big\}$ satisfying (\ref{zpq}).  Once we write down a parent Hamiltonian for $\Psi_{\text{b}}$, we will have other ground states, which will all be of the form (\ref{gsHstd}) except with nontrivial holonomy constraints.


Because it comes from gauging a global $\Z_2$ symmetry in a short range entangled state, the shadow model wave function $\Psi_{\text{b}}$ describes a toric code topological order.  Furthermore, since $\Psi_{\text{SPT}}$ is $\tG$ symmetric, $\Psi_{\text{b}}$ is $G$ symmetric, and hence the shadow model wave function describes a $G$-symmetry enriched toric code. One can also explicitly check that
\begin{align} \label{eq:Gpverify}
\hatG_p |\Psi_{\text{b}}\ra = |\Psi_{\text{b}}\ra
\end{align}
for all $p$, so that $|\Psi_{\text{b}}\ra$ contains only fermion excitations, without any unbound $e$ excitations or $m$ excitations, in the sense defined above.  We will also verify (\ref{eq:Gpverify}) below by writing down a finite depth circuit of local unitaries which commutes with all of the $\hatG_p$, and constructs $\Psi_{\text{b}}$ from a state which trivially lies in the $\hatG_p=+1$ eigenspace of each $\hatG_p$.

\subsection{Bosonic shadow model Hamiltonian from a finite depth circuit} \label{shadowhamiltonian}


Our ultimate aim is to use the fermionization duality of Ref. [\onlinecite{Yu-an17}] to turn the bosonic shadow model wave function into the ground state of a fermionic SPT.  However, as this fermionization duality is defined at the level of local operators, we must first write down a local parent Hamiltonian for $|\Psi_{\text{b}}\ra$ on which we can apply the duality.

One way to obtain such a parent Hamiltonian is to simply start with the form of the bosonic $\tG$ SPT parent Hamiltonian written down in Ref. [\onlinecite{Chen2013}] and directly couple it to a lattice $\Z_2$ gauge field.  We outline this approach in appendix \ref{shadowgsderivation}, but for our purposes, we will find it more useful to construct a different parent Hamiltonian for $|\Psi_{\text{b}}\ra$.

Our choice of parent Hamiltonian is based on the insight that $|\Psi_{\text{b}}\ra$, as defined by the wavefunction in (\ref{gsHstd}), can be obtained by applying an appropriate finite depth circuit of local unitaries to a ground state of the following Hamiltonian, which describes a trivial generalized $G$-spin paramagnet and a decoupled copy of the toric code:
\begin{align}\label{toriccode}
    \hat{H}^0_{\text{b}}=-\sum_p \hatP_p^{\text{sym}}-\sum_{p}\prod_{\la st \ra \ni p}\hatX_{st}-\sum_{\la pqr \ra}\hatW_{pqr}.
\end{align}
\noindent Here, $\hatP_p^{\text{sym}}$ is the projector onto the symmetric state $\frac{1}{\sqrt{|G|}}\sum_{g_p\in G}|g_p\ra$ at vertex $p$ tensored with the identity on the remaining sites, and $\hatW_{pqr}$, which was defined in (\ref{defW}), measures the $\Z_2$ flux on $\la pqr \ra$.  One ground state of (\ref{toriccode}) is 
\begin{align}\label{tc}
     \Psi_{\text{t.c.}}(\{g_p, Z_{pq}\})= 
      \left(\prod_{\la pqr \ra}  \delta_{Z_{pq}Z_{qr}Z_{pr},1}\right)h(\{Z_{pq}\}),
\end{align}
\noindent where the holonomy constraint $h(\{Z_{pq}\})$ was defined below (\ref{gsHstd}).

We now claim that
\begin{align} \label{eq:tcSET}
|\Psi_{\text{b}}\ra=\hatU_{\text{b}} |\Psi_{\text{t.c.}}\ra,
\end{align}
\noindent where $\hatU_{\text{b}}$ is the following finite depth circuit of local unitaries:
\begin{align}\label{ucirc}
   \hatU_{\text{b}}=\prod_{\la pqr \ra} \left(\hat{\nu}_{pqr}^{o_{pqr}}  \hatZ_{pq} ^{\hat{n}_{qr}}\right) 
    \prod_{\la pq \ra}\hatX_{pq}^{\hat{n}_{pq}}\prod_{\la pqr \ra}\hatW_{ pqr }^{\hat{n}_{pr}}.
\end{align}
\noindent Here, $\hat{n}_{pq}$ is the operator defined by 
\begin{align}
    \hat{n}_{pq}\left|\{g_t\}\right\ra=n(g_p,g_q,\id)|\{g_t\}\ra,
\end{align} 
and $\hat{\nu}_{pqr}^{o_{pqr}}$ is given by
\begin{align}
    \hat{\nu}^{o_{pqr}}_{pqr}|\{g_t\}\ra=\nu(g_p,g_q,g_r,\id)^{o_{pqr}}|\{g_t\}\ra.
\end{align} 

To see that $|\Psi_{\text{b}}\ra=\hatU_{\text{b}} |\Psi_{\text{t.c.}}\ra$, first note that the all of the configurations appearing with non-zero amplitude in $|\Psi_{\text{t.c.}}\ra$ have trivial $\Z_2$-flux through all triangles, while the states in (\ref{gsHstd}) have nontrivial $\Z_2$-flux at triangles $\la pqr \ra$ for which $(-1)^{n(g_p,g_q,g_r)}=-1$.  This difference is remedied by the term $\prod_{\la pq \ra}\hatX_{pq}^{\hat{n}_{pq}}$ in (\ref{ucirc}).  The cocycle condition $\delta n=0$ guarantees that the nontrivial $\Z_2$-fluxes are put into the correct positions by this term.  Second, the term
$\prod_{\la pqr \ra} \hat{\nu}_{pqr}^{o_{pqr}}  \hatZ_{pq}^{\hat{n}_{qr}}$ is simply to ensure that the phases assigned to configurations match those in $|\Psi_{\text{b}}\ra$.  

It is proved in Appendix \ref{ap:symH} that $\hatU_{\text{b}}$ is nearly $G$-symmetric - conjugating it by any global symmetry generator yields $\hatU_{\text{b}}$ multiplied by a product of some $\hatG_p$ operators.  This property of $\hatU_{\text{b}}$ in particular relies on the term $\prod_{\la pqr \ra}\hatW_{ pqr }^{\hat{n}_{pr}}$ in (\ref{ucirc}), which may have seemed unnecessary at first since it acts trivially on the toric code ground states.

Together with the manifest $G$ and $\hatG_p$ invariance of $\hat{H}^0_{\text{b}}$, this property of $\hatU_{\text{b}}$ implies that 
\begin{align} \label{Hbos}
    \hat{H}_{\text{b}}=\hatU_{\text{b}}\hat{H}^0_{\text{b}}\hatU_{\text{b}}^\dag,
\end{align}
is a $G$-symmetric parent Hamiltonian for $|\Psi_{\text{b}}\ra$.  We will see in section \ref{sec:fSPT} that $ \hatU_{\text{b}}$ also commutes with all $\hatG_p$, so that the $\hat{H}_{\text{b}}$ does as well.  We have thus constructed, using group supercohomology data, a bosonic shadow model Hamiltonian that commutes with all of the $\hatG_p$, and whose ground states all satisfy $\hatG_p=+1$.  This bosonic shadow model describes a $G$-symmetry enriched toric code phase.

\begin{figure}
\centering
\includegraphics[scale=.5,trim={0cm 7cm 23cm 9cm},clip]{fSPT_figures2/z2amplitude.pdf}
\caption{Here we show the evaluation of $\Psi^\pm_\text{b}$ on a specific configuration with one vertex in the $|0\ra$ state and all other vertices, including those not pictured, in the $|1\ra$ state along with $|1\ra$ states at every link Hilbert space.  The amplitude of this configuration is zero because there is no configuration $\big \{ g_p^{(m_p)} \big \}$ such that (\ref{zpq}) is satisfied.  This can be seen by the fact that the product of $Z_{pq}$ around either one of the two shaded triangles is $1$ while the product of $(-1)^{\ep\big( \big(g^{(m_p)}_p \big)^{-1} g^{(m_q)}_q \big)}$ around either of these triangles is $(-1)^{n(g_p,g_q,g_r)}=-1$.}
\label{fig:z2amplitude2}
\end{figure}

\subsection{Example: $G=\mz_2$} \label{Z2sec}

Let us describe the above constructions for the simplest nontrivial examples of supercohomology phases, which occur for $G=\Z_2$ (i.e. total symmetry group $\Z_2 \times \Z_2^{\text{f}}$).  In contrast to the case of general $G$, where we used multiplicative notation for the group law, in the case $G=\Z_2$, we will use additive notation and denote $\Z_2$ elements by $s=0,1$.

For $G=\Z_2$, there are four inequivalent supercohomology classes.  Two of these have trivial $n$ and correspond to the trivial phase and the purely bosonic $\Z_2$ SPT.  The other two both have the same nontrivial $n$:
\begin{align}\label{n2z2}
n(s_0,s_1,s_2) =
    \begin{cases}
     1 & (s_0,s_1,s_2)=(0,1,0)\\
     1&  (s_0,s_1,s_2)=(1,0,1) \\
    0 & \text{otherwise,}
    \end{cases}
\end{align}
\noindent but different $\nu$:
\begin{align}\label{nu3z2}
    \nu_{\pm}(s_0,s_1,s_2,s_3) = 
    \begin{cases}
    \pm i & (s_0,s_1,s_2,s_3)=(1,0,1,0) \\
    \pm i & (s_0,s_1,s_2,s_3)=(0,1,0,1) \\
    1 & \text{otherwise.}
    \end{cases}
\end{align}

\noindent This data defines two possible phases according to the choice of sign in (\ref{nu3z2}), which turn out to be the index $2$ and $6$ members of the $\Z_8$ interacting classification in this symmetry class [\onlinecite{Qi, Ryu_Zhang, Gu_Levin}].  The $\Z_2$ extension of $G=\Z_2$ defined by $n$ is $\tG=\Z_4$, and $\ep$ is $\ep(0)=\ep(3)=0$, $\ep(1)=\ep(2)=1$.  Explicitly computing the cocycle $\alpha$ defined in (\ref{def_alpha}), we obtain
\begin{align}\label{alpha3Z2}
\alpha_{\pm}(\ts_0,\ts_1,\ts_2,\ts_3)=
(\pm i)^{(\ts_0-\ts_1)(\overline{\ts_1-\ts_2}) (\overline{\ts_2-\ts_3})},
\end{align}
where $\ts \in \mz_4$ and the overline denotes reduction modulo $2$.

The corresponding auxiliary $\Z_4$ SPT wave function is:
\begin{align} 
\Psi^{\pm}_{\text{SPT}}(\{\ts_p\}) =  \prod_{\lan pqr \ran} \alpha_{\pm}(\ts_p,\ts_q,\ts_r,0)^{o_{ pqr }}.
\end{align}

\noindent The bosonic shadow Hilbert space has $\Z_2$ degrees of freedom $s_p=0,1$ on vertices $p$ and spin-$\frac{1}{2}$ degrees of freedom $Z_{pq} = \pm 1$ on links $\la pq \ra$.  The shadow model ground states are (see FIG. \ref{fig:z2amplitude} and \ref{fig:z2amplitude2})
\begin{align}\label{setgs2}
     &\Psi^{\pm}_{\text{b}}(\{s_p\},\{ Z_{pq}\})= \\ \nonumber
     &\prod_{\la pqr \ra} \nu_{\pm}\left(s_p,s_q,s_r,0 \right)^{o_{pqr}}  Z_{pq}^{n \left(s_q,s_r,0 \right)} \\ \nonumber
     \times  &\left( \prod_{\la pqr \ra}  \delta_{Z_{pq}Z_{qr}Z_{pr},(-1)^{n\left(s_p,s_q,s_r\right)}} \right)  h(\{ Z_{pq}(-1)^{n(1,s_p,s_q)}\}).
\end{align}
\noindent Here, $h(\{Z_{pq}\})=0,1$ is a function that projects onto a choice of holonomy of the $\Z_2$ gauge field.

Using the explicit form of the supercohomology data $n(s_p,s_q,0) = (1-s_p)s_q$ and $\nu_{\pm}(s_p,s_q,s_r,0)=(\pm i)^{s_p(1-s_q)s_r}$, we see that the circuit (\ref{ucirc}) becomes
\begin{align}\label{z2circuit}
   \hatU^{\pm}_{\text{b}}=&\prod_{\la pqr \ra} (\pm i)^{o_{pqr}\hats_p(1-\hats_q)\hats_r}  \hatZ_{pq} ^{(1-\hats_q)\hats_r} \\ \nonumber
    \times &\prod_{\la pq \ra}\hatX_{pq}^{(1-\hats_p)\hats_q}\prod_{\la pqr \ra}\hatW_{ pqr }^{(1-\hats_p)\hats_r}.
\end{align}
From this circuit, we obtain the Hamiltonian
\begin{align}
\hat{H}^{\pm}_{\text{b}} = \hatU^{\pm}_{\text{b}}\hat{H}^0_{\text{b}}(\hatU^{\pm}_{\text{b}})^\dag
\end{align}
\noindent for the gauged model.

For completeness, we note that the global $\Z_2$ symmetry generator in the gauged model acts by 
\begin{align}
|\{s_p,Z_{pq} \} \ra \rightarrow |\{1-s_p,Z_{pq} \} \ra.
\end{align}
This is just the descendant of the $\Z_4$ generator in the $\Z_4$ SPT.

\section{Fermionizing the shadow model} \label{sec:duality}

In the previous section, we used the supercohomology data to construct a bosonic shadow model ${\hat{H}}_{\text{b}}$ on a Hilbert space consisting of generalized $G$-spin degrees of freedom on vertices $p$ and spin-$\frac{1}{2}$ degrees of freedom on links $\lan pq \ran$. In this section, we describe how this bosonic model may be fermionized, i.e. rewritten in terms of local fermionic operators.  This fermionization is effectively a procedure for `un-gauging' fermion parity symmetry. Equivalently, it can be viewed as a prescription for a lattice level fermion condensation (see Appendix \ref{ap:fermioncondensation} for further detail).  We emphasize that this is the only point at which a choice of spin structure enters the construction.

Focusing just on the spin-$\frac{1}{2}$ link degrees of freedom, we utilize the fermionization prescription developed in Ref. [\onlinecite{Yu-an17}], reviewed in the next three subsections, which provides an exact duality between the local operator algebra of a bosonic model and that of a fermionic model.  To define this duality, one must specify some combinatorial data, which we show amounts to a choice of spin structure for the spatial manifold $M$.  We will first define the local bosonic and fermionic operator algebras $\cal{A}_\text{bos}$ and $\cal{A}_\text{fer}$, respectively, and then construct the spin-structure dependent duality between them. Finally, we apply this duality to ${\hat{H}}_{\text{b}}$ to produce our fermionic Hamiltonian ${\hat{H}}_{\text{f}}$ and demonstrate that it describes an SPT.

\subsection{Bosonic operator algebra $\cal{A}_{\text{bos}}$}

On the bosonic side, we consider the spin-$\frac{1}{2}$ degrees of freedom living on links, with Pauli algebra generated by $\hatX_{pq}$ and $ \hatZ_{pq}$. $\cal{A}_{\text{bos}}$ is defined as the operator algebra generated by the subset of local operators that commute with all the ${\hat G}_p$ defined in (\ref{def:modifiedgausslaw}):
\begin{align}
{\hat G}_p = \prod_{\substack{\la tqr \ra \\ t=p}} {\hat{Z}}_{tq} {\hat{Z}}_{qr} {\hat{Z}}_{tr} \prod_{\la st \ra \ni p} {\hat X}_{st}
\end{align}
and modulo the relations ${\hat G}_p = 1$ for all $p$.\footnote{Note that on manifolds $M$ with nontrivial $H_1(M)$ global relations need to be specified to ensure that the duality is consistent.  These additional relations can be seen as coming from operator identities on the fermionic side of the duality - certain products of fermionic `hopping' operators and parity operators along nontrivial $1$-cycles are equivalent to the identity.}  Thus we may think of $\cal{A}_{\text{bos}}$ as the algebra of operators generated by the subset of local operators which are gauge invariant with respect to the modified Gauss's law $\hatG_p=1$.

\begin{figure}
\centering
\includegraphics[scale=.35,trim={4cm 2cm 2cm 0cm},clip]{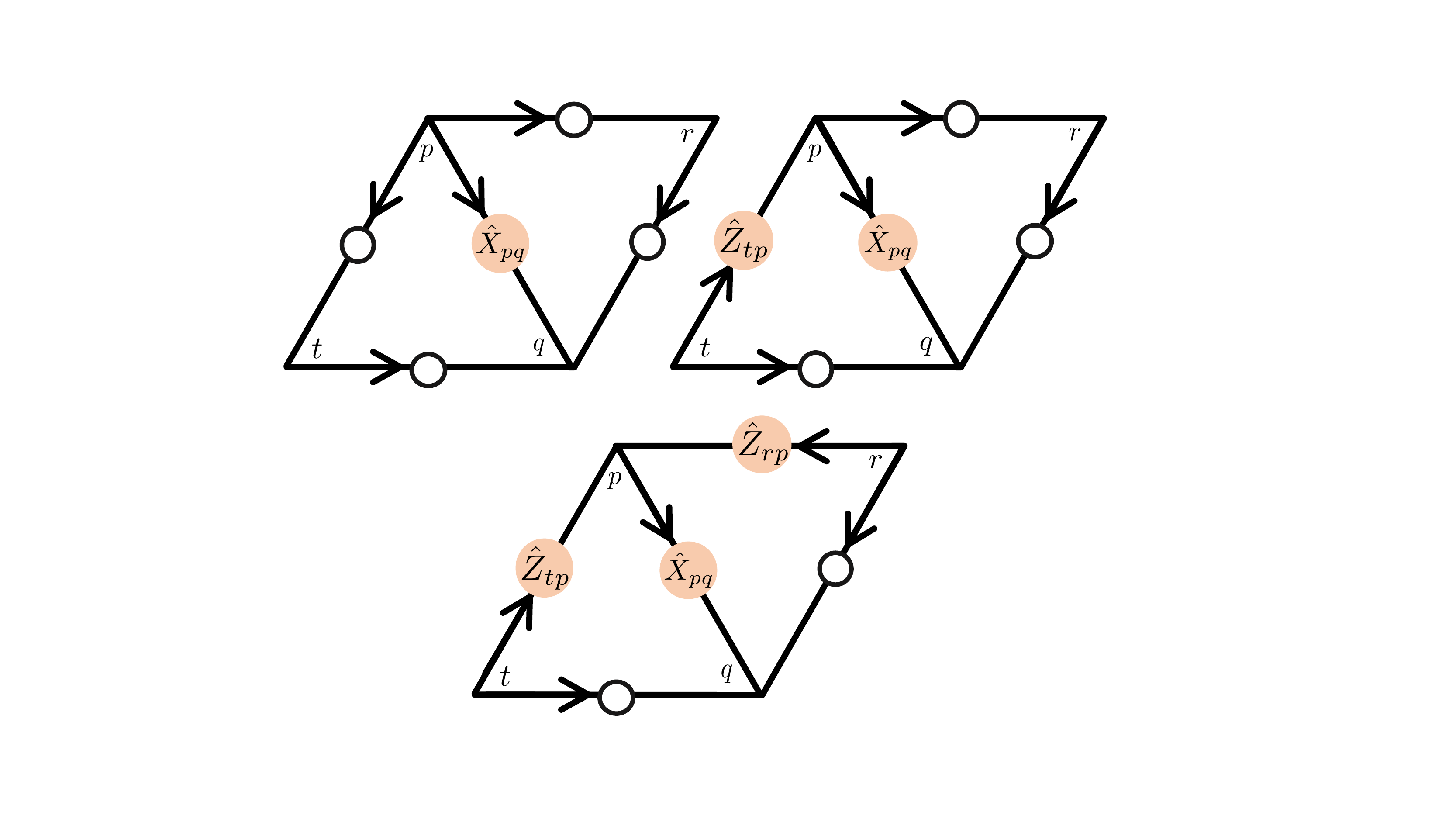}
\caption{The action of $\hatU_{pq}$ on link $\la pq \ra$ depends on the branching structure of the neighboring triangles.  $\hatX_{pq}$ is always applied to $\la pq \ra$, but a Pauli $\hat{Z}$ acts on the link connecting the first and second vertex of the neighboring triangle if and only if $\la pq \ra$ is the link connecting the second and third vertices of that triangle.}
\label{fig:Ueaction}
\end{figure}

We now identify two sets of local, modified Gauss's law invariant operators which generate all of $\cal{A}_{\text{bos}}$ [\onlinecite{Yu-an17}]. The first is $\hatW_{pqr} = \hatZ_{pq} \hatZ_{qr} \hatZ_{pr}$.  The second is $\hatU_{pq}$, defined as:
\begin{align} \label{def:Ue}
    \hatU_{pq} \equiv \hatX_{pq}\hat{K}_{L_{pq}}\hat{K}_{R_{pq}},
\end{align}
\noindent with $\hat{K}_{R_{pq}}$ and $\hat{K}_{L_{pq}}$ defined as follows.  The action of $\hat{K}_{R_{pq}}$ is dependent upon the triangle $R_{pq}$ to the right of $\la pq \ra$. If the triangle to the right of $\la pq \ra$ has vertex ordering $\la rpq \ra$, with $p$ and $q$ being the second and third vertices, respectively, then $\hat{K}_{R_{pq}}$ acts as $\hatZ_{rp}$. Otherwise, $\hat{K}_{R_{pq}}=\id$.  The action of $\hat{K}_{L_{pq}}$ is defined similarly but with `right' replaced with `left'.  Some examples of the action of $\hatU_{pq}$ are depicted in FIG. \ref{fig:Ueaction}. Intuition for this seemingly contrived definition can be obtained by recalling that the modified Gauss's law is a constraint that binds a $\Z_2$ flux on a triangle to a $\Z_2$ charge at the first vertex of that triangle. The operator $\hatU_{pq}$ then hops a $\Z_2$ flux across the link $\lan pq \ran$, and also rearranges the $\Z_2$ charges in such a way that the modified Gauss's law remains enforced.

As shown in Ref. [\onlinecite{Yu-an17}], the only nontrivial relations among the $\hatU_{pq}$ and $\hatW_{pqr}$ operators are captured in the following operator identity. For any vertex $p$,
\begin{align} \label{bosidentity}
\prod_{\substack{\la tq \ra \\ t=p}} \hatU_{tq} \prod_{\substack{\la qt \ra \\ t=p}} \hatU_{qt} = {\hat G}_p \prod_{\substack{\la tqr \ra \\ t=p}} {\hat W}_{tqr}
\prod_{\substack{\la qrt \ra \\ t=p}} {\hat W}_{qrt}.
\end{align}

\noindent Note that in the first product on the left hand side all the links are oriented away from $p$, while in the second product all the links are oriented towards $p$.

\subsection{Fermionic operator algebra $\cal{A}_\text{fer}$}

On the fermionic side, the degrees of freedom are complex fermions - one at the center of each triangle $\la pqr \ra$. We use the pair of Majorana operators $\gamma_{pqr}$ and $\gammabar_{pqr}$ to represent the operator algebra for this complex fermion.  The fermion parity at triangle $\la pqr \ra$ is measured by
\begin{align}
(-1)^{{\hat F}_{pqr}} \equiv -i \gamma_{pqr} \gammabar_{pqr},
\end{align}
\noindent and an operator is fermion parity even if it commutes with $\prod_{\la pqr \ra}(-1)^{{\hat F}_{pqr}}$.  The algebra $\cal{A}_{\text{fer}}$ of fermion parity even operators is generated by the $(-1)^{{\hat F}_{pqr}}$ and a certain set of `hopping operators', which transfer fermion parity across a link $\la pq \ra$. Specifically, we define the hopping operator
\begin{align}
{\hat S'}_{pq} \equiv i\gamma_{L_{pq}} {\gammabar}_{R_{pq}},
\end{align}
\noindent where we have again denoted the triangles to the left and right of $\lan pq \ran$ by $L_{pq}$ and $R_{pq}$, respectively.

\begin{figure}
\centering
\includegraphics[scale=.28,trim={2cm 5cm 0cm 0cm},clip]{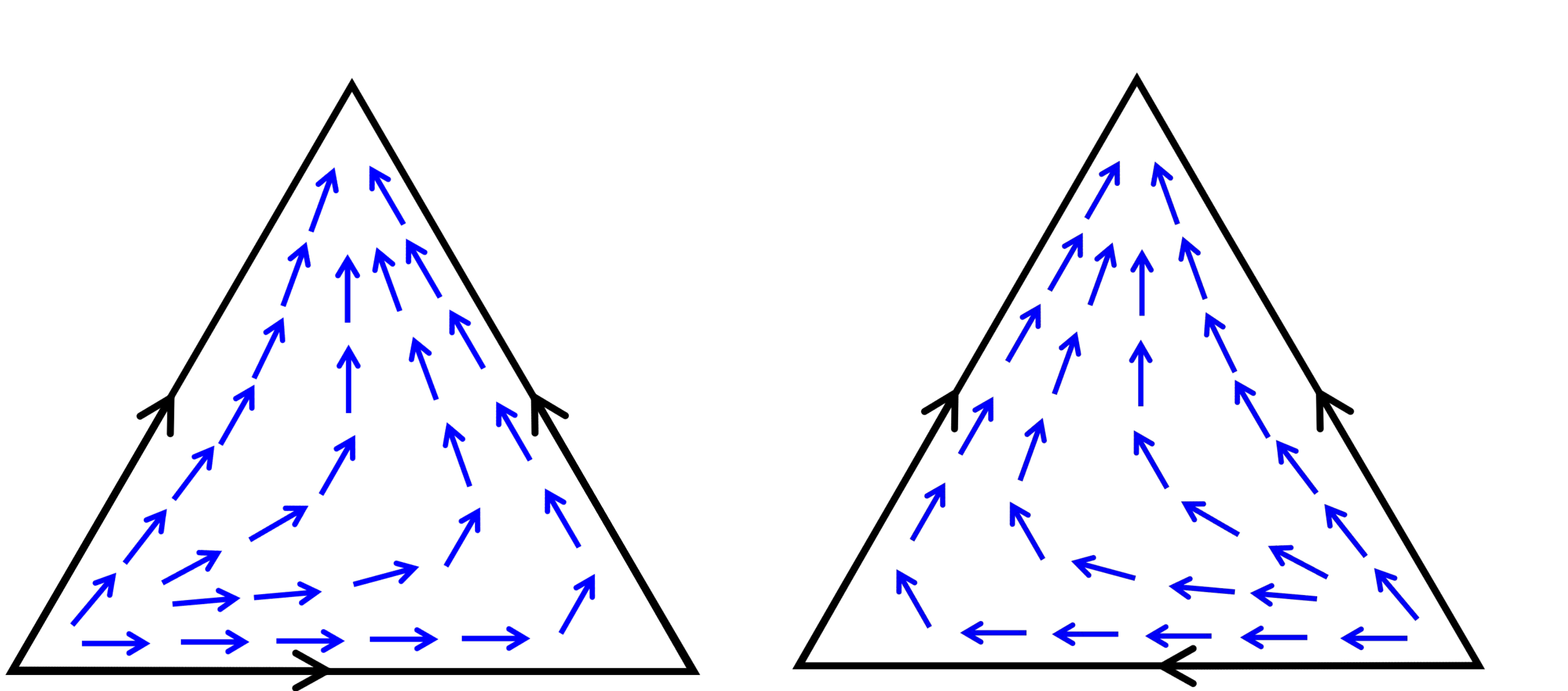}
\caption{The interpolating vector field lies parallel to the branching structure for both $o_{pqr}=+1$ and $o_{pqr}=-1$ triangles.  There are no singularities of the vector field away from the vertices.}
\label{fig:vecfieldorientation}
\end{figure}

The $(-1)^{{\hat F}_{pqr}}$ and ${\hat S'}_{pq}$ satisfy nearly the same algebraic relations with each other as do the bosonic operators $\hatW_{pqr}$ and $\hatU_{pq}$. The only difference is that $(-1)^{{\hat F}_{pqr}}$ and ${\hat S'}_{pq}$ satisfy an algebraic relation that is similar to but not exactly the same as (\ref{bosidentity}) [\onlinecite{Yu-an17}]:
\begin{align}\label{def:cp}
 \prod_{\substack{\la tq \ra \\ t=p}}  {\hat S'}_{tq} \prod_{\substack{\la qt \ra \\ t=p}}  {\hat S'}_{qt} = c(p) 
 \prod_{\substack{\la tqr \ra \\ t=p}} (-1)^{{\hat F}_{tqr}} 
 \prod_{\substack{\la qrt \ra \\ t=p}} (-1)^{{\hat F}_{qrt}}.
\end{align}
\noindent In (\ref{def:cp}), $c(p)$ is a sign factor  determined solely by the branching structure near $p$.
We prove (\ref{def:cp}) in Appendix \ref{ap:cpproof}, where we also derive the following graphical method for explicitly calculating $c(p)$.  First, we interpolate the branching structure to the interiors of the triangles to give a continuous non-vanishing vector field [\onlinecite{GK}] ${\cal V}$ (see FIG. \ref{fig:vecfieldorientation}).  Singularities in this vector field can occur only at vertices, and $c(p)=-1$ if the vertex $p$ has a singularity with odd winding number and $c(p)=1$ otherwise.


\subsection{Spin structure dependent duality between  $\cal{A}_\text{bos}$ and $\cal{A}_\text{fer}$}
\label{sec:spinstructure}

\begin{figure}
\centering
\includegraphics[scale=.25,trim={0cm 0cm 0cm 0cm},clip]{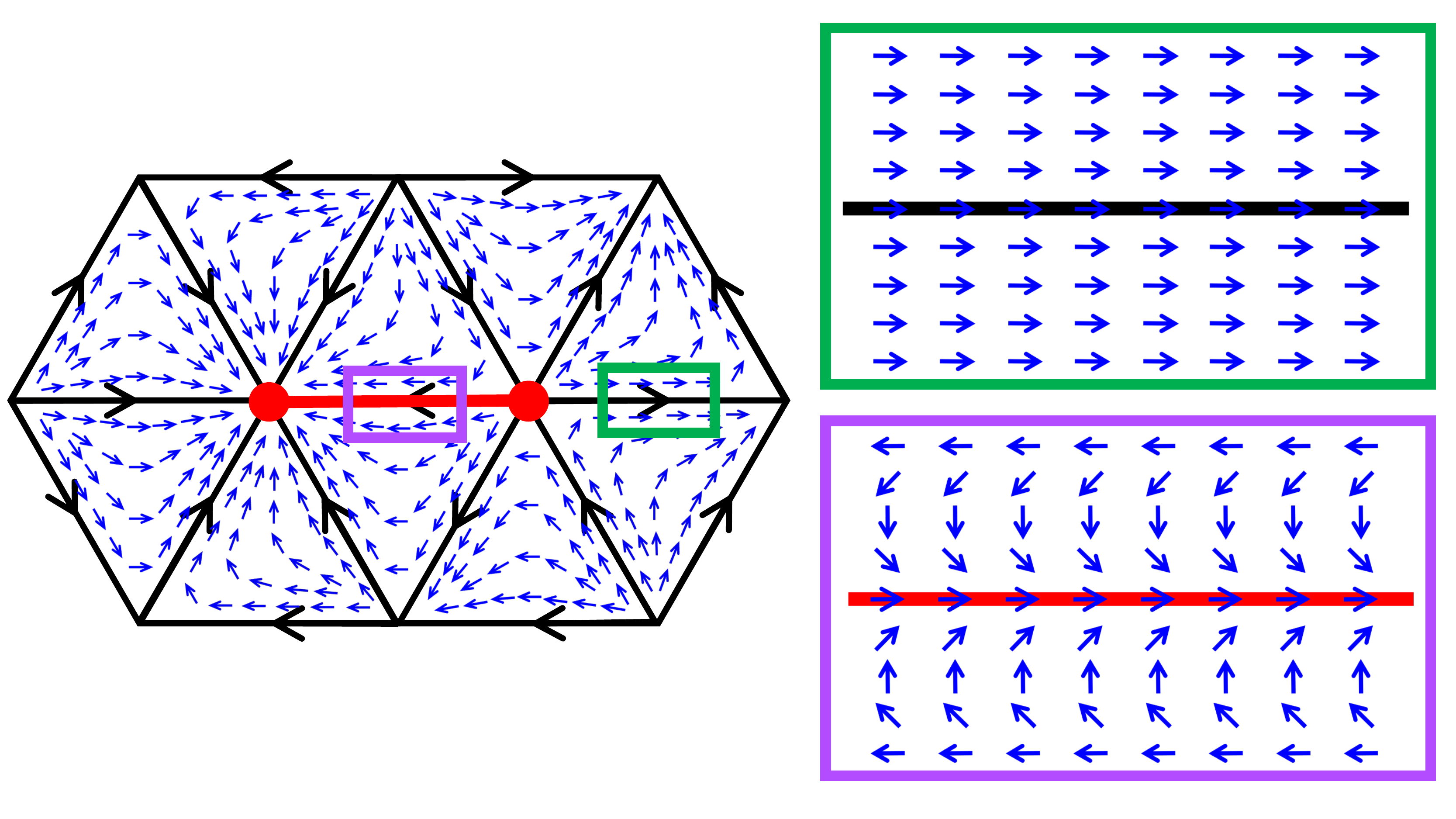}
\caption{The red vertices mark singularities of the interpolating vector field with odd winding numbers, and the red link gives a choice of $\cal{E}$.  The green inset shows the interpolating vector field near a link not belonging to $\cal{E}$, while the purple inset shows the $2\pi$ twist of the vector field near a link in $\cal{E}$.}
\label{fig:singularedge}
\end{figure}

The geometric interpretation of the sign $c(p)$ in (\ref{def:cp}) as counting the singularities of a vector field ${\cal V}$ immediately points to a possible modification of the operators generating $\cal{A}_{\text{fer}}$ that makes $\cal{A}_{\text{fer}}$ manifestly isomorphic to $\cal{A}_{\text{bos}}$.  To make this modification, first note that there are an even number of vertices with $c(p)=-1$. \footnote{The fact that the number of vertices $p$ with $c(p)=-1$ is even is just a consequence of the fact that the winding number of singularities is additive: a contour that encloses several singularities has a winding number equal to the sum of the winding numbers of those singularities.  On a compact manifold, a small contour enclosing no singularities can equivalently be thought of as a large contour enclosing all the singularities (by exchanging the notion of `inside' and `outside' the contour).}  Thus, we can find a set ${\cal E}$ of links such that the vertices in the boundary of ${\cal E}$ (the boundary being defined as the set of vertices which are endpoints of an odd number of links in ${\cal E}$) are precisely the vertices with $c(p)=-1$. Then, we can modify the vector field ${\cal V}$ by giving it an extra $2 \pi$ winding as it crosses a link in ${\cal E}$ (see FIG. \ref{fig:singularedge}). The result is a new vector field with even singularities only.  It is known that in  2 dimensions a vector field with only even singularities defines a spin structure [\onlinecite{Cimasoni07}].  Hence, a choice of ${\cal E}$ corresponds to a choice of spin structure.

Having made a choice of ${\cal E}$, we now define modified hopping operators
\begin{align}
{\hat S}_{pq} \equiv (-1)^{{\cal E}_{pq}} {\hat S'}_{pq}
\end{align}
\noindent where ${\cal E}_{pq} = 0,1$ is the indicator function for ${\cal E}$, i.e. ${\cal E}_{pq} = 1$ if $\lan pq \ran \in {\cal E}$ and ${\cal E}_{pq}=0$ otherwise.  These modified operators then satisfy
\begin{align} \label{Antonrelation}
\prod_{\substack{\la tq \ra \\ t=p}} {\hat S}_{tq}  \prod_{\substack{\la qt \ra \\ t=p}} {\hat S}_{qt} = \prod_{\substack{\la tqr \ra \\ t=p}} (-1)^{\hat{F}_{tqr}} \prod_{\substack{\la qrt \ra \\ t=p}} (-1)^{\hat{F}_{qrt}}.
\end{align}
Now, comparing with (\ref{bosidentity}), we see that the correspondence given by
\begin{align}\label{dictionary}
{\hat W}_{pqr} &\longleftrightarrow (-1)^{F_{pqr}}\\ \nonumber
{\hat U}_{pq} &\longleftrightarrow {\hat S}_{pq} 
\end{align}
\noindent defines an explicit isomorphism of operator algebras between $\cal{A}_\text{bos}$ and $\cal{A}_\text{fer}$.  We emphasize that this correspondence depends on a choice of spin structure, via the choice of ${\cal E}$.  

The fermionization duality reviewed here admits an intuitive description in terms of a `condensation of fermions'. We elaborate on this point in Appendix \ref{ap:fermioncondensation}.



\subsection{Fermionic SPT Hamiltonian} \label{sec:fSPT}

Let us now use the dictionary given in (\ref{dictionary}) to rewrite each local term in the shadow model Hamiltonian
\begin{align} \label{Hbos2}
    \hat{H}_{\text{b}}=\hatU_{\text{b}}\hat{H}^0_{\text{b}}\hatU_{\text{b}}^\dag,
\end{align}
defined in (\ref{Hbos}), in terms of local fermionic operators.  This can be carried out by fermionizing $\hat{H}_{\text{b}}^0$, defined in (\ref{toriccode}), and $\hatU_b$, defined in (\ref{ucirc}), independently.  To fermionize $\hat{H}_{\text{b}}^0$, we first use the definition of $\hatG_p$ to rewrite it as
\begin{align}\label{prefermionize}
     \hat{H}^0_{\text{b}}=-\sum_p \hatP_p^{\text{sym}} 
     -\sum_{p}\Bigg( \hatG_p\prod_{\substack{\la tqr \ra \\ t=p}}\hatW_{tqr} \Bigg)-\sum_{\la pqr \ra}\hatW_{pqr}.
\end{align}

\noindent Then, according to the dictionary in (\ref{dictionary}), $\hat{H}^0_{\text{b}}$ fermionizes to 
\begin{align}\label{atomicinsulator}
    \hat{H}_{\text{f}}^0=-\sum_p \hatP_p^{\text{sym}}-\sum_{\la pqr \ra}(-1)^{\hat{F}_{pqr}},
\end{align}
after using the gapped and unfrustrated property of the Hamiltonian to remove the fermionization of the second term in (\ref{prefermionize}).  This Hamiltonian describes a trivial atomic insulator, and the unique ground state $|\Psi_{\text{f}}^{0}\ra$ is a product state of symmetrized states at the vertices and zero fermion occupancy on the triangles. 

To fermionize $\hatU_{\text{b}}$, we note that the product 
\begin{align}
   \prod_{\la pqr \ra} \hatZ_{pq} ^{\hat{n}_{qr}} \prod_{\la pq \ra}\hatX_{pq}^{\hat{n}_{pq}}
\end{align}
\noindent in (\ref{ucirc}) can be rearranged into
\begin{align}\label{Uedgesprod}
   \hat{\kappa}\prod_{\la pq \ra}\hatU_{pq}^{\hat{n}_{pq}}.
\end{align}
\noindent where $\hat{\kappa}$ is a certain diagonal operator in the $\{g_p\}$ configuration basis with eigenvalues $\pm 1$.  The eigenvalue is locally determined, in that it is a product of signs, each of which is dependent upon only the $G$-configuration within a disk of finite radius around some point.  These signs result from commuting $\hatZ_{pq}$ past $\hatX_{pq}$ and hence the eigenvalues are dependent on the choice of ordering of the $\hatU_{pq}$ operators in (\ref{Uedgesprod}).  Although the operator $\hat{\kappa}$ is complicated to write out for general $G$, we note that the locality property above makes it a finite depth circuit of local unitaries.  Furthermore, we will see below that in the example $G=\Z_2$ the situation simplifies considerably: $\hat{\kappa}$ is trivial in that case, and all of the terms in the product in (\ref{Uedgesprod}) commute.  Also, in Appendix \ref{ap:spinH} we present another way of circumventing the issue posed by the unwieldy form of $\hat{\kappa}$, by introducing ancillary spin-$\frac{1}{2}$ degrees of freedom on the triangles. This allows for a more canonical finite depth circuit that does not require an arbitrary choice of ordering.

\begin{figure}
\centering
\includegraphics[scale=.28,trim={5cm 3cm 3cm 2cm},clip]{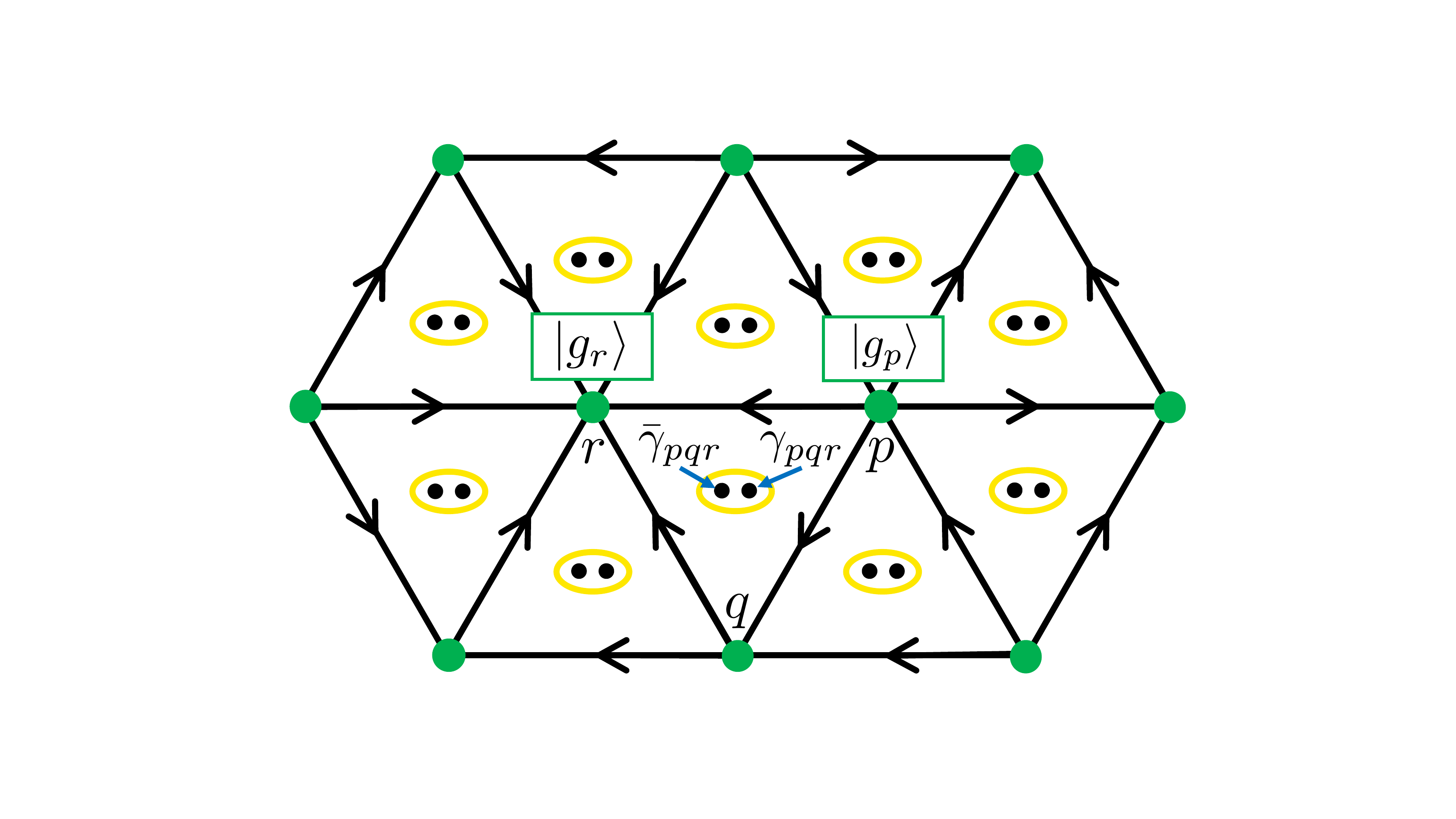}
\caption{The fermionic SPT Hamiltonian acts on a Hilbert space with generalized $G$-spin degrees of freedom on vertices and a single complex fermion degree of freedom for each triangle. Explicitly, the total Hilbert space is $\left( \bigotimes_{p}\mathbb{C}_p^{|G|} \right)\otimes^{\mz_2} \left( \bigotimes^{\mz_2}_{\la pqr \ra}\mathbb{C}_{pqr}^{1|1} \right)$, where $\bigotimes^{\mz_2}$ is a $\mz_2$ graded tensor product. We use $\gamma_{pqr}$ and $\gammabar_{pqr}$ as a basis for the operator algebra of the complex fermion at $\la pqr \ra$.}
\label{fig:fermionicdof}
\end{figure}

We now use (\ref{dictionary}) to map (\ref{Uedgesprod}) to  fermionic operators.  The result of fermionizing $\hatU_{\text{b}}$ is the finite depth circuit of local unitaries
\begin{align}\label{ufer}
   \hatU_{\text{f}}= \hat{\kappa} \prod_{\la pqr \ra} \hat{\nu}_{pqr}^{o_{pqr}} \prod_{\la pq \ra}\hatS_{pq}^{\hat{n}_{pq}}\prod_{\la pqr \ra}\left((-1)^{\hat{F}_{ pqr }}\right)^{\hat{n}_{pr}}.
\end{align}
\noindent Therefore, fermionization turns $\hat{H}_{\text{b}}$ into 
\begin{align}
    \hat{H}_{\text{f}}=\hatU_{\text{f}}\hat{H}_{\text{f}}^0\hatU_{\text{f}}^\dag.
\end{align}

$\hat{H}_{\text{f}}$ is comprised of two types of terms.  First, we have the conjugates of the terms in the second sum in (\ref{atomicinsulator}), namely:
\begin{align}
    -\hatU_{\text{f}}(-1)^{\hat{F}_{pqr}}\hatU_{\text{f}}^\dag=-(-1)^{\hat{n}_{pqr}}(-1)^{\hat{F}_{pqr}}. 
\end{align}
\noindent These energetically enforce fermions to occupy the triangles $\la pqr \ra$ with nontrivial $n(g_p,g_q,g_r)$.  Second, we have the conjugates of the terms in the first sum in (\ref{atomicinsulator}):
\begin{align}
    -\hatU_{\text{f}}\hatP^{\text{sym}}_p\hatU_{\text{f}}^\dag.
\end{align}
\noindent These fluctuate the $G$-configuration at vertex $p$ and move the neighboring fermions so that the fermion occupancy conforms to the first term.  We will see the action of $\hat{H}_{\text{f}}$ more explicitly below when we treat the case $G=\mz_2$.

$\hat{H}_{\text{f}}$ describes a fermionic SPT phase because (1) it is gapped (2) it has a unique, SRE ground state, and (3) it is symmetric. It is gapped because it is an unfrustrated commuting projector Hamiltonian.  The unique ground state is $\hatU_{\text{f}}|\Psi_{\text{f}}^0 \ra$, and since $\hatU_{\text{f}}$ is a finite depth circuit of local unitaries, the ground state is SRE.  Lastly, it is $G$-symmetric because ${\hat{H}}_b$ is $G$-symmetric, and the fermionization procedure commutes with the global action of $G$.

\subsection{Example: $G=\Z_2$} \label{Z2cont}

Recall that in the $G=\Z_2$ case, (\ref{z2circuit}) is
\begin{align} \label{z2circuit_duplicate}
   \hatU^{\pm}_{\text{b}}=&\prod_{\la pqr \ra} (\pm i)^{o_{pqr}\hats_p(1-\hats_q)\hats_r}  \hatZ_{pq} ^{(1-\hats_q)\hats_r} \\ \nonumber
    \times &\prod_{\la pq \ra}\hatX_{pq}^{(1-\hats_p)\hats_q}\prod_{\la pqr \ra}\hatW_{ pqr }^{(1-\hats_p)\hats_r}.
\end{align}
To avoid confusion, we will for the remainder of this section focus on the case $\hatU^{+}_{\text{b}}$ and drop the $+$ superscript; the case $\hatU^{-}_{\text{b}}$ can be treated similarly.

To fermionize $\hatU_{\text{b}}$, we first recognize that it may be written in terms of the local operators $\hatU_{pq}$ of section \ref{sec:duality}.  The product
\begin{align}
    \prod_{\la pqr \ra} \hatZ_{pq}^{(1-\hats_q)\hats_r} \prod_{\la pq \ra}\hatX^{(1-\hats_p)\hats_q}_{pq}
\end{align}
\noindent in (\ref{z2circuit_duplicate}) is exactly equal to 
\begin{align}
    \prod_{\la pq \ra}\hatU^{(1-\hats_p)\hats_q}_{pq}
\end{align}
\noindent without any additional factor of ${\hat{\kappa}}$.  This is due to the fact that $(1-s_q)s_r$ and $(1-s_p)s_q$ cannot simultaneously be $1$, so that we never have to move anti-commuting operators past each other to go from one expression to the other.  Therefore, the fermionization duality applied to $\hatU_{\text{b}}$ yields
\begin{align}\label{z2Uf}
    \hatU_{\text{f}}=&\prod_{\la pqr \ra} i^{{o}_{pqr}\hats_p(1-\hats_q)\hats_r}\\ \nonumber
    \times&\prod_{\la pq \ra}\hatS^{(1-\hats_p)\hats_q}_{pq} \prod_{\la pqr \ra}\left((-1)^{\hat{F}_{pqr}}\right)^{(1-\hats_p)\hats_r}
\end{align}
\noindent with $\hatS_{pq}$ and $(-1)^{\hat{F}_{pqr}}$ defined in section \ref{sec:duality}. Hence, ${\hat{H}}_{\text{b}}$ explicitly fermionizes to 
\begin{align}
    \hat{H}_{\text{f}}=\hatU_{\text{f}} \hat{H}^0_{\text{f}} \hatU_{\text{f}},
\end{align}
where
\begin{align}\label{atomicinsulator2}
    \hat{H}_{\text{f}}^0=-\sum_p \hatP_p^{\text{sym}}-\sum_{\la pqr \ra}(-1)^{\hat{F}_{pqr}}.
\end{align}
\noindent We have thus constructed a $\Z_2$-symmetric fermionic SPT Hamiltonian for the supercohomology data specified in (\ref{n2z2}) and (\ref{nu3z2}).

\begin{figure}
\centering
\includegraphics[scale=.28,trim={3cm 0cm 0cm 0cm},clip]{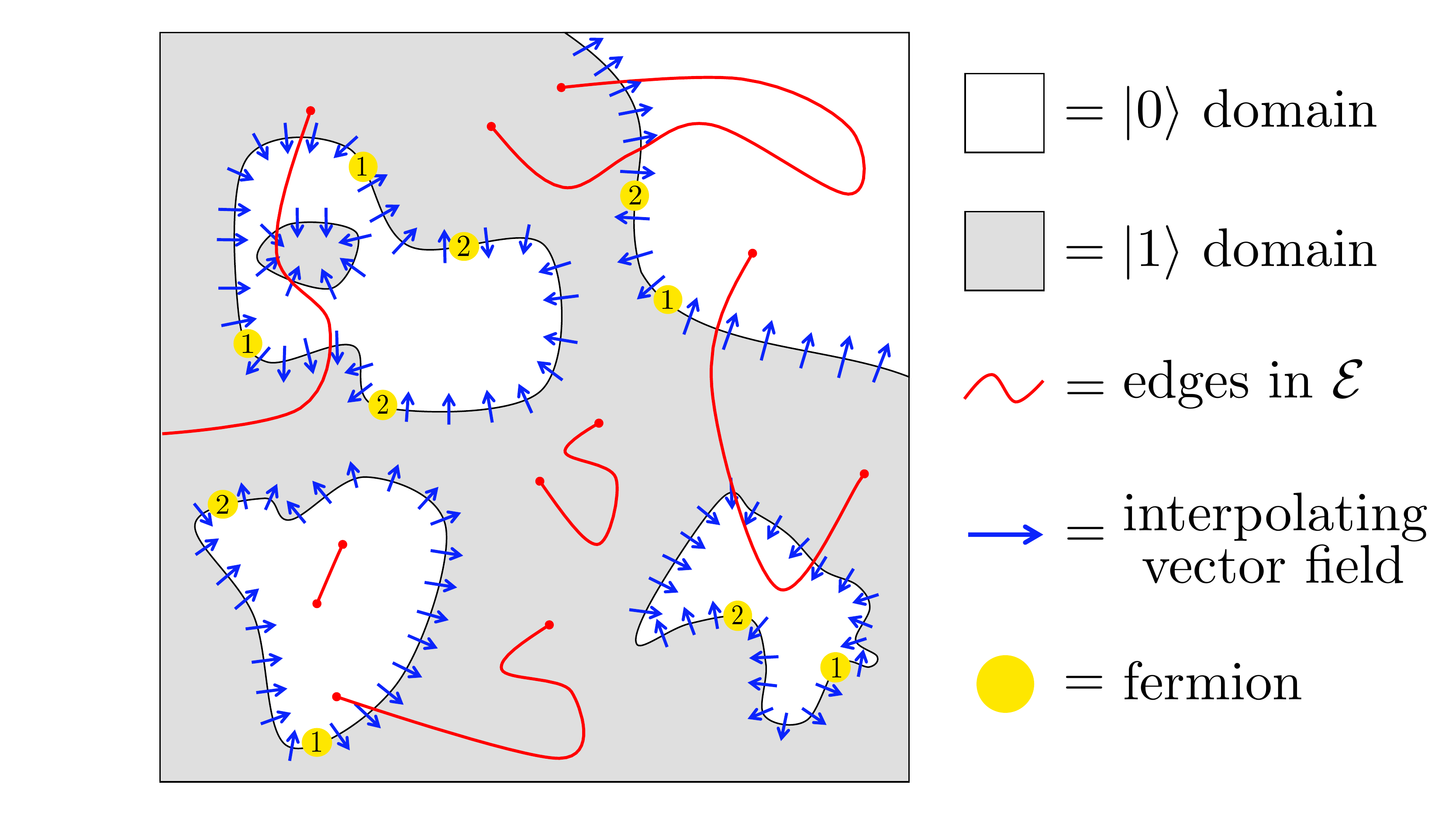}
\caption{Pictured here is the effect of $\prod_{\la pq \ra}\hatS^{(1-\hats_p)\hats_q}_{pq}$ on a domain wall configuration.  For clarity the lattice is suppressed and we have only drawn the interpolating vector field along the domain walls.  The edges of $\cal{E}$, introduced in section \ref{sec:duality}, are shown to illustrate their affect on the ordering of the fermions in the figure.}
\label{fig:z2gs}
\end{figure}

{\bf{Picture of the ground state: }}
The finite depth circuit of local unitaries $\hatU_\text{f}$ in (\ref{z2Uf}) allows us to explicitly construct the ground state $|\Psi_\text{f}\ra$ of $\hat{H}_\text{f}$. This is accomplished by applying $\hatU_\text{f}$ to $|\Psi_\text{f}^0 \ra$, the ground state of $\hat{H}_\text{f}^0$.  $|\Psi_\text{f}^0\ra$ is a product state with the $\mz_2$-symmetric state $\frac{1}{\sqrt{2}}(|0\ra+|1\ra)$ at each vertex $p$ and zero fermion occupancy at every triangle $\la pqr \ra$.  Expressed in the configuration basis, $|\Psi_\text{f}^0\ra$ is an equal amplitude superposition of domain configurations -- domains containing states $|0\ra$ or $|1\ra$ at vertices.  Note that the domain walls between the $|0\ra$ and $|1\ra$ domains run along the edges of the dual lattice.  The ground state of $\hat{H}_\text{f}$ is 
\begin{align}
    |\Psi_\text{f}\ra=\hatU_\text{f}|\Psi_\text{f}^0\ra=\sum_{\substack{\text{configs} \\ \vcenter{\hbox{\includegraphics[scale=.02,trim={7cm 0cm 7cm 0cm},clip]{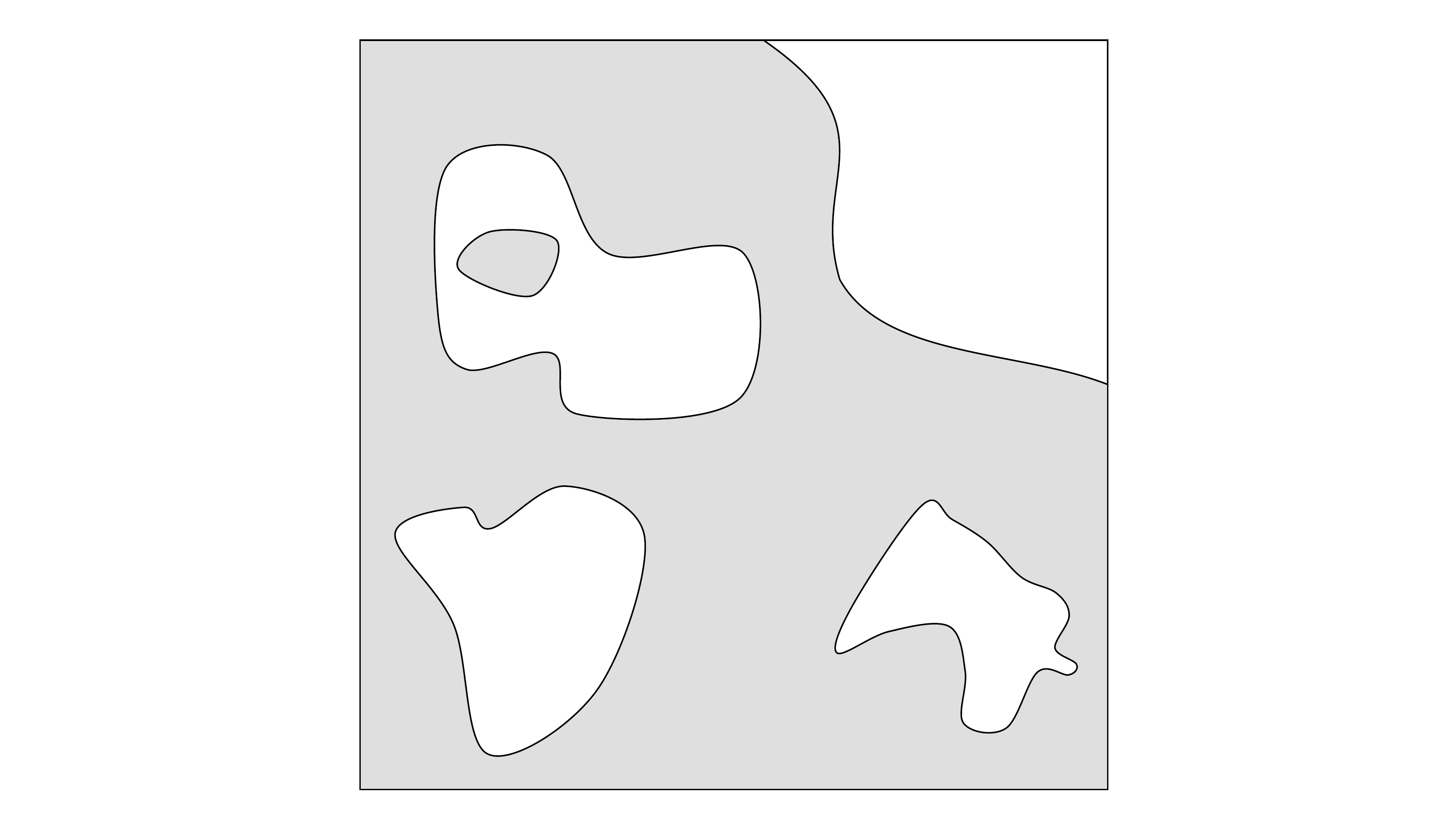}}}}}\hatU_\text{f} \left|\vcenter{\hbox{\includegraphics[scale=.08,trim={7cm 0cm 7cm 0cm},clip]{fSPT_figures2/domains.pdf}}} \right\ra.
\end{align}
The above sum is over all $\Z_2$-spin domain configurations tensored with the empty fermionic state.  The operator $\hatU_\text{f}$ decorates fermions onto each such domain configuration and multiplies by a configuration-dependent phase, but it does not alter the shape of the domains.

We can break the action of $\hatU_\text{f}$ on a domain configuration up into three steps.  In the first step, we apply 
\begin{align}\label{fermionparityterm}
    \prod_{\la pqr \ra}\left((-1)^{\hat{F}_{pqr}}\right)^{(1-\hats_p)\hats_r}.
\end{align}
As the domain configurations in $|\Psi_\text{f}^0\ra$ have no fermions, they are $+1$ eigenvectors of the fermion parity operators in (\ref{fermionparityterm}).  Thus, this term does not affect the state.

\begin{figure}
\centering
\includegraphics[scale=.25,trim={0cm 3cm 0cm 0cm},clip]{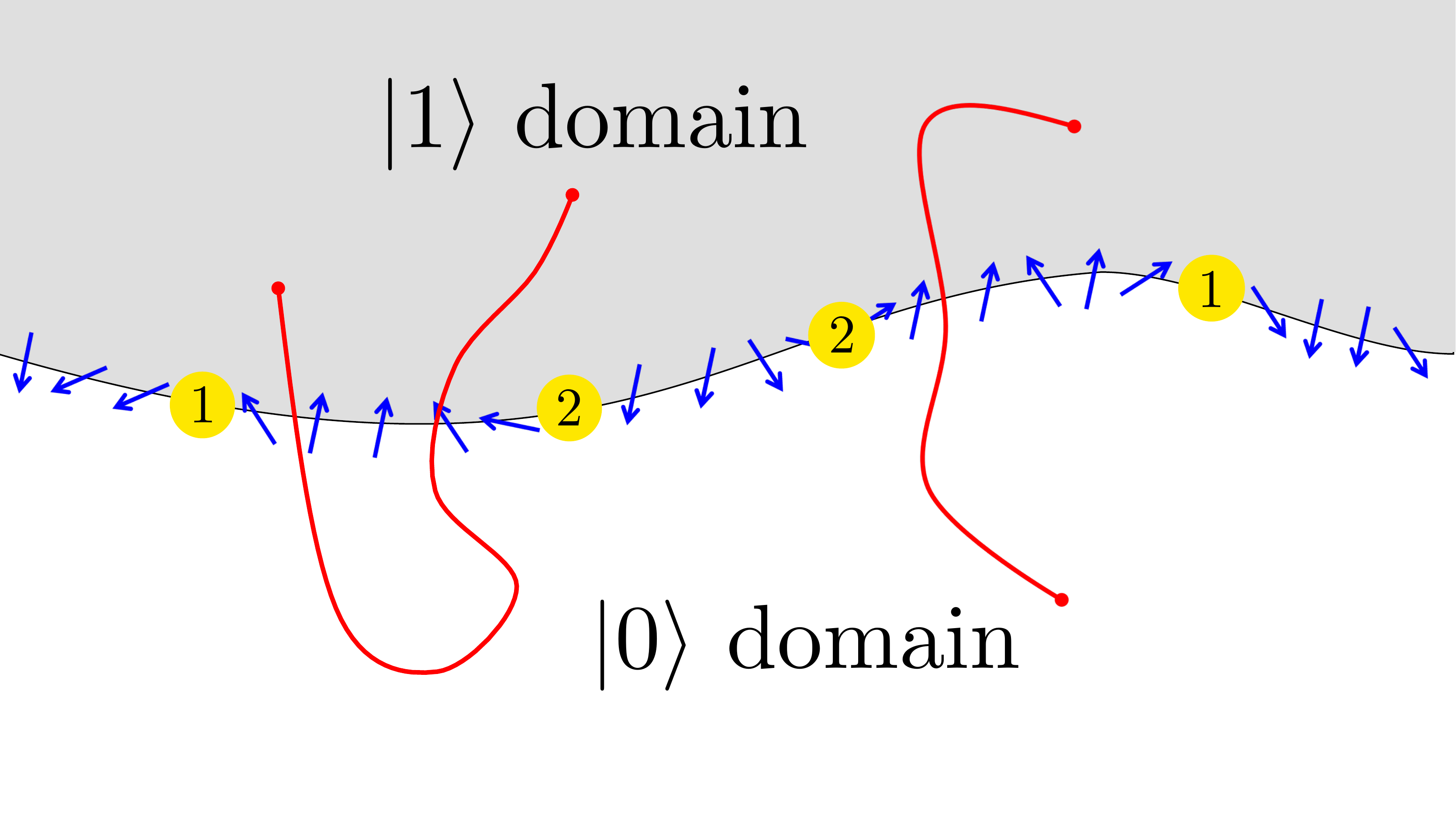}
\caption{The order in which the fermions are created along the domain wall is based on the spin structure.  Fermions are created in pairs -- one on either side of the regions for which the interpolating vector field points from $|0\ra$ (white) to $|1\ra$ (gray).  These fermions are created left to right if there are an even number of edges in $\cal{E}$ (pictured in red) between them and created right to left order otherwise.  The ordering is labeled above.  Note that equivalently, the fermions can be ordered from left to right across every $|0\ra$ to $|1\ra$ pointing region as long as a $-1$ sign is picked up for each edge in $\cal{E}$ oriented from $|0\ra$ to $|1\ra$.}
\label{fig:z2gsordering}
\end{figure}

In the second step, we act on the domain configuration with 
\begin{align}\label{fermionterm}
    \prod_{\la pq \ra}\hatS^{(1-\hats_p)\hats_q}_{pq}.
\end{align}
The exponent in (\ref{fermionterm}) is $1$ precisely when the link $\la pq \ra$ points from a $|0\ra$ domain to a $|1\ra$ domain.  As a result, Majorana operators are applied to the two triangles on either side of the link $\la pq \ra$, and in this way, fermions are only created along the domain wall.  The result is a pair of fermions at the two endpoints of each portion of the domain wall where the interpolating vector field points from the $|0\ra$ to the $|1\ra$ domain (see FIG. \ref{fig:z2gs} and  \ref{fig:z2gsordering}).  The order in which these two fermions are created depends on the spin structure $\cal{E}$ as follows. First, we locally orient the domain wall so that it runs horizontally with the $|0\ra$ domain below and the $|1\ra$ domain above, as illustrated in FIG. \ref{fig:z2gsordering}.  If there are an even number of edges in $\cal{E}$ crossing the $|0\ra$ to $|1\ra$ pointing portion of the domain wall, then we create the fermion on the left endpoint first, followed by the fermion on the right endpoint.  When there are an odd number of edges in $\cal{E}$ crossing the region, the fermions are created in the opposite order (FIG. (\ref{fig:z2gsordering})).  Since the difference between these two procedures is just a minus sign, we can alternatively always create the fermions from left to right, and at the end multiply by $-1$ for every edge of ${\cal E}$ that points from the $|0\ra$ to the $|1\ra$ domain.  

\begin{figure}
\centering
\includegraphics[scale=.25,trim={0cm 3cm 0cm 0cm},clip]{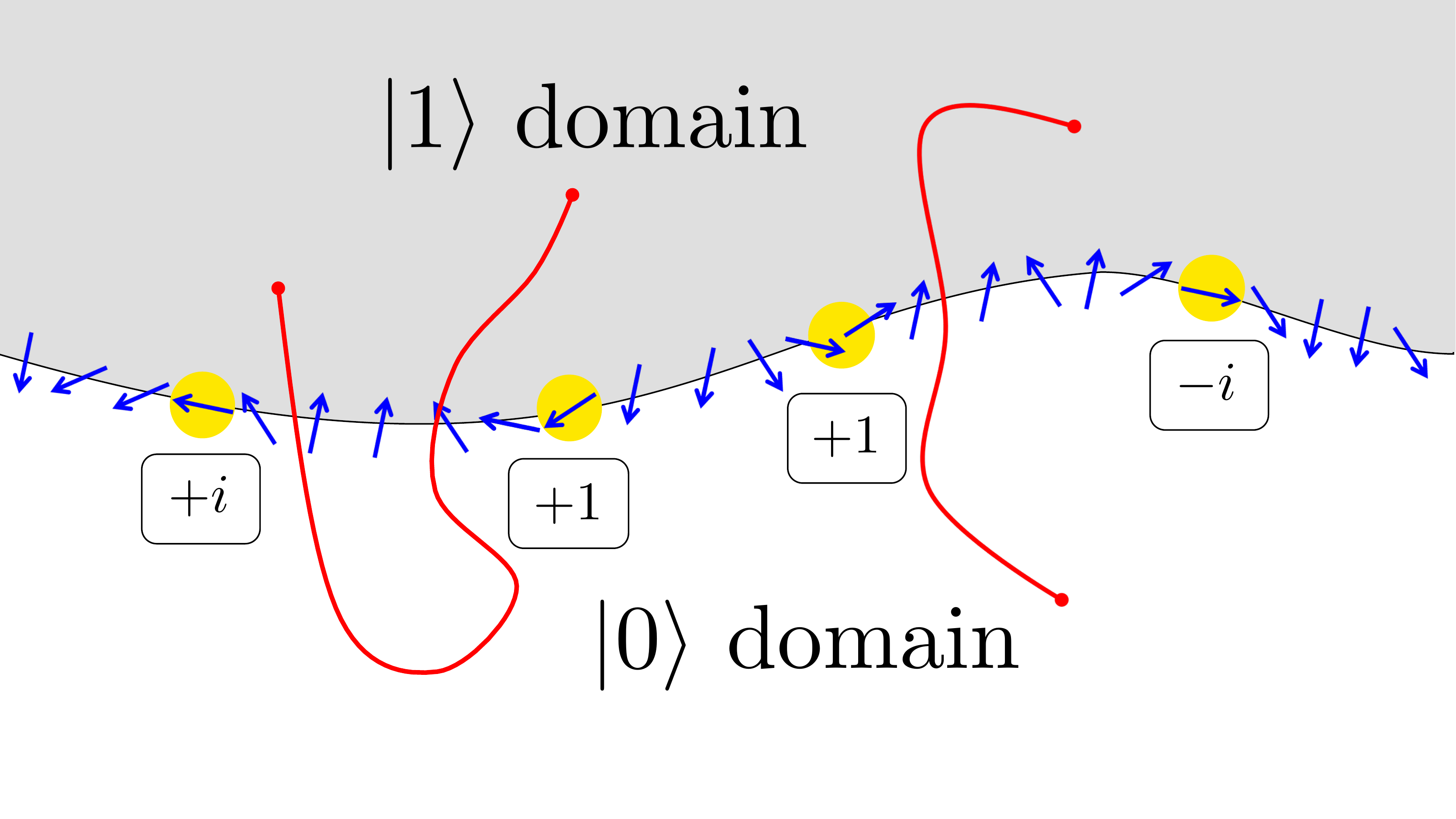}
\caption{In the figure above, as we move left to right along the domain wall, the interpolating vector field (blue arrows) rotates.  When it rotates clockwise from pointing towards the $|0\ra$ domain (white) to pointing towards the $|1\ra$ domain (gray), we get a phase of $+ i$.  When it rotates clockwise from pointing towards the $|1\ra$ domain to pointing towards the $|0\ra$ domain, we obtain a phase of $- i$.  Counterclockwise rotations of the vector field along the domain wall give a trivial phase.  The twist of the vector field near an edge in $\cal{E}$ (red edges), as displayed in FIG. \ref{fig:singularedge}, does not affect the calculation of this phase.}
\label{fig:z2gsphase}
\end{figure}

Lastly, we act with 
\begin{align}
    \prod_{\la pqr \ra} i^{o_{pqr}\hats_p(1-\hats_q)\hats_r}.
\end{align}
This term assigns a phase to each configuration, which can be thought of as a product of contributions associated to points of tangency of the vector field with the domain wall, or, equivalently, associated to the fermions.  These contributions can be determined as follows.  Moving from left to right along a domain wall with the $|0\ra$ domain below and the $|1\ra$ domain above, we track the interpolating vector field.  If the interpolating vector field rotates clockwise, from initially pointing in the direction of the $|0\ra$ domain to finally pointing in the direction of the $|1\ra$ domain, then we accrue a phase of $i$.  If the interpolating vector field rotates clockwise from initially pointing towards the $|1\ra$ domain to finally pointing towards the $|0\ra$ domain, then a phase of $-i$ is picked up (see FIG. \ref{fig:z2gsphase}).  For the two other possible rotations, no phase is picked up.  

We would like to emphasize that the ground state constructed according to this prescription admits a continuum interpretation.  Namely, in the continuum we can think of the spin structure being encoded in a smooth vector field together with a set of smooth segments $\cal{E}$ connecting the odd singularities of this vector field. The ground state is a superposition over smooth domain wall configurations decorated with fermions.  The fermions appear precisely at the locations where the vector field is tangent to a domain wall, and the above prescription gives a specific ordering of fermion creation operators used to create this fermionic state from the empty fermionic state.  Finally, the amplitude for each decorated domain wall is multiplied by products of $\pm i$ as determined by the rotation of the vector field at the points of tangency, as detailed above.  

\section{Classification}\label{sec:classification}


Thus far, we have used a choice of supercohomology data $(n,\nu)$ together with a spin structure on a 2d triangulated spatial manifold with branching structure to construct a zero correlation length fermionic SPT Hamiltonian. The strategy was to first construct a bosonic shadow model using the group supercohomology data. This led us to a finite depth circuit $\hatU_\text{b}$ (\ref{ucirc}):
\begin{align}
   \hatU_{\text{b}}=\prod_{\la pqr \ra} \left(\hat{\nu}_{pqr}^{o_{pqr}}  \hatZ_{pq} ^{\hat{n}_{qr}}\right) 
    \prod_{\la pq \ra}\hatX_{pq}^{\hat{n}_{pq}}\prod_{\la pqr \ra}\hatW_{ pqr }^{\hat{n}_{pr}},
\end{align}
which, when applied to an ordinary toric code ground state, produced the ground state of the bosonic shadow model.  Furthermore, the fermionization of $\hatU_\text{b}$ yielded $\hatU_\text{f}$ (defined in (\ref{ufer})) - a fermionic finite depth circuit that builds a fermionic SPT ground state from a trivial product state. 

In this section, we show that the composition of these circuits gives insight into the group structure of fermionic SPT phases.  First, we clarify the physical meaning behind composing finite depth circuits. Then, we give a physically motivated definition of equivalence for sets of supercohomology data. Lastly, we use this notion of equivalence to establish group supercohomology classes as topological invariants for lattice fermionic SPT Hamiltonians.  


\subsection{Stacking as composition of circuits}

The additive group structure on the set of SPT phases is given by stacking.  To stack two SPT Hamiltonians, let us imagine that they are defined on identical lattices extending in the $x,y$ directions, and let us put one lattice directly over the other, i.e. separated in the $z$ direction.  Then, grouping pairs of vertically separated sites with the same $x,y$-coordinates into supersites, the sum of the two decoupled SPT Hamiltonians for the two layers defines another $2$d gapped SPT Hamiltonian.  This stacking operation respects the notion of phase equivalence and thus defines an additive structure on the set of SPT phases.

We can reinterpret the stacking operation as composition of finite depth circuits of local unitaries that create the corresponding SPT ground states from a product state. To see this, suppose that $\hatU$ and $\hatU'$ are two such circuits that act on identical Hilbert spaces made out of sites which form identical $G$-representations.  The ground state of the stacked system is
\begin{align}
\left(\hatU \otimes \hatU'\right) \left(|0\ra \otimes |0\ra'\right)=
\left(\hatU |0\ra\right) \otimes \left(\hatU' |0\ra'\right).
\end{align}
Now let $\hatV$ be the unitary operator which exchanges the two layers.  Note that $\hatV$ can be defined as a tensor product of finite dimensional unitaries acting on the individual supersites, where they just swap the two sites in each supersite.  $\hatV$ clearly commutes with the action of the global symmetry, and we have
\begin{align}\label{eq:stacked}
\hatV&\left(\hatU \otimes 1\right)\hatV^\dag \left( 1 \otimes \hatU' \right) \left(|0\ra \otimes |0\ra'\right) \\ \nonumber
&= \left(1 \otimes \hatU \hatU'\right) \left(|0\ra \otimes |0\ra'\right) \\ \nonumber
&= |0\ra \otimes \left(\hatU \hatU' |0\ra'\right).
\end{align}
Hence (\ref{eq:stacked}) is equivalent to the state obtained by composing the two circuits.  Notice, $\hatV$ can be continuously connected to the identity via a path in the space of symmetric finite depth circuits.  To construct such a path, one just needs to find a path connecting the swap unitary to the identity for a single supersite and tensor these over all the supersites.  For a single supersite, the problem is straightforward. This is because, in a finite dimensional Hilbert space, any symmetric unitary is connected to the identity through a path in the space of symmetric unitaries, as can be seen by breaking up the Hilbert space into irreducible representations of $G$ and applying Schur's lemma.

We now use this equivalence between stacking and composing circuits to derive the stacking rule for our supercohomology SPT models.  In particular, this will show that the supercohomology SPT phases form a closed subgroup under stacking.

\subsection{Computation of stacking rules by composing circuits}

Let $(n,\nu)$ and $(n',\nu')$ be two sets of supercohomology data.  Further, denote the bosonic finite depth circuits obtained from $(n,\nu)$ and $(n',\nu')$ via our construction by $\hatU_\text{b}^{n\nu}$ and $\hatU_\text{b}^{n'\nu'}$, respectively.  The composition of $\hatU_\text{b}^{n\nu}$ with $\hatU_\text{b}^{n'\nu'}$ yields a finite depth circuit corresponding to yet another set of supercohomology data.  This can be seen by explicit computation.  The product of $\hatU_\text{b}^{n\nu}$ with $\hatU_\text{b}^{n'\nu'}$ is
\begin{align}
    \hatU_\text{b}^{n\nu}\hatU_\text{b}^{n'\nu'}=&\prod_{\la pqr \ra} \left(\hat{\nu}_{pqr}^{o_{pqr}}  \hatZ_{pq} ^{\hat{n}_{qr}}\right) 
    \prod_{\la pq \ra}\hatX_{pq}^{\hat{n}_{pq}}\prod_{\la pqr \ra}\hatW_{ pqr }^{\hat{n}_{pr}}\\ \nonumber
    \times &\prod_{\la pqr \ra} \left({\hat{\nu}}'^{o_{pqr}}_{pqr}  \hatZ_{pq} ^{{\hat{n}}'_{qr}}\right) 
    \prod_{\la pq \ra}\hatX_{pq}^{{\hat{n}}'_{pq}}\prod_{\la pqr \ra}\hatW_{ pqr }^{{\hat{n}}'_{pr}}.
\end{align}
To obtain an expression in the same form as $\hatU_\text{b}$, and thus reveal the group structure of the fermionic circuits, we group similar terms.  In doing so, the only non-trivial signs arise when we move \begin{align}
    \prod_{\la pqr \ra}\hatW_{ pqr }^{\hat{n}_{pr}}\text{ past }\prod_{\la pq \ra}\hatX_{pq}^{{\hat{n}}'_{pq}}
\end{align}
and
\begin{align}
    \prod_{\la pqr \ra}\hatX_{ pq }^{\hat{n}_{pq}}\text{ past }\prod_{\la pq \ra}\hatZ_{pq}^{{\hat{n}}'_{qr}}.
\end{align}
Using $\delta n = 0$, we can write the resulting sign as:
\begin{align} \label{cuponesign}
    \prod_{\la pqr \ra}(-1)^{\hat{n}_{pr}\hat{n}'_{pqr}+\hat{n}_{pq}\hat{n}'_{qr}}.
\end{align}
We then have
\begin{align} \label{compositioncircuit}
\hatU_\text{b}^{n\nu}\hatU_\text{b}^{n'\nu'}=&\prod_{\la pqr \ra} \hat{\nu}^{o_{pqr}}_{pqr}{\hat{\nu}}'^{o_{pqr}}_{pqr}(-1)^{\hat{n}_{pr}\hat{n}'_{pqr}+\hat{n}_{pq}\hat{n}'_{qr}}  \\ \nonumber
    \times &\prod_{\la pqr \ra} \hatZ_{pq} ^{\hat{n}_{qr}+{\hat{n}}'_{qr}} \prod_{\la pq \ra}\hatX_{pq}^{\hat{n}_{pq}+\hat{n}'_{pq}}\prod_{\la pqr \ra}\hatW_{ pqr }^{\hat{n}_{pr}+\hat{n}'_{pr}}.
\end{align}
This is precisely the circuit $\hatU_\text{b}$ formed from the input supercohomology data $(n+n',\nu \nu' (-1)^{n \cup_1 n'})$, where\setcounter{footnote}{400}\footnote{As a reminder, the cup product \unexpanded{$\cup$} between homogeneous functions \unexpanded{$f:G^{\ell+1} \to A$} and \unexpanded{$h:G^{k+1} \to A$} (for abelian group $A$) is
\begin{align}
    (f \cup h)(g_0,...,g_{\ell+k-1}) = f(g_0,...,g_\ell)h(g_\ell,...,g_{\ell+k-1}). 
\end{align}
The \unexpanded{$\cup_1$} product of $f$ and $h$ is \begin{align}
    \unexpanded{&(f \cup_1 h)(g_0,...,g_{\ell+k-1})= \\ 
    &\sum_{i=0}^{\ell-1}f(g_0,...,g_i,g_{k+i},g_{\ell+k-1})h(g_i,...,g_{k+i})}. 
\end{align}} 
\begin{align}
    (n \cup_1 n'&) (g_p,g_q,g_r,1) = \\ \nonumber &n(g_p,g_r,1)n'(g_p,g_q,g_r)+n(g_p,g_q,1)n'(g_q,g_r,1)
\end{align}
agrees with the sign in (\ref{cuponesign}) and (\ref{compositioncircuit}).

Therefore, stacking the fermionic SPT phases corresponding to $(n,\nu)$ and $(n',\nu')$ results in the fermionic SPT phase corresponding to $(n+n',\nu \nu' (-1)^{n \cup_1 n'})$, or
\begin{align}\label{stackingrule}
    (n,\nu)*(n',\nu')=(n+n',\nu \nu' (-1)^{n \cup_1 n'})
\end{align}
with $*$ denoting the stacking operation.  This is in accord with the supercohomology data group law found in Ref. [\onlinecite{Bhardwaj}] through continuum space-time methods.


\subsection{Equivalence relation on supercohomology data} \label{sec:superequivalencerelation}

The stacking rules allow us to define a physically motivated notion of equivalence between two sets of supercohomology data, which agrees with the mathematical one given in e.g. Ref. [\onlinecite{KapustinThorngren}]. We will say that two sets of supercohomology data are equivalent if the corresponding fermionic SPT Hamiltonians $\hat{H}_\text{f}$, constructed in section \ref{sec:fSPT}, are in the same phase.  

Consider the supercohomology data\cite{Note401} 
\begin{align} \label{eq:trivialdata}
(n_0,\nu_0)=(\delta \beta, (-1)^{\beta \cup \delta \beta}\delta \omega),
\end{align}
where $\beta:G\times G \to \mz_2$ and $\omega:G \times G\times G \to U(1)$ are both homogeneous.
We claim that this set of data gives a finite depth circuit $\hatU_\text{b}^{n_0 \nu_0}$ built from symmetric local unitaries (up to factors of $\hatG_p$), i.e. the fermionic SPT phase corresponding to this set of data is trivial [\onlinecite{Chen_localunitaries}]. In Appendix \ref{ap:trivialcircuit}, we compute $\hatU_\text{b}^{n_0 \nu_0}$ in detail, and we simply state the result here:
\begin{align}
    \hatU_\text{b}^{n_0\nu_0}= &\prod_{\la pqr \ra}\hat{\omega}^{- {o}_{pqr}}_{pqr}(-1)^{\hat{\beta}_{pq}\hat{\beta}_{qr}}\\ \nonumber
    \times &\prod_{\la pqr \ra}\hatZ_{pq}^{\hat{\beta}_{qr}} \prod_{\la pq \ra}\hatX_{pq}^{\hat{\beta}_{pq}}\prod_{\la pqr \ra}\hatW_{pqr}^{\hat{\beta}_{pr}}\prod_p \hatG_p^{\hat{\beta}_p}.
\end{align}
Above, $\hat{\omega}_{pqr}$, $\hat{\beta}_{pq}$, and $\hat{\beta_{p}}$ are defined by
\begin{align} \label{omega}
    \hat{\omega}_{pqr}|\{g_t\}\ra &= \omega(g_p,g_q,g_r)|\{g_t\}\ra \\ \label{betapq}
    \hat{\beta}_{pq}|\{ g_t \}\ra&= \beta(g_p,g_q)|\{ g_t \}\ra \\ \label{betap}
    \hat{\beta}_p |\{ g_t \}\ra&= \beta(g_p,1)|\{ g_t \}\ra.
\end{align}
The local unitary operators in $\hatU_\text{b}^{n_0\nu_0}$ (besides $\hatG_p^{\hat{\beta}_p}$) are then manifestly symmetric due to the homogeneity properties of $\beta$ and $\omega$. Fermionization maps $\hatG_p^{\hat{\beta}_p}$ to the identity, so the finite depth circuit $\hatU_\text{f}^{n_0 \nu_0}$ obtained from fermionization is indeed built from symmetric local unitaries.  Hence, $\hatU_\text{f}^{n_0 \nu_0}$ applied to a trivial product state gives us a trivial SPT.  

Stacking a trivial SPT phase leaves the system in the same phase. Therefore, composition of $\hatU_\text{b}^{n_0 \nu_0}$ with $\hatU_\text{b}^{n \nu}$ should give us a circuit corresponding to some supercohomology data that is equivalent to $(n,\nu)$.  According to the composition rules (\ref{stackingrule}) in the previous subsection, the product $\hatU_\text{b}^{n_0 \nu_0}\hatU_\text{b}^{n \nu}$ is the circuit corresponding to the supercohomology data\footnote{Here we use the definition of $\cup_1$ (see Appendix A of [\unexpanded{\onlinecite{KapustinThorngren}}]) to write 
\begin{align}
    \unexpanded{n\cup_1 \delta \beta=n \cup \beta + \beta \cup n + \delta(n \cup_1 \beta)}.
\end{align}}
\begin{align}\label{superrel0}
    \left(n+\delta \beta, \nu (-1)^{\beta \cup \delta \beta + n \cup \beta + \beta \cup n}\delta \left[\omega(-1)^{n \cup_1 \beta}\right]\right).
\end{align}
$\omega(-1)^{n \cup_1 \beta}$ in (\ref{superrel0}) is some homogeneous function, which we will denote as $\eta$, from $G \times G \times G$ to $U(1)$.  Therefore, two sets of supercohomology data $(n,\nu)$ and $(n',\nu')$ are equivalent if there exists a homogeneous function $\beta:G \times G\to \mz_2$ and homogeneous function $\eta:G \times G \times G\to U(1)$ such that 
\begin{align} \label{eq:equiv}
    n'&=n+\delta \beta \\ \nonumber
    \nu'&=\nu (-1)^{\beta \cup \delta \beta + n \cup \beta + \beta \cup n}\delta \eta.
\end{align}
It can be checked that this is a symmetric and transitive relation, and hence defines an equivalence relation.  In what follows, we will show that two sets of group supercohomology data that are inequivalent with respect to this relation necessarily give rise to distinct SPT phases.

\subsection{Quantized invariants for fermionic SPT phases}\label{sec:quantizedinvariants}

We are now in a position to establish group supercohomology data as quantized invariants for fermionic SPT phases at the level of gapped lattice Hamiltonians. In the previous subsection, two sets of supercohomology data were said to be equivalent if they correspond to the same fermionic SPT phase.  Therefore, we need only argue that inequivalent sets of supercohomology data necessarily correspond to distinct fermionic SPT phases. 

Suppose $(n',\nu')$ and $(n'',\nu'')$ are inequivalent choices of group supercohomology data with respect to the equivalence relation (\ref{eq:equiv}).  We will show that the corresponding models are in distinct SPT phases.  First, we stack the phase corresponding to $(n',\nu')$ with the inverse of the phase corresponding to $(n'',\nu'')$.  Then, using the fact that SPT phases form an abelian group under stacking, the two phases will be distinct if and only if
\begin{align}
    (n,\nu)\equiv(n',\nu')*(n'',\nu'')^{-1}
\end{align} 
gives rise to a nontrivial fermionic SPT phase.  In other words, we need to demonstrate that $(n,\nu)$ corresponds to a nontrivial phase whenever it is not of the form (\ref{eq:trivialdata}).  

To show that the phase corresponding to $(n,\nu)$ is nontrivial, we bosonize it, i.e. reverse the fermionization procedure described above.  This should simply return our bosonic shadow model.  However, because the bosonization dictionary is many-to-one, in the sense that all the $\hatG_p$ operators map to the identity on the fermionic side, we have to define our bosonization procedure carefully to avoid ambiguities.  We do this by dressing each local term on the bosonic side with a projector onto the $\hatG_p=1$ Hilbert space everywhere in the vicinity of that term and by adding a term $-\sum_p\hatG_p$ to ensure that the ground state is in the $\hatG_p=1$ subspace.  It is important to note that this bosonization can be performed for any gapped fermionic Hamiltonian defined on our Hilbert space, not just on our specific fixed point model.  Now, having mapped the fermionic SPT Hamiltonian corresponding to $(n,\nu)$ to a bosonic symmetry enriched toric code Hamiltonian, we  look for quantized invariants of the symmetry enriched model that can then be pulled back to give fermionic SPT invariants.

If $(n,\nu)$ is nontrivial, i.e. not of the form (\ref{eq:trivialdata}), then there are two cases.  The first is that $n$ cannot be written in the form $n=\delta \beta$ for any choice of $\beta$ ($\beta$ defined below (\ref{eq:trivialdata})).  The second is that $n$ can be written as $\delta \beta$, but $\nu$ is nontrivial (clarified below).  We treat these cases in turn.

{\bf{Case 1:}} 
Assume that $n$ cannot be written as $\delta \beta$.  Then, after bosonizing, we will show that the fermion parity flux excitations ($e$ or $m$ excitations of the bosonic shadow model) carry the nontrivial fractionalization class $n \in H^2(G,\Z_2)$.
Starting with the ground state of the bosonic shadow model $|\Psi_\text{b}\ra$, we can create a pair of $e$ excitations at some well separated vertices $a$ and $b$ by applying a string operator.  From this state, a low energy Hilbert space ${\cal{H}}_{\text{L}}$ is obtained by projecting onto fixed values of the $G$-spins $g_a$ and $g_b$ at vertices $a$ and $b$, respectively.  ${\cal{H}}_{\text{L}}$ has dimension $|G|^2$, with a natural basis $\{|\Psi^{ee}_\text{b};g_a, g_b\ra\}$.  Explicitly,
\begin{align}
|\Psi^{ee}_\text{b};g_a, g_b\ra=\hatU_\text{b}^{n \nu} |\Psi^{ee}_\text{t.c.};g_a,g_b\ra,
\end{align}
where $|\Psi^{ee}_\text{t.c.};g_a,g_b\ra$ is the toric code state consisting of two $e$ excitations at $a$ and $b$ respectively, tensored with a trivial $G$-spin paramagnet on all vertices $p \neq a,b$ and $G$-spins at $a$ and $b$ fixed to $g_a$ and $g_b$, respectively.

Letting $\hat{V}(g)$ be the global on-site symmetry operator corresponding to the group element $g$, we now compute $\hat{V}(g)|\Psi^{ee}_\text{b};g_a, g_b\ra$:
\begin{align}
\hat{V}(g)|\Psi^{ee}_\text{b};g_a, g_b\ra = \hat{V}(g) \hatU_\text{b}^{n \nu} |\Psi^{ee}_\text{t.c.};g_a,g_b\ra.
\end{align}
Using the fact (proved in Appendix \ref{ap:symH}) that $\hatU_\text{b}^{n \nu}$ is symmetric up to factors of $\hatG_p$:
\begin{align}\label{ucircsymmetry}
    \hat{V}(g)\hatU_\text{b}^{n \nu}=\hatU_\text{b}^{n \nu} \prod_p \hatG_p^{\hat{n}^g_p} \hat{V}(g),
\end{align}
where $\hat{n}^g_p$ is defined by
\begin{align}\label{neigenvalues}
    \hat{n}^g_p|\{ g_t \}\ra=n(g_p,1,g)|\{ g_t \}\ra,
\end{align}
we have 
\begin{align} \label{longeq}
\hat{V}(g)|\Psi^{ee}_\text{b};g_a, g_b\ra &= \hatU_\text{b}^{n \nu}\prod_p \hatG_p^{\hat{n}^g_p}|\Psi^{ee}_\text{t.c.};gg_a,gg_b\ra \nonumber \\ &= \hatU_\text{b}^{n \nu} 
(-1)^{\hat{n}^g_a} (-1)^{\hat{n}^g_b}|\Psi^{ee}_\text{t.c.};gg_a,gg_b\ra \nonumber \\
&=(-1)^{n(gg_a,1,g)+n(gg_b,1,g)} |\Psi^{ee}_\text{b};g g_a,gg_b\ra.
\end{align}
\noindent Focusing on just the $a$ vertex, we see from (\ref{longeq}) that the local effective action of $\hatV(g)$ near $a$ is given by the operator:
\begin{align}\label{localsymmetry}
    \hatV_a(g)|\Psi^{ee}_\text{b};g_a,g_b\ra = (-1)^{n(gg_a,1,g)}|\Psi^{ee}_\text{b};gg_a,g_b\ra.
\end{align}
With $\hatV_b(g)$ defined analogously, we recover $\hatV_a(g) \hatV_b(g)=\hatV(g)$, as required.  Note that there is a $g$ dependent sign ambiguity in the definition of this local effective action.  (A possible phase ambiguity is restricted to just an ambiguity in sign by the $\mz_2$ fusion rules of the $e$ excitations [\onlinecite{Maissam2014,Tarantino2015}].)

The fractionalization class captures the failure of the symmetry group law to be satisfied by the effective symmetry action on a single anyon.  To compute this fractionalization class, we therefore compute the phase difference between $\hatV_a(g) \hatV_a(h)$ and $\hatV_a(gh)$.  For $\hatV_a(g) \hatV_a(h)$, we have
\begin{align}\label{VgVh}
    \hatV_a(g) &\hatV_a(h) |\Psi^{ee}_\text{b};g_a,g_b\ra = \nonumber \\
    &= \hatV_a(g) (-1)^{n(hg_a,1,h)}|\Psi^{ee}_\text{b};hg_a,g_b\ra \nonumber \\
    &= (-1)^{n(ghg_a,1,g)+n(hg_a,1,h)}|\Psi^{ee}_\text{b};g h g_a,g_b\ra,
\end{align}
while for $\hatV_a(gh)$, we have
\begin{align}\label{Vgh}
    \hatV_a(gh)|\Psi^{ee}_\text{b};g_a,g_b\ra = (-1)^{n(ghg_a,1,gh)} |\Psi^{ee}_\text{b};g h g_a,g_b\ra.
\end{align}
Using $\delta n=0$ and the homogeneity of $n$, we see that the difference in sign between the far right hand side of (\ref{VgVh}) and the right hand side of (\ref{Vgh}) is precisely $(-1)^{n(1,g,gh)}$.  Thus, the fractionalization class of the local symmetry action is indeed given by $n$.  Accounting for the $g$ dependent sign ambiguity in the local symmetry action noted just below (\ref{localsymmetry}), one can show [\onlinecite{Maissam2014,Tarantino2015,Else14}] that the symmetry fractionalization is well defined with $n \in H^2(G,\mz_2)$.

The nontrivial symmetry action on the fermion parity fluxes indicates that the bosonic shadow model corresponding to $(n,\nu)$ is in a nontrivial symmetry enriched phase [\onlinecite{Maissam2014,Cheng15}].  Pulling back via bosonization, this implies that the fermionic SPT corresponding to $(n,\nu)$ is nontrivial.  Hence, the fermionic SPT phases given by $(n',\nu')$ and $(n'',\nu'')$ are distinct.

Alternatively, the nontrivial symmetry fractionalization can be seen more informally by recalling that the shadow model comes from gauging the $\Z_2$ subgroup of $\tG$, with $\tG$ the $\Z_2$-extension of $G$ determined by $n$.  Therefore, the $G$ group law relations close only modulo a $\Z_2$ gauge transformation, and the fermion parity flux, being charged under this gauged $\Z_2$, acquires minus signs corresponding to the fractionalization class $n$ when acted on by global $G$ symmetry.

{\bf{Case 2:}} Now, suppose instead that $n$ is trivial, i.e. $n=\delta \beta$. Then using the equivalence relation (\ref{eq:equiv}), we can `gauge' $n$ away entirely, so that the supercohomology data $(n,\nu)$ is equivalent to $(0,\tilde{\nu})$, with $\delta \tilde{\nu}=0$. For $(0,\tilde{\nu})$ to be nontrivial, it must be that there does not exist an $\omega$ (as defined below (\ref{eq:trivialdata})) such that $\delta \omega=\tilde{\nu}$. That is to say, $\tilde{\nu}$ must be nontrivial in $H^3(G,U(1))$. 

The fixed point fermionic circuit $\hatU_\text{f}^{0\tilde{\nu}}$ corresponding to $(0,\tilde{\nu})$ acts trivially on the fermionic degrees of freedom, whereas the portion of it that acts on the bosonic $G$-spin degrees of freedom is precisely the circuit that constructs a group cohomology SPT ground state from a trivial product state [\onlinecite{Chen2013}].  To see that this system is nontrivial as a fermionic SPT, we bosonize the system. The result is a trivial toric code phase stacked with the bosonic group cohomology phase corresponding to $\tilde{\nu}$.
This symmetry enriched toric code is precisely what one obtains from gauging the $\Z_2$ subgroup of $G \times \Z_2$ in the ordinary bosonic SPT of $G \times \Z_2$ with cocycle $\tilde{\nu} \otimes 1$. $\tilde{\nu} \otimes 1$ is nontrivial in $H^3(G\times\mz_2,U(1))$ by K{\"u}nneth's theorem [\onlinecite{Munkres}] and the assumption that $\tilde{\nu}$ is nontrivial.

We have thus shown that (\ref{eq:equiv}) generates the maximal possible set of equivalence relations on supercohomology data, with inequivalent data necessarily giving rise to distinct phases. A subtle point is that the fermionic phases corresponding to inequivalent sets of supercohomology data $(n',\nu')$ and $(n'',\nu'')$ might still bosonize into the same symmetry enriched toric code phase [\onlinecite{Cheng15}]. Hence, it was important in the above argument to bosonize the model corresponding to $(n',\nu')*(n'',\nu'')^{-1}$, rather than bosonizing those corresponding to $(n',\nu')$ and $(n'',\nu'')$ individually.  This subtlety arises in the the $G=\Z_2$ example, which we discuss below.

\subsection{Example: $G=\Z_2$}

For $G=\Z_2$, we have
\begin{align}
   \hatU^{\pm}_{\text{b}}=&\prod_{\la pqr \ra} (\pm i)^{o_{pqr}\hats_p(1-\hats_q)\hats_r}  \hatZ_{pq} ^{(1-\hats_q)\hats_r} \\ \nonumber
    \times &\prod_{\la pq \ra}\hatX_{pq}^{(1-\hats_p)\hats_q}\prod_{\la pqr \ra}\hatW_{ pqr }^{(1-\hats_p)\hats_r}.
\end{align}
Let us square this circuit.  Then the sign in (\ref{cuponesign}) is just $+1$, so that, according to (\ref{compositioncircuit}), we get
\begin{align}
\left(\hatU_{\text{b}}^{\pm}\right)^2 = \prod_{\la pqr \ra} \hat{\nu}^{2}_{pqr}.
\end{align}
But
\begin{align}
\hat{\nu}^{2}_{pqr}=(-1)^{\hats_p (1-\hats_q) \hats_r}
\end{align}
is just the nontrivial cocycle in $H^3(\Z_2,U(1))$ evaluated on $(s_p,s_q,s_r,0)$.  Therefore the circuit $\left(\hatU_{\text{b}}^{\pm}\right)^2$ builds the nontrivial bosonic $\Z_2$ SPT [\onlinecite{LevinGu,Chen2013}].  Thus, stacking two identical copies of either the $+$ or $-$ group supercohomology phase results in the nontrivial bosonic $\Z_2$ SPT phase, and in this sense, these group supercohomology phases are `square roots' of the bosonic phase.

Note that bosonizing the $+$ and $-$ phases actually results in the same symmetry enriched topological order.  Indeed, after gauging the $\Z_2$ global symmetry, the resulting twisted $\Z_4$ topological orders are the same.  This can be seen from the fact that both can be obtained by gauging $\Z_4$ in the corresponding auxiliary $\Z_4$ bosonic SPTs, and the $3$-cocycles defining these SPTs differ by the generator of $H^3(\Z_2,U(1))$ pulled back to $H^3(\Z_4,U(1))$, which is trivial.  Thus, the $+$ and $-$ phases cannot be distinguished in this simple way; nevertheless, we know they correspond to distinct fermionic SPT phases by the argument in the previous section.

\section{Discussion} \label{sec:discussion}


We have shown how to use group supercohomology data $(n,\nu)$, together with a choice of spin structure on a 2d oriented manifold $M$, to construct a corresponding lattice fermionic SPT Hamiltonian on $M$.  Our procedure cleanly disentangles the roles of the supercohomology data and spin structure.  The former is used to build a bosonic `shadow' model, and the latter to fermionize this model.  Another advantage of our procedure is that it explicitly builds the finite depth circuit of local unitaries $\hatU_{\text{f}}$ that creates the desired fermionic SPT ground state from a product state.  Our SPT Hamiltonian is then
\begin{align}
    \hat{H}_{\text{f}}=\hatU_{\text{f}}\hat{H}_{\text{f}}^0\hatU_{\text{f}}^\dag,
\end{align}
\noindent where $\hat{H}_{\text{f}}^0$ is the trivial fermionic Hamiltonian - an atomic insulator tensored with a trivial $G$-spin paramagnet whose ground state is a product state with zero fermion occupancy. 
\noindent Key to this approach is the fact that the circuit $\hatU_{\text{f}}$ is $G$-symmetric.  This is the case despite the fact that the individual local unitaries that make up the circuit cannot all be $G$-symmetric, for otherwise the fermionic SPT would be trivial.  Note that while we have assumed that the global action of $G$ is unitary, we expect our construction to generalize to anti-unitary symmetries with only minor modifications.

Our commuting projector Hamiltonians suffice to show that the supercohomology phases protected by abelian groups [\onlinecite{Potter16}] can be many-body localized. The couplings in the Hamiltonian\footnote{Strictly speaking, the argument holds for a slightly modified \unexpanded{$\hat{H}^0_{\text{f}}$} (see appendix A of Ref. [\href{https://arxiv.org/abs/1506.00592}{23}]).} $\hat{H}^0_{\text{f}}$ can be disordered (or made quasi-periodic), leading to a many-body localized Hamiltonian [\onlinecite{1d_MBL_SPT_1, 1d_MBL_SPT_2, Chandran1, Bauer1, Potter_Vishwanath}], or at least one that has a long thermalization time scale.  Since  $\hat{H}_{\text{f}}$ is just the conjugate of $\hat{H}^0_{\text{f}}$ by a finite depth circuit, the same is true of $\hat{H}_{\text{f}}$.

The fermionization duality that was used to construct our zero correlation length lattice models can also be reversed and used to bosonize fermionic SPT Hamiltonians.  This, together with a reinterpretation of the stacking structure of SPT phases in terms of composition of the corresponding finite depth circuits, allows well established invariants of the bosonic symmetry enriched toric code to be pulled back to these fermionic SPT Hamiltonians.  The result is a classification of fermionic supercohomology SPT phases, with inequivalent supercohomology data necessarily defining distinct phases.   


One may ask whether a similar construction is possible for the so-called beyond supercohomology phases [\onlinecite{Tarantino16, Gu_Wang, KapustinThorngren, Bhardwaj, Cheng15, Chenjie_Gu}] in $2+1$D.  That is, for these phases, can an exactly solvable model be obtained by conjugating a trivial fermionic Hamiltonian by a symmetric finite depth circuit of local unitaries?  We argue that, in contrast to the supercohomology phases, the answer is no. In particular, {we claim that} the ground states of beyond supercohomology phases cannot be constructed from a trivial product state by applying a globally symmetric finite depth circuit of local unitaries.  See Appendix \ref{beyondsuper} for further discussion. 

It is worth discussing the relation of our work to previous work. Supercohomology models were introduced in the pioneering work of Ref. [\onlinecite{Gu_Wen}], where wave functions for these models were written from a lattice path integral.  However, the wave functions were only explicitly constructed on a specific planar lattice and required seemingly arbitrary choices to account for a spin structure.  In Ref. [\onlinecite{GWW}], a related wave function, for the so-called fermionic toric code, was written down; this is the topological order that would result from gauging the global $\Z_2$ in our $G=\Z_2$ models.  The ground states were defined by graphical rules, but again, the spin structure was encoded in these rules in a non-manifest way.  The roles of the spin structure and group supercohomology data were disentangled in Ref. [\onlinecite{GK}], but only in a lattice spacetime formalism.  Ref. [\onlinecite{Bhardwaj}] extended this to beyond-supercohomology models, and also made the connection between the supercohomology data and the algebraic data defining the shadow models.  Insofar as lattice Hamiltonians, Refs. [\onlinecite{Tarantino16}, \onlinecite{Ware}] clarified the role of spin structure in beyond-supercohomology models, and Ref. [\onlinecite{Gu_Wang}] extended this to include supercohomology models; however, Ref. [\onlinecite{Gu_Wang}] did not write down explicit Hamiltonians, but rather defined the ground states implicitly using certain self-consistent lattice-deforming local rules.  The present work builds on these developments by constructing explicit Hamiltonians, as well as building the ground states explicitly using finite depth circuits, on oriented $2d$ manifolds of any topology.  It uses in an essential way the 2+1D bosonization duality introduced in Ref. [\onlinecite{Yu-an17}].

There are many possible avenues for future work.  One would be to extend this formalism to group supercohomology models in three spatial dimensions.  
Another avenue is to extend the present formalism to more complicated groups than $G \times \Z_2^{\text{f}}$, such as ones where the fermion parity symmetry forms a nontrivial subgroup of the overall symmetry.  Yet another possibility is to extend the quantum circuit formalism to beyond-supercohomology models, both in two and three spatial dimensions.  It may also be fruitful to understand our work in terms of tensor network states and operators. Indeed, preliminary investigations suggest that the bosonization duality can naturally be interpreted as a tensor network operator.  It would then be nice to understand the relation between the present work and the fermionic models written down in Ref. [\onlinecite{Bultinck17}]. {Futher}, our finite depth circuits could be used to study the edge theories of these fermionic SPT phases. Finally, it would be interesting to study the classification of symmetry enriched phases using finite depth circuits applied to the ground states of fixed point Hamiltonians. The circuits $\hatU_\text{b}$, introduced in section \ref{shadowhamiltonian}, construct ground states of symmetry enriched toric code phases from a trivial toric code state, and thus provide a nontrivial realization of such a construction.

\vspace{0.1in}
\noindent{\it Acknowledgements -- } We are grateful to Sujeet Shukla, Anton Kapustin, Frank Verstraete, Max Metlitski, Dan Freed, Ryan Thorngren, Dave Aasen, and especially Ashvin Vishwanath for useful conversations.  LF is supported by NSF DMR-1519579.

\begin{appendix}



\section{Derivation of the bosonic shadow theory ground state and the `standard' parent Hamiltonian} \label{shadowgsderivation}

In this appendix, we provide a derivation of the bosonic shadow theory ground state $\Psi_{\text{b}}$ introduced in section \ref{tGSPT}.  Recall that the first step of the construction is to form an auxiliary $\tG$ bosonic SPT from a choice of normalized\cite{Note1} supercohomology data ($n$,$\nu$), where $\tG$ is a $\mz_2$ extension of $G$ by $n$. The next step is to `gauge' the $\mz_2$ subgroup of $\tG$ in the standard way by minimally coupling the SPT Hamiltonian to a $\mz_2$ gauge field.  We will implement this procedure explicitly and argue that the symmetry enriched Hamiltonian obtained from this procedure - which we refer to as the `standard' symmetry enriched Hamiltonian - can in principle be fermionized since it commutes with the modified Gauss's law $\hatG_p$ for all $p$. 

\subsection{Gauging the $\mz_2 \subset \tG$}

As stated in (\ref{gsSPT}), the auxiliary $\tG$ SPT ground state wave function in the configuration basis is
\begin{align} \label{gsagain}
    \Psi_{\text{SPT}}\Big(\Big\{g^{(m_p)}_p\Big\}\Big)= \prod_{\la pqr \ra}\alpha\Big(g^{(m_p)}_p,g^{(m_q)}_q,g^{(m_r)}_r,\id\Big)^{o_{pqr}}.
\end{align}
(Recall that $\alpha$ can be expressed in terms of $n$, $\nu$, and $\ep$ using (\ref{def_alpha}).)  A simple Hamiltonian $\Hspt$ with this ground state is 
\begin{align} \label{HZ4}
\Hspt = - \hatU_{\text{SPT}} \left(\sum_p \hat{\tilde{P}}^{\text{sym}}_p \right) {\hat U}^{\dag}_{\text{SPT}}
\end{align}
where $\hatU_{\text{SPT}}$ is the finite depth circuit of local unitaries defined by matrix elements
\begin{align}\label{eq:UtG}
\big \lan& \big\{ h^{(\ell_p)}_p \big\}\big| \hatU_{\text{SPT}} \big|\big\{g^{(m_p)}_p \big\} \big \ran = \\ \nonumber  &\delta_{\big\{ h^{(\ell_p)}_p \big\},\big\{g^{(m_p)}_p \big\}}  \prod_{\la pqr \ra} \alpha\big(g^{(m_p)}_p, g^{(m_q)}_q, g^{(m_r)}_r, \id \big)^{o_{pqr}},
\end{align}
and $\hat{\tilde{P}}^{\text{sym}}_p$ is the projector onto the symmetric state at vertex $p$
\begin{align}
\frac{1}{\sqrt{|\tG|}}\sum_{g^{(m_p)}_p\in \tG} \big|g^{(m_p)}_p\big\ran
\end{align}
tensored with the identity on the remaining sites. 

We gauge the $\mz_2$ subgroup of $\tG$ using the usual algorithm as described in Ref. [\onlinecite{LevinGu}] and Ref. [\onlinecite{Chen_symfrac}]. First, we introduce at each link $\la pq \ra$ a spin-$\frac{1}{2}$ Hilbert space with Pauli operators $\mu^x_{pq}$ and $\mu^z_{pq}$, and at all sites $p$, impose the gauge constraint 
\begin{align}\label{eq:gauge1}
   \prod_{\la st \ra \ni p} \hat{\mu}_{st}^x = \hat{e}_p^{\mz_2}.
\end{align}
\noindent Here, the product runs over all links starting or ending on $p$, and ${\hat{e}}_p^{\mz_2}$ is the operator that on vertex $p$ takes
\begin{align} \label{def_gpz2}
\big|g^{(m_p)}_p \big\ra \rightarrow \big|g^{(m_p+1)}_p\big\ra
\end{align}
and acts as the identity on all other sites.  In other words, the action of ${\hat{e}}_p^{\mz_2}$ in the configuration basis is multiplication by the generator of the $\mz_2$ subgroup, $1^{(1)}$, with the assumption that $n$ is normalized. 

Second, we minimally couple each term in (\ref{HZ4}) to the $\Z_2$ gauge field degrees of freedom.  In order to make this gauging procedure unambiguous, we multiply each term by a projector onto trivial $\mz_2$ flux on triangles in the vicinity of that term, and add the term
\begin{align}\label{eq:noflux}
    -J\sum_{\la pqr \ra} \hat{\mu}^z_{pq}\hat{\mu}^z_{qr}\hat{\mu}^z_{pr}
\end{align}
\noindent with $J$ large enough to ensure that the ground state is in the trivial $\Z_2$ flux sector.  The result is a Hamiltonian $\hat{H}_\text{gauged}$ which is invariant under the gauge constraints in (\ref{eq:gauge1}). 

A ground state of $\hat{H}_\text{gauged}$ can be written as
\begin{align}\label{gsgauge}
     &\Psi_{\text{gauged}}\big(\big\{g^{(m_p)}_p\big\},\{ \mu^z_{pq}\}\big)= \\ \nonumber
     &\prod_{\la pqr \ra} \nu(g_p,g_q,g_r,\id)^{o_{pqr}}  \left(\mu^z_{pq}(-1)^{\ep\big(\big(g^{(m_p)}_p\big)^{-1} g^{(m_q)}_q\big)}  \right)^{n(g_q,g_r,\id)} \\ \nonumber
     &\times  \left( \prod_{\la pqr \ra} \delta_{W'_{pqr},1} \right ) h(\{\mu^z_{pq}\}).
\end{align}
The function $h$ determines the holonomy of the particular ground state. The ground state with trivial holonomy, for example, is obtained with the choice of $h$: 
\begin{align}
    h(\{\mu^z_{pq}\})=
\begin{cases}
1 & \text{if $\{\mu^z_{pq}\}\sim \{\mu^z_{pq}=+1\}$} \\
0 & \text{otherwise}
\end{cases}
\end{align}
\noindent where $\sim$ means `gauge equivalent to'.  For the ground states with nontrivial holonomy, $h$ is defined similarly.

The $\delta$ function in (\ref{gsgauge}) is a consequence of the flux penalizing term in the gauging procedure.  The $\mu^z$-flux is  
\begin{align}
    W'_{pqr}=\mu^z_{pq}\mu^z_{qr}\mu^z_{pr},
\end{align}
\noindent so the delta function 
\begin{align}
    \delta_{W'_{pqr},1}=
    \begin{cases}
    1 & \text{if $W'_{pqr}=1$}\\
    0 & \text{otherwise}
    \end{cases}
\end{align}
\noindent  ensures that all configurations in the ground states have trivial $\mu^z$-flux.  

In going from (\ref{gsagain}) to (\ref{gsgauge}), we have also multiplied $(-1)^{\ep\big(\big(g^{(m_p)}_p\big)^{-1} g^{(m_q)}_q\big)}$ by $\mu_{pq}^z$. This guarantees that $\Psi_{\text{gauged}}$ is gauge invariant  and reduces to $\Psi_\text{SPT}$ when all $\mu^z_{pq}=+1$.

\subsection{Mapping to unconstrained variables}

To obtain $\Psi_{\text{b}}$ as expressed in section \ref{tGSPT}, we must rewrite the system in terms of unconstrained variables.  To this end,
we define an isomorphism of operator algebras below. This isomorphism will allow us to convert $\hat{H}_\text{gauged}$ into $\hat{H}'_\text{b}$, a Hamiltonian acting on an unconstrained Hilbert space with ground state $\Psi_{\text{b}}$.  On one side of the isomorphism, we have the algebra $\cal{A}_\text{constrained}$ appearing in the previous subsection and consisting of gauge invariant operators, modulo the Gauss's law relation. On the other side of the isomorphism, we have $\cal{A}_\text{unconstrained}$, an operator algebra naturally represented on a tensor product Hilbert space with degrees of freedom matching those of the bosonic shadow theory.  We now define $\cal{A}_\text{constrained}$ and $\cal{A}_\text{unconstrained}$ more carefully and write an explicit isomorphism between the two algebras.


\subsubsection{Algebra of constrained operators $\cal{A}_\text{constrained}$}

$\cal{A}_\text{constrained}$ admits a representation on the Hilbert space discussed in the previous subsection, i.e. $\tG$ degrees of freedom on vertices and spin-$\frac{1}{2}$ degrees of freedom on links. 
It can generated by $\hat{e}_p^{\mz_2}$, $\hatg^{(0)}_p$, $\hat{\tilde{P}}_p^{g}$, $\hat{\mu}^x_{pq}$, and $(-1)^{\hat{\ep}_{pq}}\hat{\mu}^z_{pq}$ obeying the relation
\begin{align}\label{gaugerelations}
    \prod_{\la st \ra \ni p} \hat{\mu}_{st}^x = \hat{e}_p^{\mz_2}
\end{align}
for all $p$.  Here, $\hatg^{(0)}_p$ and $\hat{\tilde{P}}_p^{g}$ are the defined by their action on a configuration state: 
\begin{align}
    \hatg^{(0)}_p\big|h^{(\ell_p)}_p\big\ra=\big|(gh)^{(\ell_p)}_p\big\ra
\end{align}
and 
\begin{align}
    \hat{\tilde{P}}_p^{g}\big|\big\{h^{(\ell_q)}_q\big\}\big\ra=\delta_{g_p,h_p}\big|\big\{h^{(\ell_q)}_q\big\}\big\ra.
\end{align}
In words, $\hatg^{(0)}_p$ is the operator that multiplies by ${g^{(0)}}$ at vertex $p$ and acts as the identity elsewhere, while $\hat{\tilde{P}}_p^{g}$ is the projector onto the subspace spanned by states with configuration $g^{(0)}$ or $g^{(1)}$ at vertex $p$.  

Finally, $\hat{\ep}_{pq}$, appearing in the generator $(-1)^{\hat{\ep}_{pq}}\hat{\mu}^z_{pq}$, is given by
\begin{align}
    \hat{\ep}_{pq}\big|\big\{g^{(m_r)}_r\big\}\big\ra=\ep\big(\big(g^{(m_p)}_p\big)^{-1}g^{(m_q)}_q\big)\big|\big\{g^{(m_r)}_r\big\}\big\ra.
\end{align}
\noindent It can be checked that products of $\hatg^{(0)}_p$, $\hat{\tilde{P}}_p^{g}$ $\hat{\mu}^x_{pq}$, and $(-1)^{\hat{\ep}_{pq}}\hat{\mu}^z_{pq}$ span all gauge invariant operators.


\subsubsection{Algebra of unconstrained operators $\cal{A}_\text{unconstrained}$}

We will represent $\cal{A}_\text{unconstrained}$ on a tensor product Hilbert space comprised of $G$ degrees of freedom $|g_p\ra$ on vertices and spin-$\frac{1}{2}$ degrees of freedom on links.  The generators of this operator algebra acting on vertex Hilbert spaces are $\hatg_p$ and $\hatP_p^g$ defined by
\begin{align}
    \hatg_p|h_p\ra=|(gh)_p\ra
\end{align}
and
\begin{align}
    \hatP_p^g|\{ h_q \}\ra=\delta_{g_p,h_p}|\{ h_q \}\ra.
\end{align}
We take generators acting on the link Hilbert spaces to be the Pauli operators $\hatX_{pq}$ and $\hatZ_{pq}$.  

\subsubsection{Isomorphism of $\cal{A}_\text{constrained}$ with $\cal{A}_\text{unconstrained}$}

An isomorphism of $\cal{A}_\text{constrained}$ with $\cal{A}_\text{unconstrained}$ is given by the map of generators:
\begin{align}\label{constrainedisomorphism}
\begin{split}
\hat{e}^{\mz_2}_p &\longleftrightarrow \prod_{\la st \ra \ni p}\hatX_{st}\\
  \hatg^{(0)}_p &\longleftrightarrow \hatg_p \prod_{\substack{\la tq \ra \\ t=p}}\hatX_{tq}^{\hat{\xi}^{g_p}_{tq}}\prod_{\substack{\la qt \ra \\ t=p}}\hatX_{qt}^{\hat{\xi}^{g_p}_{qt}} \\
  \hat{\tilde{P}}_p^{g}&\longleftrightarrow \hatP_p^g \\
  \hat{\mu}^x_{pq} &\longleftrightarrow \hatX_{pq} \\
  (-1)^{\hat{\ep}_{pq}}\hat{\mu}_{pq}^z&\longleftrightarrow \hatZ_{pq}
\end{split}
\end{align}
where $\hat{\xi}^{g_p}_{pq}$ and $\hat{\xi}^{g_p}_{qp}$ are defined by
\begin{align}
    \hatg^{(0)}_p(-1)^{\hat{\ep}_{pq}}=(-1)^{\hat{\xi}^{g_p}_{pq}}(-1)^{\hat{\ep}_{pq}}\hatg^{(0)}_p
\end{align}
and
\begin{align}
    \hatg^{(0)}_p(-1)^{\hat{\ep}_{qp}}=(-1)^{\hat{\xi}^{g_p}_{qp}}(-1)^{\hat{\ep}_{qp}}\hatg^{(0)}_p.
\end{align}
Explicitly, $\hat{\xi}^{g_p}_{pq}$ and $\hat{\xi}^{g_p}_{qp}$ act on configuration states as
\begin{align}
    &\hat{\xi}^{g_p}_{pq}|\{ h_t \}\ra=\\ \nonumber &(n(1,h_p,h_q)+n(1,g_p,g_ph_p)+n(1,g_ph_p,h_q))|\{ h_t \} \ra
\end{align}
and
\begin{align}
    &\hat{\xi}^{g_p}_{qp}|\{ h_t \}\ra=\\ \nonumber &(n(1,h_q,h_p)+n(1,g_p,g_ph_p)+n(1,h_q,g_ph_p))|\{ h_t \} \ra.
\end{align}
$\hat{\xi}^{g_p}_{pq}$ and $\hat{\xi}^{g_p}_{qp}$ defined in this way ensure that the commutation relations exhibited by $\hatg^{(0)}_p$ and $(-1)^{\hat{\ep}_{pq}}\hat{\mu}_{pq}^z$ are mirrored on the right hand side of the mapping (\ref{constrainedisomorphism}).  Note that the isomorphism is well defined since, for all $p$, the relation 
\begin{align}
    \prod_{\la st \ra \ni p} \hat{\mu}_{st}^x = \hat{e}_p^{\mz_2}
\end{align}
is mapped to the identity.

Now, given a system described in terms of the operators in $\cal{A}_\text{constrained}$, one can rewrite it as a system in terms of the operators belonging to $\cal{A}_\text{unconstrained}$.  In particular, we can apply this isomorphism to $\hat{H}_\text{gauged}$ to obtain $\hat{H}'_\text{b}$ acting on an unconstrained Hilbert space.  $\hat{H}'_\text{b}$ has the ground state 
\begin{align}
     &\Psi_{\text{b}}(\{g_p\},\{ Z_{pq}\})= \\ \nonumber
     &\prod_{\la pqr \ra} \nu^{o_{pqr}}(g_p,g_q,g_r,\id)  Z_{pq}^{n(g_q,g_r,\id)} \\ \nonumber
     \times  &\left( \prod_{\la pqr \ra}  \delta_{W_{pqr},(-1)^{n(g_p,g_q,g_r)}} \right)  h(\{ Z_{pq}(-1)^{n(1,g_p,g_q)}\}),
\end{align}
\noindent which is precisely the ground state of the bosonic shadow theory identified in section \ref{tGSPT}.

\subsection{Fermionizability of `standard' Hamiltonian}

To conclude this appendix, we prove that $\Hstd$ is fermionizable, i.e. $\Hstd$ commutes with the modified Gauss's law $\hatG_p$ for all sites $p$.  To do this, we first note that $\hat{H}_\text{SPT}$ in (\ref{HZ4}) commutes with 
\begin{align} \label{eq:Z2flip}
 \hatU_{\text{SPT}} {\hat{e}}_p^{\mz_2} {\hat U}^{\dag}_{\text{SPT}}
\end{align}
which follows simply from the fact that ${\hat e}_p^{\mz_2}$ commutes with ${\hat P}^{\text{sym}}_p$.  After gauging the $\mz_2$ subgroup of $\tG$, $\Hstd$ must commute with the gauged version of (\ref{eq:Z2flip}), which, using the definition of $\alpha$ along with $\delta n=0$ we find to be equal to
\begin{align} \label{eq:term2}
\prod_{\substack{\la tqr \ra \\ t=p}} (-1)^{\hat{n}_{pqr}} \prod_{\la st \ra \ni p} {\hat X}_{st}.
\end{align}
\noindent Next, we see that (\ref{eq:noflux}), the term penalizing $\mu^z$-flux in triangle $\lan pqr \ran$, turns into  
\begin{align} \label{eq:term1}
J(-1)^{\hat{n}_{pqr}} {\hat{Z}}_{pq} {\hat{Z}}_{qr} {\hat{Z}}_{pr}. 
\end{align}
when written in terms of the unconstrained variables (here we used the definition of the group law of $\tG$ to simplify the expression).

Then, multiplying the operator in (\ref{eq:term2}) by the product of the terms in (\ref{eq:term1}) over all triangles whose first vertex is $p$ yields an operator proportional to 
\begin{align}
\prod_{\substack{\la tqr \ra \\ t=p}} \hatW_{tqr} \prod_{\la st \ra \ni p} {\hat X}_{st},
\end{align}
which is just $\hatG_p$, as defined in (\ref{def:modifiedgausslaw}).  Since the Hamiltonian $\Hstd$ commutes with both (\ref{eq:term1}) and (\ref{eq:term2}), it must commute with ${\hat G}_p$ as well.  Thus $\Hstd$ is fermionizable.  However, explicitly fermionizing it is unwieldy in general, as we do not have an explicit expression for it in terms of the modified Gauss's law invariant operators $\hatW_{pqr}$ and $\hatU_{pq}$.  For this reason, we constructed and worked with the parent Hamiltonian $\hat{H}_\text{b}$ in section \ref{shadowhamiltonian} instead.

\section{Symmetry of the shadow model Hamiltonian} \label{ap:symH}

In this appendix we prove that $\hat{H}_{\text{b}}$ is $G$-symmetric, as claimed in section \ref{shadowhamiltonian}.  It follows that the fermionic SPT Hamiltonian $\hat{H}_{\text{f}}$ constructed in section \ref{sec:fSPT} is also $G$-symmetric, since the fermionization procedure commutes with the global $G$-symmetry action. Concretely, letting $\hat{V}(g)$ be the global symmetry operator representing $g \in G$ and acting as
\begin{align}
|g_p\ra \rightarrow |gg_p\ra
\end{align}
\noindent on every vertex degree of freedom, we will show that $\hat{V}(g)$ commutes with $\hat{H}_{\text{b}}$.

Recall that
\begin{align}
\hat{H}_{\text{b}} = \hatU_{\text{b}}\hat{H}^0_{\text{b}} \hatU^\dag_{\text{b}}
\end{align}
\noindent so that
\begin{align}\label{conjH}
   \hat{V}(g) \hat{H}_{\text{b}}\hat{V}^\dag(g)=\left(\hat{V}(g) \hatU_{\text{b}}\hat{V}^\dag(g)\right)
    \hat{H}^0_{\text{b}}
   \left(\hat{V}(g) \hatU^\dag_{\text{b}}\hat{V}^\dag(g)\right)
\end{align} 
\noindent where we have used that $\hat{H}^0_{\text{b}}$ is symmetric.  

Let us now compute $\hat{V}(g) \hatU_{\text{b}}\hat{V}^\dag(g)$.  In (\ref{ucirc}),  $\hatU_{\text{b}}$ was defined as
\begin{align}
   \hatU_{\text{b}}=\prod_{\la pqr \ra} \left(\hat{\nu}_{pqr}^{o_{pqr}}  \hatZ_{pq} ^{\hat{n}_{qr}}\right) 
    \prod_{\la pq \ra}\hatX_{pq}^{\hat{n}_{pq}}\prod_{\la pqr \ra}\hatW_{ pqr }^{\hat{n}_{pr}},
\end{align}
\noindent with the operators $\hat{\nu}_{pqr}^{o_{pqr}}$ and $\hat{n}_{pr}$ defined just below (\ref{ucirc}).  To proceed, it is useful to first re-express $\hatU_{\text{b}}$ in terms of the generators of the bosonic algebra ${\cal A}_{\text{bos}}$.  Following (\ref{Uedgesprod}), the result is
\begin{align}
   \hatU_{\text{b}}= {\hat{\kappa}} \prod_{\la pq \ra}\hatU_{pq}^{\hat{n}_{pq}}\prod_{\la pqr \ra}\hatW_{pqr}^{\hat{n}_{pr}}.
\end{align}
\noindent Here, $\kappa$ is a unitary operator that acts as multiplication by a $\{|g_p\ran \}$-dependent eigenvalue and whose explicit form will not be required.

Conjugating by $\hat{V}(g)$ gives
\begin{align}\label{conjugated}
   \hat{V}(g) \hatU_{\text{b}}\hat{V}^\dag(g)=\hat{\phi}\hat{\kappa}\prod_{\la pq \ra}\hatU_{pq}^{\hat{n}^g_{pq}}\prod_{\la pqr \ra}\hatW_{pqr}^{\hat{n}^g_{pr}}.
\end{align}
\noindent Here $\hat{\phi}$ and $\hat{n}^g_{pq}$ are operators that act as multiplication by a $\{|g_p\ran \}$-dependent eigenvalue.  $\hat{\phi}$ is unitary and its explicit form will again not be required, whereas the eigenvalue of $\hat{n}^g_{pq}$ is $n(g^{-1}g_p,g^{-1}g_q,\id)=n(g_p,g_q,g)$.  The cocycle condition $\delta n=0$ gives
\begin{align}\label{somens}
n(g_p,g_q,g) =
    n(g_p,\id,g)+n(g_q,\id,g)+n(g_p,g_q,\id).
\end{align}
\noindent Thus, $\hat{n}^g_{pq}$ decomposes into three diagonal operators corresponding to the terms in (\ref{somens}), i.e. 
\begin{align}
    \hat{n}^g_{pq}=\hat{n}^g_p+\hat{n}^g_q+\hat{n}_{pq}.
\end{align}
\noindent If we substitute for $\hat{n}^g_{pq}$ and do some rearranging, the right hand side of (\ref{conjugated}) becomes
\begin{align}\label{conjugated2}
\hat{\phi}' \hatU_{\text{b}} \prod_{\la pq \ra}\hatU_{pq}^{\hat{n}^g_p+\hat{n}^g_q}\prod_{\la pqr \ra}\hatW_{pqr}^{\hat{n}^g_p+\hat{n}^g_r}
\end{align}
\noindent where, again, $\hat{\phi}'$ multiplies by a $\{|g_p\ran \}$-dependent eigenvalue whose precise form will not be required.  It is a combination of $\hat{\phi}$ and a phase picked up in commuting the $\hatU_{pq}$ operators.

Next, the product of $\hatU_{pq}$ in (\ref{conjugated2}) can be re-organized so that (\ref{conjugated2}) is 
\begin{align}
\hat{\phi}'' \hatU_{\text{b}} 
  \prod_p\left(\prod_{\substack{\la tq \ra \\ t=p}} {\hatU}_{tq}  \prod_{\substack{\la qt \ra \\ t=p}} {\hatU}_{qt}\right)^{\hat{n}^g_p} \prod_{\la pqr \ra}\hatW_{pqr}^{\hat{n}^g_p+\hat{n}^g_r}.
\end{align}
\noindent $\hat{\phi}''$ is yet another diagonal operator in the configuration basis. Employing the identity (\ref{bosidentity}), we thus find that $\hat{V}(g) \hatU_{\text{b}}\hat{V}^\dag(g)$ is equal to
\begin{align}\label{justastep}
\hat{\phi}'' \hatU_{\text{b}} 
  \prod_p
  \left(\hatG_p \prod_{\substack{\la tqr \ra \\ t=p}} \hatW_{tqr}  \prod_{\substack{\la qrt \ra \\ t=p}} \hatW_{qrt} \right)^{\hat{n}^g_p} \prod_{\la pqr \ra}\hatW_{pqr}^{\hat{n}^g_p+\hat{n}^g_r}.
\end{align}
\noindent The flux operators $\hatW_{pqr}$ in (\ref{justastep}) cancel, so we conclude
\begin{align}\label{almostthere}
    \hat{V}(g)\hatU_{\text{b}}\hat{V}^\dag(g)=
    \hat{\phi}''(\{g_p\}) \hatU_{\text{b}}\prod_p \hatG_p^{\hat{n}^g_p}.
\end{align}

Next, we show that $\hat{\phi}''$ in (\ref{almostthere}) must be $\id$.  Let us denote the ground state of $\hat{H}_{\text{b}}^0$ with trivial holonomy by $|\Psi_{\text{b}}^0\ra$.  It is a tensor product of trivial symmetric states
\begin{align}
\frac{1}{|G|} \sum_{g_p} |g_p\ran
\end{align}
at vertices $p$ with the trivial holonomy toric code ground state for the $Z_{pq}$ degrees of freedom.  The latter is just a superposition of all trivial holonomy $Z_{pq}$ configurations with trivial $\Z_2$-flux $W_{pqr}$ at every triangle. We then have the following chain of equalities
\begin{align}\label{punchline}
    |\Psi_{\text{b}}^0\ra
    &=\hat{V}(g)|\Psi_{\text{b}}^0\ra \\ \nonumber
    &=\hat{V}(g)\hatU_{\text{b}}{}^\dag\hatU_{\text{b}}|\Psi_{\text{b}}^0\ra \\ \nonumber
    &=\hat{V}(g)\hatU_{\text{b}}{}^\dag\hat{V}^\dag(g)\hatU_{\text{b}}|\Psi_{\text{b}}^0\ra \\ \nonumber
    &=\hat{\phi}''^*\hatU_{\text{b}}{}^\dag\hatU_{\text{b}}|\Psi_{\text{b}}^0\ra \\ \nonumber
    &=\hat{\phi}''^*|\Psi_{\text{b}}^0\ra.
\end{align}
\noindent In the first equality we used that $|\Psi_{\text{b}}^0\ra$ is symmetric, and in the third equality we used that $\hatU_{\text{b}}|\Psi_{\text{b}}^0\ra$ is symmetric. The fourth equality uses (\ref{almostthere}) and the fact that $\hatU_{\text{b}}|\Psi_{\text{b}}^0\ra$ belongs to the $\hatG_p=\id$ eigenspace for all $p$.

Now, comparing the far left hand side and the far right hand side of (\ref{punchline}), we can see that $\hat{\phi}''$ is trivial as follows.  $\hat{\phi}''$ is a diagonal operator in the configuration basis, while $|\Psi_{\text{b}}^0\ra$ contains an equal amplitude superposition over all $G$ configurations at vertices. For the equality to hold, it must be that $\hat{\phi}''$ has eigenvalue $1$ on all configurations. Hence, $\hat{\phi}''=\id$.  Looking back at (\ref{almostthere}), we therefore have
\begin{align}\label{conjU}
    \hat{V}(g)\hatU_{\text{b}}\hat{V}^\dag(g)=
     \hatU_{\text{b}}\prod_p \hatG_p^{\hat{n}^g_p}.
\end{align}
\noindent Substituting (\ref{conjU}) into (\ref{conjH}) and using the fact that $\hat{H}^0_{\text{b}}$ commutes with $\hatG_p$ to cancel the factors of $\hatG_p$, we find
\begin{align}
   \hat{V}(g) \hat{H}_{\text{b}}\hat{V}^\dag(g)=\hat{H}_{\text{b}}.
\end{align} 
\noindent Therefore, $\hat{H}_{\text{b}}$ is symmetric, and $\hat{H}_{\text{f}}$ is symmetric since fermionization commutes with the global $G$ symmetry.

\section{Graphical interpretation of spin structure dependent relation}\label{ap:cpproof}

\begin{figure}
\centering
\includegraphics[scale=.28,trim={4cm 0cm 3cm 0cm},clip]{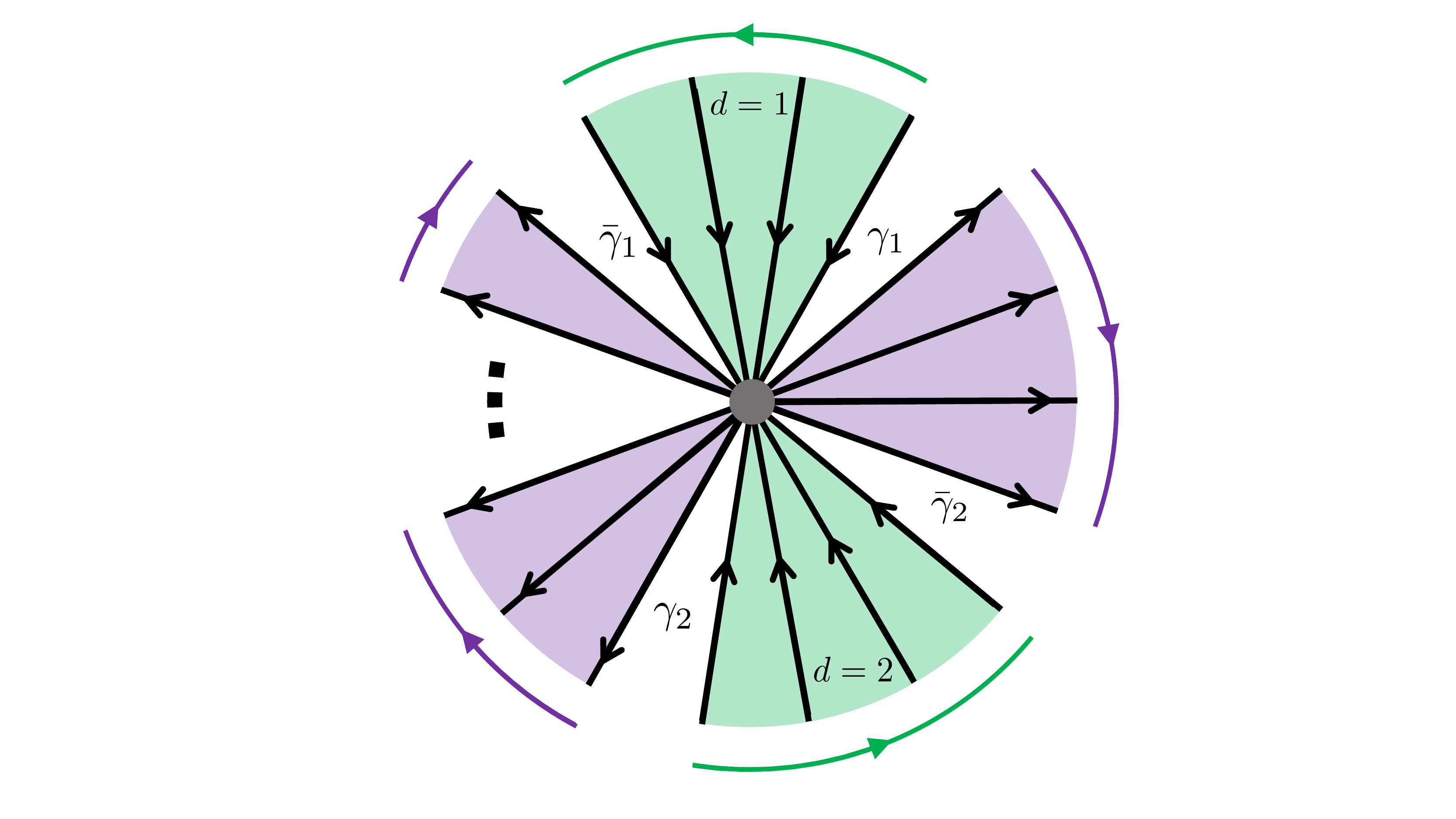}
\caption{The inward oriented domains consisting of $\la qrp \ra$-triangles are shaded in green, and the outward oriented domains consisting of $\la pqr \ra$-triangles are shaded in purple.  The green (purple) arrows show the ordering of the product over inward (outward) oriented links in (\ref{eq:edgeorder}).  We have also shown one of the two Majorana operators associated with each $\lan qpr \ran$-triangle, and re-labeled it with the subscript $d$ that labels inward pointing domains in order to make subsequent notation convenient.  The reason for only showing one of the two Majorana operators is that the one we have shown is the only one that enters into the computation of $c(p)$.}
\label{fig:S_eordering}
\end{figure}

Here we prove that $c(p)$, as defined in (\ref{def:cp}) and restated here for convenience:
\begin{align}
 \prod_{\substack{\la tq \ra \\ t=p}}  {\hat S'}_{tq} \prod_{\substack{\la qt \ra \\ t=p}}  {\hat S'}_{qt} = c(p) 
 \prod_{\substack{\la tqr \ra \\ t=p}} (-1)^{{\hat F}_{tqr}} 
 \prod_{\substack{\la qrt \ra \\ t=p}} (-1)^{{\hat F}_{qrt}},
\end{align}
is proportional to the identity operator, and equal to $\pm 1$ depending on whether the interpolating vector field ${\cal V}$, illustrated in FIG.\ref{fig:vecfieldorientation} has an even or odd winding number about $p$.  To see this, we first examine the two types of links around $p$.  There are links that are oriented towards $p$ and links that are oriented away from $p$.  These two types of links form domains as seen in FIG. \ref{fig:S_eordering}.  Domains of outward pointing links are separated from domains of inward pointing links by a triangle $\edge{qpr}$ where $p$ is the second vertex in the ordering.  There are necessarily an even number of triangles around $p$ for which $p$ is the second vertex.  We will call these types of triangles $\edge{qpr}$-triangles, and we will think of them in pairs -- the two $\edge{qpr}$-triangles on either side of an inward pointing domain forming a pair.  Moving counter-clockwise around $p$, we see that each pair results in a $2\pi$ clockwise rotation of the vector field ${\cal V}$, relative to the outward normal.  Without any $\edge{qpr}$-triangles, all the links are oriented towards $p$ or they are all oriented away from $p$, and the vector field rotates by $2\pi$. Therefore, the winding number of the interpolating vector field around $p$ is, modulo $2$, equal to $1-\frac{N_{qpr}}{2}$, where $N_{qpr}$ is the (even) number of triangles for which $p$ is the second vertex in the ordering.  

We will now show that $c(p)$ is $-(-1)^{\frac{N_{qpr}}{2}}$.  In terms of Majorana operators, the equation for $c(p)$ is
\begin{align} \label{eq:c1}
   c(p) = &\left ( \prod_{\substack{\la tq \ra \\ t=p}} i \gamma_{L_{tq}} \gammabar_{R_{tq}} \right) \left ( \prod_{\substack{\la qt \ra \\ t=p}} i \gamma_{L_{qt}} \gammabar_{R_{qt}} \right) \\ \nonumber
   \times &\left ( \prod_{\substack{\la tqr \ra \\ t=p}} -i \gamma_{{tqr}} \gammabar_{{tqr}} \right) \left ( \prod_{\substack{\la qrt \ra \\ t=p}} -i \gamma_{{qrt}} \gammabar_{{qrt}} \right).
\end{align}
\noindent Each term in the products over links (the first two products on the right hand side of (\ref{eq:c1})) has a factor of $i$.  The number of such factors of $i$ is equivalent to the total number of triangles having $p$ as a vertex.  We can thus assign each of these factors of $i$ to a different triangle having $p$ as a vertex.  Each term in the product over $\edge{pqr}$-triangles and $\edge{qrp}$-triangles (the last two products on the right hand side of (\ref{eq:c1})) contains a factor of $-i$. After multiplying out all of the factors of $i$ and $-i$ we are thus left only with an $i$ for each $\lan {qpr} \ran$-triangle.  Since these come in pairs we have
\begin{align}\label{eq:edgeorder}
   c(p) = &(-1)^{\frac{N_{qpr}}{2}} \\ \nonumber
   \times&\left ( \prod_{\substack{\la tq \ra \\ t=p}}  \gamma_{L_{tq}} \gammabar_{R_{tq}} \right) \left ( \prod_{\substack{\la qt \ra \\ t=p}}  \gamma_{L_{qt}} \gammabar_{R_{qt}} \right) \\ \nonumber
   \times &\left ( \prod_{\substack{\la tqr \ra \\ t=p}}  \gamma_{{tqr}} \gammabar_{{tqr}} \right) \left ( \prod_{\substack{\la qrt \ra \\ t=p}}  \gamma_{{qrt}} \gammabar_{{qrt}} \right).
\end{align}

\noindent Next, we notice that the terms in the product over inward pointing links all commute with each other.  Likewise, the terms in the product over outward pointing links all commute with each other.  Therefore, we may choose any ordering of the terms within each  product.  We choose to have the inward pointing link terms to be ordered counter-clockwise around $p$ and for the outward pointing link terms to be ordered clockwise as in FIG. \ref{fig:S_eordering}.  With this ordering, the two Majorana operators corresponding to each $\edge{pqr}$ triangle end up being positioned next to each other in the product over outward pointing links.  Similarly, the two Majorana operators corresponding to each $\edge{qrp}$ triangle end up positioned next to each other in the product over inward pointing links.  This accounts for all but two of the Majorana operators in each such product.  The remaining two Majorana operators appearing at the beginning and end of each product are located respectively on the two $\edge{qpr}$-triangles bordering each domain.  Numbering the inward oriented domains by a domain index $d=1,\ldots,D$, we then (by a slight abuse of notation) re-label these $\edge{qpr}$-triangle Majorana operators by $\gamma_d$, $\gammabar_d$, as illustrated in figure \ref{fig:S_eordering}.  Note that this labeling scheme accounts for only one Majorana operator located on each $\lan qpr \ran$-triangles; the other one does not appear in the expression for $c(p)$ and hence plays no role in the following. Moving the $\gammabar_d$ operators past the (even number of) other terms in each product, we obtain    
\begin{align}
    c(p)=&(-1)^{\frac{N_{qpr}}{2}} \\ \nonumber
    \times& \left( \prod^D_{d=1} \gammabar_{d+1} \gamma_d \right) \left( \prod_{\substack{\la tqr \ra \\ t=p}} \gammabar_{tqr} \gamma_{tqr} \right)  \\ \nonumber
    \times&  \left( \prod^D_{d=1} \gammabar_{d} \gamma_d \right) \left( \prod_{\substack{\la qrt \ra \\ t=p}} \gammabar_{qrt} \gamma_{qrt} \right)  \\ \nonumber
    \times&\left( \prod_{\substack{\la tqr \ra \\ t=p}} \gamma_{tqr} \gammabar_{tqr} \right) \left( \prod_{\substack{\la qrt \ra \\ t=p}} \gamma_{qrt} \gammabar_{qrt} \right)
\end{align}
\noindent where the subscript $D+1$ is meant to be read as $1$. 

After canceling the Majorana bilinears corresponding to $\edge{pqr}$-triangles and $\edge{qrp}$-triangles (last line in the product above), we find
\begin{align}
    c(p)&=(-1)^{\frac{N_{qpr}}{2}} \left( \prod^D_{d=1} \gammabar_{d+1} \gamma_d \right)   \left( \prod^D_{d=1} \gammabar_{d} \gamma_d \right)  \\ \nonumber 
    &=-(-1)^{\frac{N_{qpr}}{2}} \gamma_D \left( \prod^{D-1}_{d=1} \gammabar_{d+1} \gamma_d \right) \gammabar_1  \left( \prod^D_{d=1} \gammabar_{d} \gamma_d \right) \\ \nonumber
    &=-(-1)^{\frac{N_{qpr}}{2}}  \left( \prod^{D}_{d=1} \gamma_{d} \gammabar_d \right)   \left( \prod^D_{d=1} \gammabar_{d} \gamma_d \right) \\ \nonumber
    &=-(-1)^{\frac{N_{qpr}}{2}}. 
\end{align}
\noindent As we argued at the beginning of this appendix, $-(-1)^{\frac{N_{qpr}}{2}}$ is precisely $-1$ raised to the power of the winding number of the interpolating vector field ${\cal V}$. Thus, we have proved the claim: $c(p)$ is $-1$ when the interpolating vector field has an odd winding number at $p$ and $c(p)$ is $1$ otherwise.

\section{Fermion condensation and fermionization} \label{ap:fermioncondensation}

In this appendix, we will illustrate that the fermionization duality outlined in section \ref{sec:duality} and first described in Ref. [\onlinecite{Yu-an17}], can be interpreted as a fermion condensation procedure for certain lattice Hamiltonians.  

Fermion condensation has been thouroughly studied using a spacetime formulation [\onlinecite{Aasen17}] and admits the following intuitive picture.  We begin with a bosonic system with emergent fermions and introduce a system with bonafide physical fermions.  Next, we pair each emergent fermion with a physical fermion and the composite excitation, having bosonic statistics, is then condensed.  This results in a fermionic theory in which all particles braiding with the emergent fermion have been confined.  As argued in Ref. [\onlinecite{Bhardwaj}], it outputs a supercohomology SPT phase when applied to the corresponding bosonic shadow theory. 

To see the relation with the fermionization duality described in the main text, we must first develop fermion condensation at the lattice Hamiltonian level. To do so, we consider Hamiltonians defined on a Hilbert space consisting of spin-$\frac{1}{2}$ degrees of freedom on links, the same as that of the bosonic shadow models defined in section \ref{sec:bosonic}, and we assume that the Hamiltonians commute with $\hatG_p$, for all vertices $p$.  The restriction to Hamiltonians that commute with $\hatG_p$ can be motivated by interpreting $\hatG_p$ as a short closed emergent fermion string operator around the vertex $p$.  Thus, the Hamiltonians considered here have a particular emergent fermion string operator, which we describe in detail below.  We note that this particular string operator creates emergent fermion excitations in the bosonic shadow models constructed in section \ref{sec:bosonic}, since those models commute with $\hatG_p$.

We define the emergent fermion string operator $\hat{\widetilde{U}}_\Gamma$ by
\begin{align}
    \hat{\widetilde{U}}_\Gamma \equiv \prod_{\la pq \ra \in \Gamma} \hat{\widetilde{U}}_{pq},
\end{align}
where $\Gamma$ is a path in the dual lattice and $\hat{\widetilde{U}}_{pq}$ is 
\begin{align}
    \hat{\widetilde{U}}_{pq} \equiv  \hatX_{pq}\hat{\widetilde{K}}_{L_{pq}}\hat{\widetilde{K}}_{R_{pq}}.
\end{align} 
The action of $\hat{\widetilde{K}}_{R_{pq}}$ is dependent upon the triangle $R_{pq}$ to the right of $\la pq \ra$. If the triangle to the right of $\la pq \ra$ has vertex ordering $\la pqr \ra$, with $p$ and $q$ being the first and second vertices, respectively, then $\hat{K}_{R_{pq}}$ acts as $\hatZ_{qr}$. Otherwise, $\hat{K}_{R_{pq}}=\id$.  The action of $\hat{K}_{L_{pq}}$ is defined similarly but with `right' replaced with `left'.  Intuitively, $\hat{\widetilde{U}}_{pq}$ creates a pair of $\Z_2$ fluxes and moves $\Z_2$ charges so that they are bound to the fluxes at the third vertex in the vertex ordering. Letting $\Gamma$ be a path in the dual lattice around the vertex $p$, we find that $\hat{\widetilde{U}}_{\Gamma}$ is equal to $\hatG_p$ up to an inconsequential sign. 

We may now describe the fermion condensation procedure for a Hamiltonian $\hat{H}$ which commutes with $\hatG_p$.  First, as $\hatU_{pq}$ and $\hatW_{pqr}$ generate all the local operators that commute with $\hatG_p$, $\hat{H}$ can be expressed in terms of $\hatU_{pq}$ and $\hatW_{pqr}$. To make this explicit, we write $\hat{H}$ as $\hat{H}(\hatU_{pq},\hatW_{pqr})$.  Next, we introduce fermionic degrees of freedom into our system by adding a complex fermion degree of freedom to each triangle. The fermion parity even operators are generated by fermion parity $(-1)^{\hat{F}_{pqr}}$ and hopping operators $\hatS_{pq}$. Here, $\hatS_{pq}$ includes the spin structure related sign for edges in $\mathcal{E}$, as described in section \ref{sec:spinstructure}. 

The next step in fermion condensation is to bind physical fermions to emergent fermion excitations and to condense the composite particles.  The binding of physical fermions to emergent fermion excitations is accomplished by replacing $\hatW_{pqr}$ with $\hatW_{pqr}(-1)^{\hat{F}_{pqr}}$ so that
\begin{align}
    \hat{H}(\hatU_{pq},\hatW_{pqr}) \longrightarrow \hat{H}(\hatU_{pq},\hatW_{pqr}(-1)^{\hat{F}_{pqr}}).
\end{align}
In the resulting system, a pair of emergent fermions can be created for free as long as there is a physical fermion attached to each emergent fermion.  As a consequence, the Hamiltonian becomes highly degenerate.  This degeneracy, however, is eliminated by adding a term that proliferates emergent fermion-physical fermion pairs.  A pair of composite excitations is created by the operator $\hat{\widetilde{U}}_{pq}\hatS_{pq}$, so the term 
\begin{align} \label{Jcondense}
    -J\sum_{\la pq \ra} \hat{\widetilde{U}}_{pq}\hatS_{pq}, \quad J>0
\end{align}
energetically prefers states where the composite excitations have been proliferated.  Adding the term in \ref{Jcondense} to the Hamiltonian, we have 
\begin{align}
    \hat{H}(\hatU_{pq},\hatW_{pqr}&(-1)^{\hat{F}_{pqr}})  \nonumber \\ 
    &\longrightarrow \hat{H}(\hatU_{pq},\hatW_{pqr}(-1)^{\hat{F}_{pqr}})-J\sum_{\la pq \ra} \hat{\widetilde{U}}_{pq}\hatS_{pq}.
\end{align}
One may be concerned that the $J$-term will not proliferate the composite excitations as promised due to possible competition with $\hat{H}(\hatU_{pq},\hatW_{pqr}(-1)^{\hat{F}_{pqr}})$.  However, it can be shown that $\hatU_{pq}$ commutes with $\hat{\widetilde{U}}_{st}$ for every $\la st \ra$ and $\hatW_{pqr}(-1)^{\hat{F}_{pqr}}$ commutes with $\hat{\widetilde{U}}_{st}\hatS_{st}$ for every $\la st \ra$.  Therefore, $\hat{H}(\hatU_{pq},\hatW_{pqr}(-1)^{\hat{F}_{pqr}})$ commutes with $-J\sum_{\la pq \ra} \hat{\widetilde{U}}_{pq}\hatS_{pq}$ and the $J$-term is indeed minimized in the ground state. 

Finally, we drive the system deep into the fermion condensed regime and consider the limit as $J \to \infty$.  In the resulting effective Hilbert space, $\hat{\widetilde{U}}_{pq}\hatS_{pq}=1$ for all $\la pq \ra$. Thus, in this effective Hilbert space, the Hamiltonian acts as 
\begin{align}
    \hat{H}(\hatU_{pq}\hat{\widetilde{U}}_{pq}\hatS_{pq}, \hatW_{pqr}(-1)^{\hat{F}_{pqr}}),
\end{align}
where we have inserted $\hat{\widetilde{U}}_{pq}\hatS_{pq}=1$ and removed the $J$-term. Relabeling $\hatW_{pqr}(-1)^{\hat{F}_{pqr}}$ as $(-1)^{\hat{\widetilde{F}}_{pqr}}$ (note the tilde above $F$) and $\hatU_{pq}\hat{\widetilde{U}}_{pq}\hatS_{pq}$ as $\hat{\widetilde{S}}_{pq}$, we have
\begin{align}
    \hat{H}(\hat{\widetilde{S}}_{pq},(-1)^{\hat{\widetilde{F}}_{pqr}}).
\end{align}
$(-1)^{\hat{\widetilde{F}}_{pqr}}$ and $\hat{\widetilde{S}}_{pq}$ satisfy the same commutation relations as fermion parity operators $(-1)^{\hat{F}_{pqr}}$ and hopping operators $\hat{{S}}_{pq}$, respectively, and it can be checked that they satisfy a relation analogous to (\ref{Antonrelation}). 

Functionally, our prescription for fermion condensation maps a Hamiltonian $\hat{H}(\hatU_{pq},\hatW_{pqr})$ to $\hat{H}(\hat{\widetilde{S}}, (-1)^{\hat{\widetilde{F}}_{pqr}})$. In effect, we have replaced $\hatU_{pq}$ with $\hat{\widetilde{S}}_{pq}$ and $\hatW_{pqr}$ with $(-1)^{\hat{\widetilde{F}}_{pqr}}$, which is precisely the result of applying the fermionization duality. Hence, we have shown that, for Hamiltonians that commute with $\hatG_p$, fermionization agrees with fermion condensation.

We expect that the steps described above can be generalized to a wider class of emergent fermion string operators. This may yield new fermionization dualities and further extend our understanding of fermion condensation at the lattice level.

\section{Ancillary spin-$\frac{1}{2}$ degrees of freedom and evaluation of the fermionic SPT Hamiltonian} \label{ap:spinH}

In section \ref{sec:fSPT}, we presented the construction of a fermionic SPT Hamiltonian obtained by conjugating a trivial fermionic Hamiltonian by $\hatU_{\text{f}}$.  However, $\hatU_{\text{f}}$ was written with an unspecified locally determined configuration dependent sign $\hat{\kappa}$.  $\hat{\kappa}$ is dependent upon the triangulation of the manifold as well as an ordering of operators.  Here we discuss a work around to calculating $\hat{\kappa}$ applicable to arbitrary triangulations.  

Recall that $\hat{\kappa}$ is a consequence of rearranging terms in 
\begin{align}\label{ucirc2}
   \hatU_{\text{b}}=\prod_{\la pqr \ra} \left(\hat{\nu}_{pqr}^{o_{pqr}}  \hatZ_{pq} ^{\hat{n}_{qr}}\right) 
    \prod_{\la pq \ra}\hatX_{pq}^{\hat{n}_{pq}}\prod_{\la pqr \ra}\hatW_{ pqr }^{\hat{n}_{pr}},
\end{align}
to make it manifestly fermionizable.  We will show that by adding ancillary spin-$\frac{1}{2}$ degrees of freedom on triangles and composing $\hatU_{\text{b}}$ with a certain trivial circuit, we can reorganize the expression into a fermionizable operator without accruing a sign. 

To this end, let us add a spin-$\frac{1}{2}$ degree of freedom to each triangle so that in total we have $G$ degrees of freedom at vertices, a spin-$\frac{1}{2}$ degree of freedom at every link, and a spin-$\frac{1}{2}$ at each triangle.  The Pauli X and Pauli Z operators acting on the spin-$\frac{1}{2}$ at $\la pqr \ra$ will be denoted as $\hat{\tau}^x_{pqr}$ and $\hat{\tau}^z_{pqr}$, respectively. 

Next, we compose the operator in (\ref{ucirc2}) with a circuit that acts trivially on the triangle spin-$\frac{1}{2}$ degrees of freedom.  Namely, we compose with
\begin{align}\label{trivialtrianglecircuit}
      1=&\left(\prod_{\la pqr \ra} \left( \hattau^z_{pqr} \right)^{\hat{n}_{qr}} \prod_{\la pqr \ra} \left( \hattau^x_{pqr} \right)^{\hat{n}_{pq}}\right)\\ \nonumber
    \times &\left( \prod_{\la pqr \ra} \left( \hattau^x_{pqr} \right)^{\hat{n}_{pq}}
     \prod_{\la pqr \ra} \left( \hattau^z_{pqr} \right)^{\hat{n}_{qr}}\right)
\end{align}
to obtain
\begin{align}
    \hatU^\tau_\text{b}\equiv &\prod_{\la pqr \ra} \left(\hat{\nu}_{pqr}^{o_{pqr}}  \hatZ_{pq} ^{\hat{n}_{qr}}\right) 
    \prod_{\la pq \ra}\hatX_{pq}^{\hat{n}_{pq}}\prod_{\la pqr \ra}\hatW_{ pqr }^{\hat{n}_{pr}}\\ \nonumber
    \times &\left(\prod_{\la pqr \ra} \left( \hattau^z_{pqr} \right)^{\hat{n}_{qr}} \prod_{\la pqr \ra} \left( \hattau^x_{pqr} \right)^{\hat{n}_{pq}}\right)\\ \nonumber
    \times &\left( \prod_{\la pqr \ra} \left( \hattau^x_{pqr} \right)^{\hat{n}_{pq}}
     \prod_{\la pqr \ra} \left( \hattau^z_{pqr} \right)^{\hat{n}_{qr}}\right).
\end{align}
The circuit in (\ref{trivialtrianglecircuit}) is equal to the identity, and as such, stacking it with $\hatU_{\text{b}}$ certainly does not affect the phase of our system.  However, we can use the anti-commutativity of $\hattau^z_{pqr}$ and $\hattau^x_{pqr}$ to make up for the anti-commutivity of $\hatZ_{pq}$ and $\hatX_{pq}$.  In particular, we arrange the $\hattau^z_{pqr}$ and $\hattau^x_{pqr}$ so as to `dress' the $\hatZ_{pq}$ and $\hatX_{pq}$ and avoid incurring the sign $\hat{\kappa}$.

First, we move $\hattau^z_{pqr}$ operators next to the $\hatZ_{pq}$ operators:
\begin{align}
    \hatU^\tau_{\text{b}}= &\prod_{\la pqr \ra} \left(\hat{\nu}_{pqr}^{o_{pqr}}  \left(\hatZ_{pq} \hattau^z_{pqr} \right)^{\hat{n}_{qr}}\right) 
    \prod_{\la pq \ra}\hatX_{pq}^{\hat{n}_{pq}}\prod_{\la pqr \ra}\hatW_{ pqr }^{\hat{n}_{pr}}\\ \nonumber
    \times &  \prod_{\la pqr \ra} \left( \hattau^x_{pqr} \right)^{\hat{n}_{pq}}\\ \nonumber
    \times &\left( \prod_{\la pqr \ra} \left( \hattau^x_{pqr} \right)^{\hat{n}_{pq}}
     \prod_{\la pqr \ra} \left( \hattau^z_{pqr} \right)^{\hat{n}_{qr}}\right).
\end{align}

Next, we rewrite the product $\prod_{\la pqr \ra} \left( \hattau^x_{pqr} \right)^{\hat{n}_{pq}}$ as a product over edges.  This will allow us to dress the $\hatX_{pq}$ terms appearing in a product over edges.  To the $\la pq \ra$ edge of triangle $\la pqr \ra$ we associate the the operator $\hattau^x_{pqr}$.  This gives
\begin{align}
    \prod_{\la pqr \ra} \left( \hattau^x_{pqr} \right)^{\hat{n}_{pq}}=\prod_{\la pq \ra} \left( \hat{A}_{L_{pq}}\hat{A}_{R_{pq}} \right)^{\hat{n}_{pq}},
\end{align}
where the action of $\hat{A}_{R_{pq}}$, appearing in the formula above, is determined as follows.  If the triangle to the right of $\la pq \ra$ is $\la pqr \ra$, where $p$ and $q$ are the first and second vertices, respectively, then $\hat{A}_{R_{pq}}$ acts as $\hattau^x_{pqr}$. Otherwise, $\hat{A}_{R_{pq}}=\id$.  The action of $\hat{A}_{L_{pq}}$ is defined analogously but we look at the triangle to the left of $\la pq \ra$ instead. 

Now we write
\begin{align}
   \hatU^\tau_{\text{b}}=
   &\prod_{\la pqr \ra} \hat{\nu}_{pqr}^{o_{pqr}} \left(\hatZ_{pq} \hattau^z_{pqr} \right)^{\hat{n}_{qr}} \\ \nonumber
    \times &\prod_{\la pq \ra}\left(\hatX_{pq}\hat{A}_{L_{pq}}\hat{A}_{R_{pq}}\right)^{\hat{n}_{pq}}\prod_{\la pqr \ra}\hatW_{ pqr }^{\hat{n}_{pr}} \\ \nonumber
    \times & \prod_{\la pqr \ra} \left( \hattau^x_{pqr} \right)^{\hat{n}_{pq}}
     \prod_{\la pqr \ra} \left( \hattau^z_{pqr} \right)^{\hat{n}_{qr}}.
\end{align}
One can check that $\hatZ_{pq} \hattau^z_{pqr}$ and $\hatX_{pq}\hat{A}_{L_{pq}}\hat{A}_{R_{pq}}$ commute.  Therefore, we are free to rearrange the $\hatZ_{pq} \hattau^z_{pqr}$ and $\hatX_{pq}\hat{A}_{L_{pq}}\hat{A}_{R_{pq}}$ operators to form a product of $\hatU_{pq}$ (defined in (\ref{def:Ue})) without picking up the sign $\hat{\kappa}$.

Explicitly, rearranging yields
\begin{align}
    \hatU^\tau_{\text{b}}= 
   &\prod_{\la pqr \ra} \hat{\nu}_{pqr}^{o_{pqr}}\\ \nonumber \times &\prod_{\la pq \ra}\left(\hatU_{pq}\hat{B}_{L_{pq}}\hat{B}_{R_{pq}}\hat{A}_{L_{pq}}\hat{A}_{R_{pq}}\right)^{\hat{n}_{pq}}\prod_{\la pqr \ra}\hatW_{ pqr }^{\hat{n}_{pr}} \\ \nonumber
    \times & \prod_{\la pqr \ra} \left( \hattau^x_{pqr} \right)^{\hat{n}_{pq}} 
    \prod_{\la pqr \ra} \left( \hattau^z_{pqr} \right)^{\hat{n}_{qr}},
\end{align}

\noindent where we have introduced $\hat{B}_{L_{pq}}$ and $\hat{B}_{R_{pq}}$. 
\noindent $\hat{B}_{R_{pq}}$ is $\hattau^z_{rpq}$ when the triangle to the right is of the form $\la rpq \ra$, i.e. $p$ is the second vertex in the ordering and $q$ is the third vertex in the ordering. $\hat{B}_{R_{pq}}$ is $\id$ otherwise. $\hat{B}_{L_{pq}}$ is defined analogously, but for the triangle to the left of $\la pq \ra$.

With a choice of spin structure to define the duality, $\hatU^\tau_{\text{b}}$ can be fermionized straightforwardly.  If we let $\hat{S}^\tau_{pq}$ be
$\hatS_{pq}\hat{B}_{L_{pq}}\hat{B}_{R_{pq}}\hat{A}_{L_{pq}}\hat{A}_{R_{pq}}$, then 
\begin{align} \label{def:Ufprime}
    \hatU^\tau_{\text{f}}=
   &\prod_{\la pqr \ra} \hat{\nu}_{pqr}^{o_{pqr}} \\ \nonumber
    \times &\prod_{\la pq \ra}\left(\hat{S}^\tau_{pq}\right)^{\hat{n}_{pq}}\prod_{\la pqr \ra}\left((-1)^{\hat{F}_{ pqr }}\right)^{\hat{n}_{pr}} \\ \nonumber
    \times & \prod_{\la pqr \ra} \left( \hattau^x_{pqr} \right)^{\hat{n}_{pq}} 
    \prod_{\la pqr \ra} \left( \hattau^z_{pqr} \right)^{\hat{n}_{qr}} .
\end{align}

A fermionic SPT Hamiltonian $\hat{H}^\tau_{\text{f}}$ can be formed by conjugating the trivial fermionic Hamiltonian 
\begin{align}
   -\sum_p \hatP_p^{\text{sym}}-\sum_{\la pqr \ra}(-1)^{\hat{F}_{pqr}}-\sum_{\la pqr \ra}\hattau^x_{pqr}
\end{align}

\noindent by $\hatU^\tau_{\text{f}}$. $\hat{H}^\tau_{\text{f}}$ acts identically to $\hat{H}_{\text{f}}$ on the vertex and complex fermion degrees of freedom.  We have simply encoded the sign $\hat{\kappa}$ appearing in $\hat{H}_{\text{f}}$ in the ordering of the Pauli operators of $\hat{H}^\tau_{\text{f}}$.  Indeed, if one were to cancel the triangle Pauli operators in (\ref{def:Ufprime}), one would obtain the sign $\hat{\kappa}$. In the end, we have arrived at an explicit form for a Hamiltonian in the same phase as $\hat{H}_{\text{f}}$ for an arbitrary triangulation of a 2+1D manifold with spin structure.

\section{Supercohomology equivalence relation: trivial fermionic finite depth circuit} \label{ap:trivialcircuit}

In section \ref{sec:superequivalencerelation}, we claimed that the supercohomology data $(n_0,\nu_0)=\left(\delta \beta, (-1)^{\beta \cup \delta \beta}\delta \omega \right)$ corresponds to a trivial fermionic SPT phase.  We prove this claim here by showing that the finite depth circuit $\hatU^{n_0\nu_0}_\text{b}$ can be written in terms of symmetric local unitaries, up to factors of $\hatG_p$.  This implies that the fermionized circuit $\hatU_\text{f}^{n_0\nu_0}$ constructs a trivial SPT ground state from a trivial product state because fermionization respects the $G$-symmetry and maps $\hatG_p$ to the identity. 

Plugging the data $(n_0,\nu_0)$ into the expression for $\hatU_\text{b}$ (\ref{ucirc}) we obtain
\begin{align}
    \hatU^{n_0\nu_0}_\text{b}=&\prod_{\la pqr \ra}(-1)^{\hat{\beta}_{pq}\delta \hat{\beta}_{qr}} \left( \delta \hat{\omega}_{pqr} \right)^{{o}_{pqr}}\\ \nonumber
    \times &\prod_{\la pqr \ra}\hatZ_{pq}^{\delta \hat{\beta}_{qr}} \prod_{\la pq \ra} \hatX_{pq}^{\delta \hat{\beta}_{pq}}\prod_{\la pqr \ra}\hatW_{pqr}^{\delta \beta_{pr}},
\end{align}
with $\delta \hat{\beta}_{pq}$ and $\delta \hat{\omega}_{pqr}$ defined by
\begin{align}
    \delta \hat{\beta}_{pq}|\{ g_t \}\ra&=\delta \beta(g_p,g_q,1)|\{ g_t \}\ra \\
    \delta \hat{\omega}_{pqr}|\{ g_t \}\ra&=\delta \omega(g_p,g_q,g_r,1)|\{ g_t \}\ra.
\end{align}
Now we notice
\begin{align}
    \prod_{\la pqr \ra} (\delta \omega(g_p,&g_q,g_r,1))^{o_{pqr}} = \\ \nonumber &=\prod_{\la pqr \ra} \left(\frac{\omega(g_q,g_r,1)\omega(g_p,g_q,1)}{\omega(g_p,g_r,1)\omega(g_p,g_q,g_r)}\right)^{o_{pqr}} \\ \nonumber
    &=\prod_{\la pqr \ra}\omega(g_p,g_q,g_r)^{-o_{pqr}}.
\end{align}
The last equality follows from treating $\omega(g_s,g_t,1)$ as corresponding to the edge $\la st \ra$ and canceling factors of $\omega(g_s,g_t,1)$ from neighboring triangles. Therefore, 
\begin{align}\label{omegatrianglecancel}
    \prod_{\la pqr \ra} \left(\delta \hat{\omega}_{pqr}\right)^{{o}_{pqr}}=\prod_{\la pqr \ra}\hat{\omega}_{pqr}^{-o_{pqr}}
\end{align}
for $\hat{\omega}_{pqr}$:
\begin{align}
    \hat{\omega}_{pqr}|\{g_t\}\ra=\omega(g_p,g_q,g_r)|\{g_t\}\ra.
\end{align}

Using $\delta \hat{\beta}_{pq}=\hat{\beta}_{pq}+\hat{\beta}_p+\hat{\beta}_q$ (with $\hat{\beta}_{pq}$ and $\hat{\beta}_p$ defined in (\ref{betapq}) and (\ref{betap})) as well as the equality in (\ref{omegatrianglecancel}), $\hatU^{n_0\nu_0}_\text{b}$ becomes
\begin{align}
    \hatU^{n_0\nu_0}_\text{b}=&\prod_{\la pqr \ra}\hat{\omega}_{pqr}^{-o_{pqr}}\prod_{\la pqr \ra}(-1)^{\hat{\beta}_{pq}(\hat{\beta}_{qr}+\hat{\beta}_q+\hat{\beta}_r)} \\ \nonumber
    \times &\prod_{\la pqr \ra}\hatZ_{pq}^{\hat{\beta}_{qr}+\hat{\beta}_q+\hat{\beta}_r} \prod_{\la pq \ra} \hatX_{pq}^{\hat{\beta}_{pq}+\hat{\beta}_p+\hat{\beta}_q}\prod_{\la pqr \ra}\hatW_{pqr}^{\hat{\beta}_{pr}+\hat{\beta}_p+\hat{\beta}_r},
\end{align}
Rearranging and keeping track of the resulting sign we have
\begin{align} \label{intermediatecircuit1}
    \hatU^{n_0\nu_0}_\text{b}=&\prod_{\la pqr \ra}\hat{\omega}_{pqr}^{-o_{pqr}} \\ \nonumber
    \times&\prod_{\la pqr \ra}(-1)^{\hat{\beta}_{pq}(\hat{\beta}_{qr}+\hat{\beta}_q+\hat{\beta}_r)}\prod_{\la pqr \ra}(-1)^{\hat{\beta}_{q}\hat{\beta}_{pq}+\hat{\beta}_{r}\hat{\beta}_{pq}} \\ \nonumber
    \times &\prod_{\la pqr \ra}\hatZ_{pq}^{\hat{\beta}_{qr}} \prod_{\la pq \ra} \hatX_{pq}^{\hat{\beta}_{pq}}\prod_{\la pqr \ra}\hatW_{pqr}^{\hat{\beta}_{pr}} \\ \nonumber
    \times &\prod_{\la pqr \ra}\hatZ_{pq}^{\hat{\beta}_q+\hat{\beta}_r} \prod_{\la pq \ra} \hatX_{pq}^{\hat{\beta}_p+\hat{\beta}_q}\prod_{\la pqr \ra}\hatW_{pqr}^{\hat{\beta}_p+\hat{\beta}_r}.
\end{align}

Next we write $\prod_{\la pq \ra} \hatX_{pq}^{\hat{\beta}_p+\hat{\beta}_q}$ in (\ref{intermediatecircuit1}) as a product over vertices:
\begin{align}
    \prod_{\la pq \ra} \hatX_{pq}^{\hat{\beta}_p+\hat{\beta}_q}=\prod_p\left( \prod_{\la st \ra \ni p}\hatX_{st}^{\hat{\beta}_p} \right).
\end{align}
Further, one can check that 
\begin{align}\label{Gptrivialcircuit}
    \left[\prod_p\left( \prod_{\la st \ra \ni p}\hatX_{st}^{\hat{\beta}_p} \right) \right]\prod_{\la pqr \ra}\hatW_{pqr}^{\hat{\beta}_p}=\prod_p\hat{G}_p^{\hat{\beta}_p}.
\end{align}
Hence, substituting (\ref{Gptrivialcircuit}) in (\ref{intermediatecircuit1}) and canceling signs we are left with
\begin{align} \label{intermediatecircuit2}
    \hatU^{n_0\nu_0}_\text{b}=&\prod_{\la pqr \ra}\hat{\omega}_{pqr}^{-o_{pqr}}\prod_{\la pqr \ra}(-1)^{\hat{\beta}_{pq}\hat{\beta}_{qr}} \\ \nonumber
    \times &\prod_{\la pqr \ra}\hatZ_{pq}^{\hat{\beta}_{qr}} \prod_{\la pq \ra} \hatX_{pq}^{\hat{\beta}_{pq}}\prod_{\la pqr \ra}\hatW_{pqr}^{\hat{\beta}_{pr}} \\ \nonumber
    \times &\prod_{\la pqr \ra}\hatZ_{pq}^{\hat{\beta}_q+\hat{\beta}_r} \prod_{\la pqr \ra}\hatW_{pqr}^{\hat{\beta}_r} \prod_p\hat{G}_p^{\hat{\beta}_p}.
\end{align}
The product $\prod_{\la pqr \ra}\hatZ_{pq}^{\hat{\beta}_q+\hat{\beta}_r}\prod_{\la pqr \ra}\hatW_{pqr}^{\hat{\beta}_r}$ above expressed in terms of $\hatZ_{pq}$ operators is
\begin{align} \label{trivialcircuitZidentity}
    \prod_{\la pqr \ra}\hatZ_{pq}^{\hat{\beta}_q+\hat{\beta}_r}\hatZ_{pq}^{\hat{\beta}_{r}}\hatZ_{qr}^{\hat{\beta}_{r}}\hatZ_{pr}^{\hat{\beta}_{r}}&=\prod_{\la pqr \ra}\hatZ_{pq}^{\hat{\beta}_q}\hatZ_{qr}^{\hat{\beta}_{r}}\hatZ_{pr}^{\hat{\beta}_{r}}\\ \nonumber 
    &=\prod_{\la pq \ra}\hatZ_{pq}^{\hat{\beta}_{q}}\hatZ_{pq}^{\hat{\beta}_{q}}=1.
\end{align}

With (\ref{trivialcircuitZidentity}) we finally have that
\begin{align} \label{trivialcircuit2}
    \hatU^{n_0\nu_0}_\text{b}=&\prod_{\la pqr \ra}\hat{\omega}_{pqr}^{-o_{pqr}}\prod_{\la pqr \ra}(-1)^{\hat{\beta}_{pq}\hat{\beta}_{qr}} \\ \nonumber
    \times &\prod_{\la pqr \ra}\hatZ_{pq}^{\hat{\beta}_{qr}} \prod_{\la pq \ra} \hatX_{pq}^{\hat{\beta}_{pq}}\prod_{\la pqr \ra}\hatW_{pqr}^{\hat{\beta}_{pr}} \prod_p\hat{G}_p^{\hat{\beta}_p},
\end{align}
as claimed in section \ref{sec:superequivalencerelation}.

The homogeneity of $\omega$ and $\beta$ guarantees that the local unitaries (not including $\prod_p\hat{G}_p^{\hat{\beta}_p}$) in (\ref{trivialcircuit2}) are symmetric.  Fermionization commutes with the global symmetry operator and it takes $\prod_p\hat{G}_p^{\hat{\beta}_p}$ to $1$.  Therefore, the resulting circuit $\hatU_\text{f}^{n_0\nu_0}$ is a finite depth circuit built from symmetric local unitaries, and consequently, creates a trivial fermionic SPT from a trivial product state.

\section{Nonexistence of symmetric quantum circuits for `beyond supercohomology' phases} \label{beyondsuper}

\begin{figure*}[t]
\centering
\includegraphics[scale=.5,trim={0cm 6.2cm 0cm 6cm},clip]{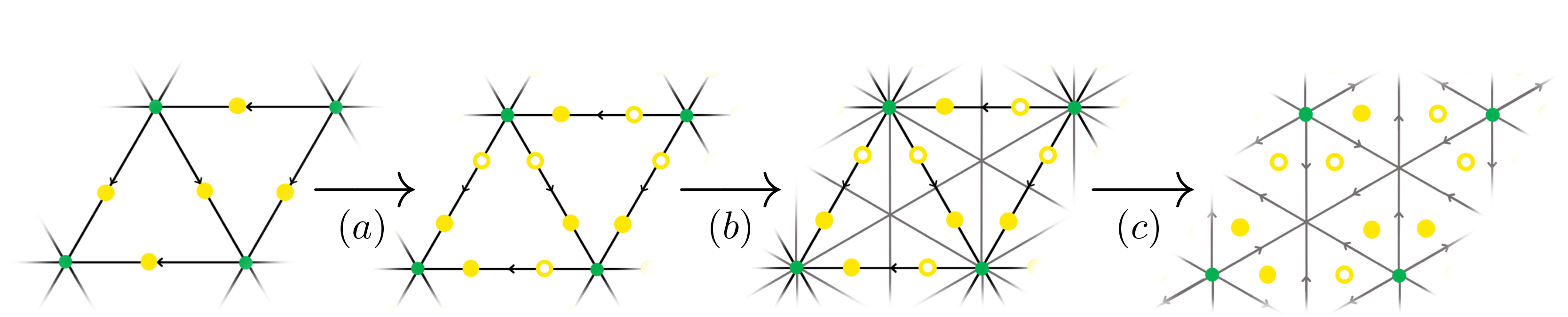}
\caption{(Far left) The beyond supercohomology models in Ref. [\onlinecite{Tarantino16}] have $G$ degrees of freedom (green dots) at vertices and a single complex fermion degree of freedom (yellow dots) at each link. (a) We add an additional complex fermion degree of freedom (hollow yellow circle) to each link and modify the Hamiltonian by adding a term that enforces zero fermion occupancy at each of these additional sites. (b) We add links to the lattice to form the barycentric subdivision of our original triangulation. (c) The links of the original lattice are removed in the bulk and we associate one complex fermion to each triangle of the resulting lattice. The branching structure is determined by the branching structure of the original lattice.  Importantly, the final Hilbert space has a single complex fermion degree of freedom for each triangle. Note that the $G$ degrees of freedom remain in place.}
\label{fig:beyondlattices}
\end{figure*}

Here, we present an argument that the ground states of the beyond supercohomology phases cannot be constructed from a trivial product state using a globally symmetric finite depth quantum circuit.  
It has been shown, using a spacetime formulation, that the shadow models for beyond supercohomology phases are symmetry enriched toric code phases with the property that certain global symmetries transform $e$ excitations into $m$ excitations and vice-versa [\onlinecite{Bhardwaj,Cheng16}]. We begin by giving an independent argument that this property has to hold for the lattice Hamiltonian shadow models associated to a beyond supercohomology SPT.  We also show how existing lattice Hamiltonian models for beyond supercohomology phases, namely those of Ref. [\onlinecite{Tarantino16}], can be slightly modified so as to be amenable to our bosonization procedure.  This argument can be generalized to show that any lattice Hamiltonian SPT model can be modified so as to be bosonizable via our procedure.  We then demonstrate that this property of a global symmetry operator exchanging $e$ and $m$ in the shadow model, together with the assumption of the existence of a globally symmetric finite depth circuit disentangling the ground state of the original fermionic SPT, lead to a contradiction. 


The beyond supercohomology models constructed in Ref. [\onlinecite{Tarantino16}] are not immediately bosonizable, because the fermionic degrees of freedom live on the links instead of at the centers of triangles.  This can be remedied as follows. First, we introduce a second complex fermion degree of freedom at each link and add a term to the Hamiltonian that energetically favors zero fermion occupancy at each of the new complex fermion degrees of freedom. Functionally, we have stacked an atomic insulator onto the original system, so it remains in the same phase. Then, we perform a barycentric subdivision of the lattice, remove the original lattice, and associate the two complex fermions per link to different triangles.  This procedure is illustrated in FIG. \ref{fig:beyondlattices}. In this process, we have not changed the dynamics of the system, and in the end, we have a beyond supercohomology model which lives in the same Hilbert space as the supercohomology models constructed in this paper, and hence is readily bosonizable. 

Now we argue that bosonizing these beyond supercohomology models gives symmetry enriched toric code models in which global symmetries convert $e$ excitations into $m$ excitations. Key to this argument is the following property of the symmetry action in beyond supercohomology models with symmetry defects.  For beyond supercohomology phases with symmetry $G$, there is an additional piece of data relative to supercohomology phases - a homomorphism \unexpanded{$\sigma: G \to \Z_2$}. According to Refs. [\onlinecite{Cheng16, Tarantino16, Potter16}], the effective symmetry action near a fermion parity defect is fermion parity odd when acting with on-site symmetry operator $\hatV(g)$ representing $g \in G$ for which $\sigma(g)$ is non-trivial.  

We analyze this effect in our beyond supercohomology models by inserting a pair of fermion parity defects at well separated vertices $a$ and $b$.  Loosely, we create the pair of fermion parity defects by choosing a path $\Gamma$ connecting $a$ and $b$ and modifying hopping operators in the Hamiltonian corresponding to links in $\Gamma$. Heuristically, the modification of the Hamiltonian makes it so that when a fermion moves around one of the fermion parity defects, it picks up an extra $-1$ sign.

To include fermion parity defects, we first write the beyond supercohomology model $\hat{H}^\text{b.s.}$ as a sum of local terms
\begin{align}\label{bshamiltonian}
    \hat{H}^\text{b.s.}=\sum_j \hath^\text{b.s.}_j,
\end{align}
with each $\hath^\text{b.s.}_j$ supported on the spatially bounded region $R_j$.  Now, we define
\begin{align}
  \hat{\mathcal{P}}_p = \frac{1}{2} \left( \id + \prod_{\substack{\la tq \ra \\ t=p}} {\hat S}_{tq}  \prod_{\substack{\la qt \ra \\ t=p}} {\hat S}_{qt}  \prod_{\substack{\la tqr \ra \\ t=p}} (-1)^{\hat{F}_{tqr}} \prod_{\substack{\la qrt \ra \\ t=p}} (-1)^{\hat{F}_{qrt}} \right)
\end{align}
and
\begin{align}
    \hat{\mathcal{P}}_{R_j}=\prod_{p \subset R_j} \hat{\mathcal{P}}_p,
\end{align}
and write
\begin{align}
    \hat{H'}^\text{b.s.} = \sum_j \hat{\mathcal{P}}_{R_j} \hath^\text{b.s.}_j \hat{\mathcal{P}}_{R_j} -\sum_p \hat{\mathcal{P}}_p.
\end{align}
$\hat{\mathcal{P}}_p$ is identically equal to $1$, so $\hat{H'}^\text{b.s.}$ is equivalent to $\hat{H}^\text{b.s.}$.  

Next, let us modify this Hamiltonian to insert a pair of defects at two well separated vertices $a$ and $b$.  Let $\Gamma$ be a path of links connecting $a$ and $b$, and let $\Gamma_{pq}$ be the indicator function
\begin{align}
    \Gamma_{pq}= \begin{cases} 
      1 & \text{if }\la pq \ra \in \Gamma \\
      0 & \text{otherwise}.
   \end{cases}
\end{align}
Now we write each local term $\hath^\text{b.s.}_j$ explicitly as a linear combination of products of $\hatS_{pq}$ and $(-1)^{\hat{F}_{pqr}}$, and we make the replacement:

\begin{align}
    \hath^\text{b.s.}_j (\hatS_{pq}, (&-1)^{\hat{F}_{pqr}}) \nonumber \\
    &\longrightarrow  \hath^\text{b.s.}_{j,\Gamma}\equiv \hath^\text{b.s.}_j((-1)^{\Gamma_{pq}}\hatS_{pq}, (-1)^{\hat{F}_{pqr}}).
\end{align}
Making the same replacement in $\hat{\mathcal{P}}_{j}$ yields a new Hamiltonian $\hat{H}^\text{b.s.}_\Gamma$.

Notice that replacing $\hatS_{pq}$ with $(-1)^{\Gamma_{pq}}\hatS_{pq}$ in the expression defining $\hat{\mathcal{P}}_R$ yields $0$ if $a$ or $b$ is contained in the region $R$.  As a consequence, Hamiltonian terms whose support contains the defects are removed from the Hamiltonian. 

Now we bosonize $\hat{H}^\text{b.s.}$ and $\hat{H}^\text{b.s.}_\Gamma$. This yields
\begin{align}
    \hat{H}^\text{b.s.}_\text{b} \equiv \hat{H'}^\text{b.s.}(\hatU_{pq},\hatW_{pqr}) 
\end{align}
and
\begin{align}
    \hat{H}^\text{b.s.}_{\text{b},\Gamma} \equiv \hat{H'}^\text{b.s.}((-1)^{\Gamma_{pq}}\hatU_{pq},\hatW_{pqr}).
\end{align}
The operators $\hat{\mathcal{P}}_R$ become projectors onto the $\hatG_p=1$ subspace for all $p \subset R$. As a consequence, the ground states of $\hat{H}^\text{b.s.}_\text{b}$ are in the $\hatG_p=1$ subspace, and away from the defects, the ground states of $\hat{H}^\text{b.s.}_{\text{b},\Gamma}$ are in the $\hatG_p=1$ subspace. We also note that, by construction, the bosonized Hamiltonians commute with $\hatG_p$ for all $p$, and thus, $\hatG_p$ can be interpreted as a small loop of an emergent fermion string operator around the vertex $p$. 

We can obtain a ground state of $\hat{H}^\text{b.s.}_{\text{b},\Gamma}$ by applying a certain string operator to a ground state of $\hat{H}^\text{b.s.}_\text{b}$.  In particular, the string operator $\hatU_\Gamma=\prod_{\la pq \ra \in \Gamma}\hatZ_{pq}$ does the job. Explicitly,
\begin{align}
    \hat{H}_{\text{b},\Gamma}^\text{b.s.}\hatU_\Gamma|\Psi_\text{b}^\text{b.s.}\ra=\hatU_\Gamma \hat{H}_\text{b}^\text{b.s.} |\Psi_\text{b}^\text{b.s.}\ra= E_\text{min}\hatU_\Gamma|\Psi_\text{b}^\text{b.s.}\ra,
\end{align}
where in the first equality, we used that $\hatU_\Gamma$ anticommutes with $\hatU_{pq}$ for $\la pq \ra \in \Gamma$.  Applying $\hatG_p$ to this ground state at either endpoint of $\Gamma$, we find that moving an emergent fermion around the endpoint produces a minus sign:
\begin{align}
    \hatG_{a/b} \hatU_\Gamma|\Psi_\text{b}^\text{b.s.}\ra = - \hatU_\Gamma |\Psi_\text{b}^\text{b.s.}\ra.
\end{align}
Hence, $\hatU_\Gamma$ creates either $e$ excitations or $m$ excitations at its endpoints.

Since ground states of $\hat{H}_{\text{b},\Gamma}^\text{b.s.}$ have a pair of $e$ or $m$ excitations relative to ground states of $\hat{H}_\text{b}^\text{b.s.}$, we can determine the effective symmetry action on a pair of $e$ or $m$ excitations at $a$ and $b$ by bosonizing the effective symmetry action on the fermionic state with fermion parity defects at $a$ and $b$.  For $g$ such that $\sigma(g)$ is non-trivial, the effective symmetry action on the state with fermion parity defects splits into a fermion parity odd operator associated to each defect.  Expressing the symmetry action in terms of local fermion parity even operators, so that it may be bosonized, requires a string of hopping operators and fermion parity operators connecting the two fermion parity odd operators. Hence, bosonization yields an effective symmetry action that includes an $em$ string connecting $a$ and $b$. This $em$ string converts an $e$ ($m$) string into an $m$ ($e$) string, and we have recovered the expected effective symmetry action in the bosonic shadow theory of a beyond supercohomology phase at the lattice level.  

Finally, to demonstrate that beyond supercohomology SPT ground states cannot be constructed by applying a symmetric finite depth quantum circuit to a trivial product state, we assume that this is indeed possible and derive a contradiction.  If such a circuit $\hatU^\text{b.s.}$ exists, then in bosonizing the circuit we obtain a circuit $\hatU_\text{b}^\text{b.s.}$ which is globally symmetric up to factors of $\hatG_p$. (The $\hatG_p$ generate the kernel of the fermionization duality.) Explicitly,
\begin{align}
    \hat{V}(g)\hatU_\text{b}^\text{b.s.}=\hatU_\text{b}^\text{b.s.} f_g(\hatG_p) \hat{V}(g)
\end{align}
for some $g$ dependent function $f_g$ of the $\hatG_p$.  In what follows, let us assume that $\hatU_\Gamma$ creates a pair of $e$ excitations. An analogous argument can be made if $\hatU_\Gamma$ instead creates $m$ excitations.  Writing a ground state of the toric code with a pair of $e$ excitations as $|\Psi^{ee}_\text{t.c.}\ra$, the ground state of $\hat{H}_\text{b}^\text{b.s.}$ with $e$ excitations is $\hatU_\text{b}^\text{b.s.}|\Psi^{ee}_\text{t.c.}\ra$.  We then compute the effective symmetry action on a pair of $e$ excitations:
\begin{align}
    \hat{V}(g)\hatU_\text{b}^\text{b.s.}|\Psi^{ee}_\text{t.c.}\ra&=\hatU_\text{b}^\text{b.s.} f_g(\hatG_p) \hat{V}(g)|\Psi^{ee}_\text{t.c.}\ra \nonumber \\
    &=\hatU_\text{b}^\text{b.s.} f_g(\hatG_p)|\Psi^{ee}_\text{t.c.}\ra.
\end{align}
As we have argued (at least for the anyons created by $\hatU_\Gamma$), the effective symmetry action should convert the $e$ excitations into $m$ excitations.  However, $f_g$ is a function of small emergent fermion loop operators.  Loops of $em$ string are unable to transform $e$ excitation into $m$ excitations. This contradicts the expected affect of global symmetry action in the bosonic shadow model for beyond supercohomology phases.

We have now shown that the ground state of a specific beyond supercohomology model cannot be constructed by applying a symmetric quantum circuit to a trivial product state. This is sufficient to argue that no ground state of any supercohomology model can by constructed from a trivial product state with a symmetric quantum circuit.  This is because, by definition, ground states of two beyond supercohomology phases can be related by a quantum circuit built of symmetric local unitaries.

While we have shown that a symmetric quantum circuit is incapable of building the ground state of a beyond supercohomology phase from a trivial product state, it would be interesting to identify a quantum circuit, albeit not symmetric, which is capable of creating the ground state of a beyond supercohomology SPT from a trivial product state.  We leave this for future work.

\end{appendix}

\bibliographystyle{unsrt}
\bibliography{references}

\end{document}